\documentclass[aps,prxquantum,reprint,twocolumn,superscriptaddress,floatfix,nofootinbib,longbibliography]{revtex4-1}
\usepackage{graphicx,amsmath,amsfonts,amssymb,amsthm,xr}
\usepackage{epsfig,amsmath,amssymb,color,dsfont,upgreek,physics}
\usepackage{mathrsfs}
\usepackage{mathtools}
\usepackage{bbold}
\usepackage{comment}
\usepackage{float}

\usepackage[bookmarks=true,colorlinks,linkcolor=OrangeRed,urlcolor=NavyBlue,citecolor=RoyalBlue]{hyperref}
\usepackage[dvipsnames]{xcolor}

\definecolor{mygold}{rgb}{0.93,0.69,0.13}

\definecolor{mypurple}{rgb}{0.49,0.18,0.56}

\definecolor{mygreen}{rgb}{0,0.5,0}

\definecolor{mygreen}{rgb}{0,0.5,0}
\definecolor{myred}{rgb}{0.7,0,0}
\definecolor{myblue}{rgb}{0,0,0.5}

\begin{document}
\title{Tuning the Topological $\theta$-Angle in Cold-Atom Quantum Simulators of Gauge Theories}
\author{Jad C.~Halimeh}
\email{jad.halimeh@physik.lmu.de}
\affiliation{Department of Physics and Arnold Sommerfeld Center for Theoretical Physics (ASC), Ludwig-Maximilians-Universit\"at M\"unchen, Theresienstra\ss e 37, D-80333 M\"unchen, Germany}
\affiliation{Munich Center for Quantum Science and Technology (MCQST), Schellingstra\ss e 4, D-80799 M\"unchen, Germany}
\author{Ian P.~McCulloch}
\affiliation{School of Mathematics and Physics, The University of Queensland, St. Lucia, QLD 4072, Australia}
\author{Bing Yang}
\affiliation{Department of Physics, Southern University of Science and Technology, Shenzhen 518055, China}
\author{Philipp Hauke}
\email{philipp.hauke@unitn.it}
\affiliation{INO-CNR BEC Center and Department of Physics, University of Trento, Via Sommarive 14, I-38123 Trento, Italy}
\affiliation{INFN-TIFPA, Trento Institute for Fundamental Physics and Applications, Trento, Italy}

\begin{abstract}
The topological $\theta$-angle in gauge theories engenders a series of fundamental phenomena, including violations of charge-parity (CP) symmetry, dynamical topological transitions, and confinement--deconfinement transitions. At the same time, it poses major challenges for theoretical studies, as it implies a sign problem in numerical simulations. Analog quantum simulators open the promising prospect of treating quantum many-body systems with such topological terms, but, contrary to their digital counterparts, they have not yet demonstrated the capacity to control the $\theta$-angle. Here, we demonstrate how a tunable topological $\theta$-term can be added to a prototype theory with $\mathrm{U}(1)$ gauge symmetry, a discretized version of quantum electrodynamics in one spatial dimension. As we show, the model can be realized experimentally in a single-species Bose--Hubbard model in an optical superlattice with three different spatial periods, thus requiring only standard experimental resources. Through numerical calculations obtained from the time-dependent density matrix renormalization group method and exact diagonalization, we benchmark the model system, and illustrate how salient effects due to the $\theta$-term can be observed. These include charge confinement, the weakening of quantum many-body scarring, as well as the disappearance of Coleman's phase transition due to explicit breaking of CP symmetry. This work opens the door towards studying the rich physics of topological gauge-theory terms in large-scale cold-atom quantum simulators.
\end{abstract}

\date{\today}
\maketitle

\tableofcontents
\section{Introduction}
\begin{figure}[t!]
    \centering
    \includegraphics[width=\columnwidth]{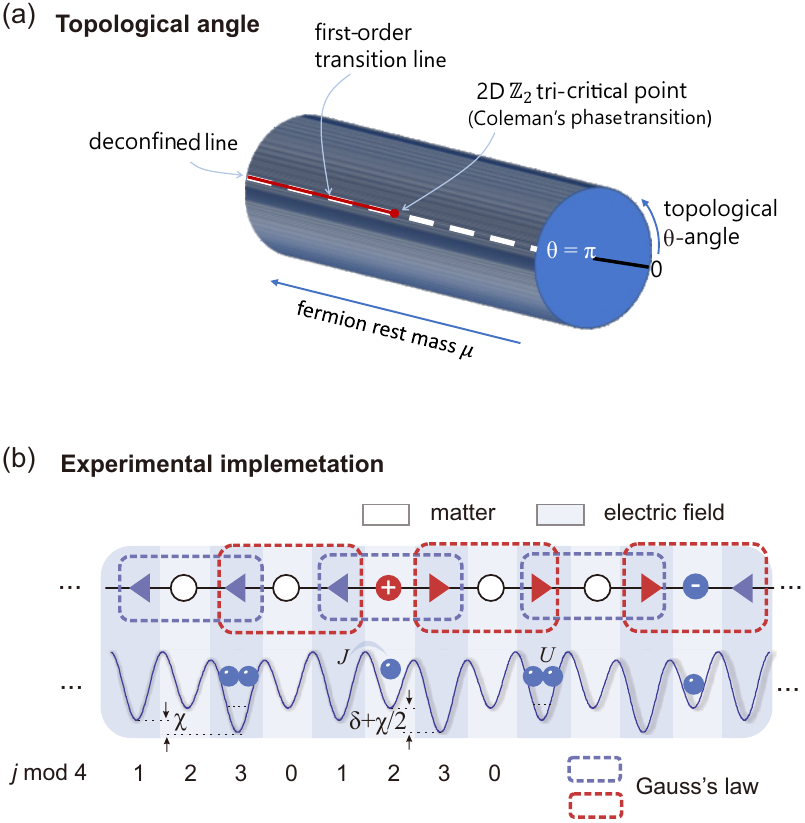}
    \caption{(Color online). (a) Topological $\theta$-angle. The ground state of quantum electrodynamics (QED) is periodic under the $\theta$-angle \cite{Adam1999,Tong_LectureNotes}. The quantum link model (QLM) formulation of the $\mathrm{U}(1)$ gauge theory loses this periodicity, but retains key features, in particular Coleman’s phase transition and the deconfined line at $\theta=\pi$.
    (b) The $\mathrm{U}(1)$ QLM can be implemented in an optical superlattice with single-species bosonic atoms, where shallow sites are associated to charged matter and deep sites to gauge fields. On-site interaction $U$, a staggered potential $\delta$, and a tilted potential $\Delta$ (not shown) provide energy penalties, which in interplay with nearest-neighbor hopping $J$ generate the desired QLM in second-order degenerate perturbation theory and enforce Gauss's law on all triple-wells centered around a matter site \cite{Yang2020,Halimeh2020e}. The topological $\theta$-angle can be realized by the straightforward addition of a superlattice with spatial period 4 and of amplitude $\chi=g^2(\theta-\pi)/(2\pi)$, where $g$ is the gauge coupling.}
    \label{fig:concept}
\end{figure}

Synthetic quantum systems \cite{Bloch2008,Georgescu_review}, i.e., well-controlled quantum many-body systems based on cold atoms, trapped ions, superconducting qubits, and photonic devices, hold the promise of a new era of scientific discovery \cite{Alexeev_review,klco2021standard}.
A particularly attractive arena is given by fundamental questions in nuclear and high-energy physics \cite{Pasquans_review,Dalmonte_review,Zohar_review,aidelsburger2021cold,Zohar_NewReview,Bauer_review}, such as the dynamics of quantum many-body systems in the presence of a topological $\theta$-angle. 
The $\theta$-angle naturally appears in the Lagrangian of certain gauge theories from the very topological nature of the vacuum, depending on the gauge group and dimensionality. 
In particular, the Lagrangian of the strong force---quantum chromodynamics (QCD) in four-dimensional spacetime---allows for a $\theta$-term \cite{tHooft1976,Jackiw1976,Callan1979}.  
Experimentally, the strength of this term has, however, been found to lie within vanishingly small bounds \cite{Chupp_review}. 
This apparent fine-tuning of nature has risen to prominence under the name of the strong CP problem (with CP standing for charge--parity symmetry), for which intriguing solutions have been proposed but not yet experimentally corroborated, such as the existence of an additional field (the axion) that couples to the $\theta$-term \cite{Peccei1977,Weinberg1978,Wilczek1978,Graham2015}. 
In practical terms, topological terms such as the $\theta$-angle imply a significant hurdle for theory investigations, as they introduce a sign problem in numerical simulations based on the Euclidean path integral formulations \cite{Unsal2012}. 
Interestingly, a less involved gauge theory---quantum electrodynamics in one spatial dimension (QED)---equally hosts a $\theta$-angle, and for this reason is often used as a prototype model for studying the effect of this topological term \cite{Byrnes2002,Buyens2014,Shimizu2014,Buyens2016}. 
The topological $\theta$-angle in $1+1$D QED gives rise to rich equilibrium and far-from-equilibrium physics [see Fig.~\ref{fig:concept}(a)], including a deconfinement--confinement transition \cite{Buyens2016,Surace2020}, the so-called Coleman's quantum phase transition in the ground state \cite{Coleman1976}, and dynamical topological phase transitions appearing after a rapid quench of the $\theta$-term \cite{Zache2019,Kharzeev2008,Fukushima2008,Kharzeev2016,Koch2017,Kharzeev2020}. 
However, getting control of this fundamental term, and thus gaining access to the rich physics it encapsulates, is an open challenge in analog quantum simulation. 

In this work, we show how a tunable topological $\theta$-term can be engineered in a cold-atom setup that has recently quantum-simulated a gauge theory of $71$ lattice sites governed by the $\mathrm{U}(1)$ gauge group that underlies QED \cite{Yang2020,Zhou2021}. 
As we show, the $\theta$-term can be realized experimentally by a surprisingly simple addition, namely an optical superlattice with a period twice that of the one already employed in the demonstrated setup; see Fig.~\ref{fig:concept}(b). 
Through numerical benchmarks using the time-dependent density matrix renormalization group method ($t$-DMRG) \cite{Vidal2004,White2004,Daley2004}, we show how this term lifts Coleman’s phase transition that has been experimentally observed in Ref.~\cite{Yang2020} (as well as in other setups \cite{Bernien2017,Kokail2019}), and how it leads to confinement of charged particles and gauge fields. 
A striking signature illustrated by our numerics is the destruction of many-body scarring due to confinement.
Detailed experimental considerations illustrate the feasibility of the approach.
Our proposal thus opens the door to studying salient effects of the topological $\theta$-angle in large engineered quantum systems. 

The rest of this work is organized as follows: In Sec.~\ref{sec:model}, we review the spin-$1/2$ quantum link formulation of $1+1$D QED on a lattice, known as the $\mathrm{U}(1)$ quantum link model, and discuss its salient features and relevance in the context of condensed matter and particle physics. In Sec.~\ref{sec:experimentalsetup}, we introduce the  experimental cold-atom setup on which we map the $\mathrm{U}(1)$ quantum link model, and discuss how the relevant initial states can be prepared in an experiment. Our main numerical results obtained from the time-dependent density matrix renormalization group method are then presented in Sec.~\ref{sec:numerics} for time evolution under adiabatic ramps and quench dynamics, allowing us to dynamically probe the deconfinement--confinement transition in the $\mathrm{U}(1)$ quantum link model. We conclude and provide future outlook in Sec.~\ref{sec:conc}. We supplement our work with Appendix~\ref{app:LSM}, which discusses in detail the quantum link formulation of $1+1$D QED on a lattice, in addition to Appendix~\ref{app:degenPT}, which provides details on the mapping of the $\mathrm{U}(1)$ quantum link model onto the bosonic system that we propose for its quantum simulation, as well as Appendix~\ref{app:InverseRamp}, which provides supporting numerical results.

\section{Model}\label{sec:model}
Fully fledged QCD in $3+1$D is still beyond the abilities of current quantum simulators. However, 
existing technology can already simulate simpler gauge theories \cite{Martinez2016,Muschik2017,Bernien2017,Klco2018,Kokail2019,Schweizer2019,Goerg2019,Mil2020,Klco2020,Yang2020,Zhou2021,Nguyen2021,Wang2021,Mildenberger2022}. 
Specifically, QED in one spatial dimension, also known as the massive Schwinger model, becomes interesting in our context as it shares with $3+1$D QCD a nontrivial topological vacuum structure, a chiral anomaly, and a CP-odd $\theta$-term \cite{Coleman1976,Hamer1982,Shimizu2014,Tong_LectureNotes}. These features make the massive Schwinger model a prototype model for $3+1$D QCD for studying effects of CP violation and a topological $\theta$-angle. 

In the temporal gauge, the Hamiltonian of the massive Schwinger model with a topological $\theta$-term can be written as \cite{Coleman1976,Byrnes2002,Tong_LectureNotes}
\begin{align}\label{eq:Hcontinuum}
\hat{H}_\mathrm{QED} = \int dx \left[  \hat{\psi}_x^\dagger \gamma^0 \left(i \gamma^1 \hat{D}_x + \mu\right) \hat{\psi}_x + \frac{1}{2}\hat{E}^2_x + \frac{g\theta} {2\pi} \hat{E}_x\right],
\end{align}
where $\hat{E}$ is the electric field, $g$ the dimensionful gauge coupling, $\hat{\psi}$ are two-component fermion operators, and $\gamma^{0/1}$ are the Dirac matrices in $1+1$D.
This Hamiltonian describes from left to right the kinetic energy of the fermions (which couples to the gauge vector potential $\hat{A}_x$ via the covariant derivative $\hat{D}_x= \partial_x + ig \hat{A}_x$), the fermion rest mass $\mu$, the electric field energy, and the topological $\theta$-term. As can be seen by the form of the latter, it is equivalent in this model to a homogeneous background field $E_\text{bg}=g\theta/(2\pi)$.

In order to make this model amenable for quantum simulators such as cold atoms in optical lattices consisting of discrete degrees of freedom, we employ here the quantum link model (QLM) formalism \cite{Chandrasekharan1997,Wiese_review,Yang2016}. 
This framework considers a lattice discretization, for which we take staggered fermions \cite{Kogut1975}, as well as a replacement of gauge fields by spin operators. 
Details of the following derivation of the Hamiltonian can be found in Appendix~\ref{app:LSM}.
The QLM lattice version of the Schwinger model with $L_\mathrm{m}$ matter sites is 
\begin{align}
    \label{eq:HQLM}
    \nonumber
    \hat{H}=&-\frac{\kappa}{2a}\sum_{\ell=1}^{L_\mathrm{m}-1}\big(\hat{\psi}^\dagger_\ell\hat{s}^+_{\ell,\ell+1}\hat{\psi}_{\ell+1}+\text{H.c.}\big)+{\mu}\sum_{\ell=1}^{L_\mathrm{m}-1}\hat{\psi}^\dagger_\ell\hat{\psi}_\ell\\
    &+\frac{ag^2}{2}\sum_{\ell=1}^{L_\mathrm{m}-1}\big(\hat{s}^z_{\ell,\ell+1}\big)^2 -a\chi\sum_{\ell=1}^{L_\mathrm{m}-1}(-1)^\ell\hat{s}^z_{\ell,\ell+1}\,.
\end{align}
In this model, matter fields live on sites $\ell$, which alternatingly represent the particle and anti-particle component of the Dirac spinor. The number of sites is $L_\mathrm{m}$. The gauge (electric) field lives on the links between sites $\ell$ and $\ell+1$. 
The fermion kinetic energy is controlled by $\kappa$, and in what follows we set the lattice spacing $a$ to unity. 
The QLM formalism has replaced the typical parallel transporter $\hat{U}_{\ell,\ell+1}=e^{i g \hat{A}_{\ell,\ell+1}}\to \hat{s}^+_{\ell,\ell+1}$ and the electric field $\hat{E}_{\ell,\ell+1}\to g\hat{s}^z_{\ell,\ell+1}$, where $\hat{s}_{\ell,\ell+1}$ are spin-$S$ operators.  
The QLM retains canonical commutation relations between $\hat{E}_{\ell,\ell+1}$ and $\hat{U}_{\ell,\ell+1}$, and controllably recovers QED in the limits of $S\to\infty$, large volume, and small lattice spacing \cite{Buyens2017,Banuls2018,Banuls2020,Zache2021achieving,Halimeh2021achieving}. 
Even more, it shares many key features with QED already for small spin representations \cite{Wiese_review,Banerjee2012,Yang2016,Surace2020}. 
Most relevant for our purposes, the QLM formalism for half-integer $S$ naturally includes a topological $\theta$-angle of $\pi$ \cite{Surace2020,Zache2021achieving}. The last term in Eq.~\eqref{eq:HQLM} accounts for this in the parameter $\chi=g^2(\theta-\pi)/(2\pi)$, which describes the deviation of $\theta$ from $\pi$. 
In what follows, we choose $S=1/2$, which is sufficient for obtaining the salient features we are interested in here and at the same time is most convenient for experimental implementation.

To connect to recent implementations with ultracold bosonic single-species atoms, we further perform a particle-hole transformation on odd matter sites ($\hat{\psi}_\ell\leftrightarrow \hat{\psi}_\ell^\dagger$ for $\ell$ odd) followed by a Jordan--Wigner transformation of fermionic matter $\hat{\psi}_\ell$ to hard-core bosons or equivalently Pauli operators $\hat{\sigma}_\ell$ \cite{Yang2016}. The resulting Hamiltonian reads (see Appendix~\ref{app:LSM})
\begin{align}\nonumber
    \hat{H}=&-\frac{\kappa}{2}\sum_{\ell=1}^{L_\mathrm{m}-1}\big(\hat{\sigma}^-_\ell\hat{s}^+_{\ell,\ell+1}\hat{\sigma}^-_{\ell+1}+\text{H.c.}\big)\\\label{eq:H}
    &+\frac{\mu}{2}\sum_{\ell=1}^{L_\mathrm{m}}\hat{\sigma}^z_\ell-\chi\sum_{\ell=1}^{L_\mathrm{m}-1}(-1)^\ell\hat{s}^z_{\ell,\ell+1}.
\end{align}

QED, as given in the Hamiltonian of Eq.~\eqref{eq:Hcontinuum}, is a gauge theory with $\mathrm{U}(1)$ gauge symmetry encoding Gauss's law. 
The Hamiltonian in Eq.~\eqref{eq:H} retains that $\mathrm{U}(1)$ gauge symmetry. It is generated by the operator
\begin{align}
    \label{eq:Gj}
    \hat{G}_\ell=(-1)^{\ell+1}\bigg[\hat{s}^z_{\ell,\ell+1}+\hat{s}^z_{\ell-1,\ell}+\frac{\hat{\sigma}^z_\ell+\mathds{1}}{2}\bigg],
\end{align}
which can be viewed as a discretized version of Gauss's law. The $\mathrm{U}(1)$ gauge symmetry of the Hamiltonian~\eqref{eq:H} is encapsulated in the commutation relations $\big[\hat{H},\hat{G}_\ell\big]=0,\,\forall \ell$, and conservation of $\hat{G}_\ell$. We will work in the \textit{physical sector} of states $\ket{\phi}$ satisfying $\hat{G}_\ell\ket{\phi}=0,\,\forall \ell$; see Fig.~\ref{fig:concept}(b).

Although in the QLM formulation leading to the Hamiltonian in Eq.~\eqref{eq:H} the infinite-dimensional gauge field is represented by a spin-$1/2$ raising operator, it inherits a richness of physical phenomena from the paradigmatic Schwinger model. This includes Coleman's phase transition at a critical mass of $\mu_\mathrm{c}=0.3275\kappa$ for a topological $\theta$-angle of $\pi$ \cite{Coleman1976}; see Fig.~\ref{fig:concept}(a). This transition is related to the spontaneous breaking of the charge conjugation and parity (CP) symmetry, which is equivalent to a global $\mathbb{Z}_2$ symmetry. 
This phase transition has been investigated in pioneering quantum simulator experiments on various platforms \cite{Bernien2017,Kokail2019,Yang2020}, where the $\theta$-angle was fixed to $\pi$.  
Upon tuning the topological $\theta$-angle away from $\pi$, the CP symmetry is explicitly broken, resulting in the vanishing of Coleman's phase transition, and the model becomes confining \cite{Surace2020}; see Fig.~\ref{fig:concept}(a). 

The $\mathrm{U}(1)$ QLM also hosts salient features that have recently received great interest in condensed matter physics, such as different types of quantum many-body scarring \cite{Bernien2017,Su2022} in which thermalization is significantly delayed despite the model being nonintegrable and disorder-free. 
Using the translation and $\mathbb{Z}_2$ symmetries of Eq.~\eqref{eq:H}, we can express its ground states on a two-link two-site unit cell as  $\ket{s^z_{0,1},\sigma^z_1,s^z_{1,2},\sigma^z_2}$ with the eigenvalues $s^z_{\ell,\ell+1}$ and $\sigma^z_\ell$ of the electric-field and matter-occupation operators $\hat{s}^z_{\ell,\ell+1}$ and $\hat{\sigma}^z_\ell$, respectively, serving as good quantum numbers. 
Scarring occurs for massless quenches at $\theta=\pi$ starting in the vacuum states $\ket{\pm1/2,-1,\mp1/2,-1}$, which are the doubly degenerate $\mathbb{Z}_2$ symmetry-broken ground states of Eq.~\eqref{eq:H} at $\mu\to\infty$ and $\theta=\pi$ \cite{Bernien2017,Turner2018}. Scarring also occurs for massive quenches starting in the charge-proliferated state $\ket{-1/2,+1,-1/2,+1}$, which is the nondegenerate $\mathbb{Z}_2$-symmetric ground state of Eq.~\eqref{eq:H} at $\mu\to\infty$ and $\theta=\pi$ \cite{Su2022,Desaules2022weak,Desaules2022prominent}. In Sec.~\ref{sec:QuenchDynamics_SS}, we investigate the effect of confinement on scarring in the quench dynamics of the vacuum state of the $\mathrm{U}(1)$ QLM.

\section{Experimental setup}\label{sec:experimentalsetup}
In this Section, we discuss how the QLM given by Eq.~\eqref{eq:H} can be engineered microscopically in state-of-the-art cold-atom setups.

\subsection{Mapping}\label{sec:mapping}
The $\mathrm{U}(1)$ QLM with the topological $\theta$-term can be obtained from strongly interacting ultracold bosons trapped in a one-dimensional tilted optical superlattice described by the Bose--Hubbard Hamiltonian (see Appendix~\ref{app:degenPT} for details) 
\begin{align}\nonumber
    \hat{H}_\text{BH}=&-J\sum_{j=1}^{L-1}\big(\hat{b}_j^\dagger\hat{b}_{j+1}+\text{H.c.}\big)+\frac{U}{2}\sum_{j=1}^L\hat{n}_j\big(\hat{n}_j-1\big)\\\label{eq:BHM}
    &+\sum_{j=1}^L\bigg[(-1)^j\frac{\delta}{2}+j\Delta+\frac{\chi_j}{2}\bigg]\hat{n}_j,
\end{align}
where $L=2L_\mathrm{m}$ is the total number of sites on the bosonic lattice. Here, the $\hat{b}_j,\hat{b}_j^\dagger$ are the bosonic ladder operators on site $j$ satisfying the canonical commutation relations $\big[\hat{b}_j,\hat{b}_l^\dagger\big]=\delta_{j,l}$ and $\big[\hat{b}_j,\hat{b}_l\big]=0$, and $\hat{n}_j=\hat{b}_j^\dagger\hat{b}_j$ is the corresponding bosonic number operator on site $j$. 
The term $\propto J$ describes hopping of bosons between neighboring wells and $U$ is an on-site interaction. The tilt $\Delta$ serves to suppress undesired second-order hopping to next-to-nearest-neighbor sites. Further, the employed superlattice generates local chemical potentials with two periodicities, a two-site periodicity due to $(-1)^j\delta/2$, and a four-site periodic term related to the topological $\theta$-angle:
\begin{align}
    \chi_j=
    \begin{cases}
      0 & \text{if}\,\,\,j\,\mathrm{mod}\,2=0,\\
      \chi & \text{if}\,\,\,j\,\mathrm{mod}\,4=1,\\
      -\chi & \text{if}\,\,\,j\,\mathrm{mod}\,4=3.
    \end{cases}  
\end{align}
As we will explain in detail in the following, in this mapping the even (shallow) sites of the bosonic superlattice now represent matter sites $\ell$ of the $\mathrm{U}(1)$ QLM on which matter fields reside, while the odd (deep) sites of the bosonic superlattice represent the links between matter sites $\ell$ and $\ell+1$ in the $\mathrm{U}(1)$ QLM where gauge and electric fields are located. See Fig.~\ref{fig:concept}(b) for an illustrative schematic of the superlattice.

The QLM can be derived in second-order perturbation theory in the limit $U,\delta\gg J,\mu$. To see the effect of the large energy scales, we collect all terms of the microscopic Hamiltonian that are diagonal in boson occupations (i.e., including the superlattice in $\chi$), to obtain
\begin{align}\nonumber
    \label{eq:Hdiag}
    \hat{H}_\mathrm{diag}=\sum_\ell \bigg\{&\frac{U}{2} \big[\hat{n}_\ell (\hat{n}_\ell +1) + \hat{n}_{\ell,\ell+1} (\hat{n}_{\ell,\ell+1} +1)\big]\\\nonumber
    &+\big[\delta + (-1)^\ell \chi\big] \hat{n}_{\ell,\ell+1}\\
    &+ \frac{\Delta}{2} \big[ 2\ell \hat{n}_\ell + (2\ell+1) \hat{n}_{\ell,\ell+1}\big]\bigg\},
\end{align}
where $\hat{n}_{\ell,\ell+1}$ denotes the boson occupation on the gauge link, i.e., the odd site $j=2\ell+1$ of the optical superlattice, between matter (even) sites $2\ell$ and $2\ell+2$. Defining generators of a ``proto-Gauss's law'',
\begin{align}
\hat{\mathcal{G}}_\ell= (-1)^\ell\bigg[\frac{1}{2}\big(\hat{n}_{\ell-1,\ell}+\hat{n}_{\ell,\ell+1}\big) + \hat{n}_\ell -1\bigg], 
\end{align}
and tuning $\delta= \mu+U/2$, this Hamiltonian can be rewritten as 
\begin{align}\nonumber
\hat{H}_\mathrm{diag}=&\sum_\ell \bigg\{ \frac{U}{4} \big[2\hat{n}_\ell (\hat{n}_\ell +1) + 2\hat{n}_{\ell,\ell+1} (\hat{n}_{\ell,\ell+1} +1)\\
&- \hat{n}_{\ell,\ell+1}\big] - \mu \hat{n}_\ell + (-1)^\ell \chi \hat{n}_{\ell,\ell+1} + c_\ell \hat{\mathcal{G}}_\ell \bigg\},
\end{align}
with $c_\ell = (-1)^\ell[\Delta \ell + (U-\delta + \Delta/2)]$.
The second and third terms are mapped to the rest mass and topological $\theta$-angle, respectively. 
When $U$ is large, the first term constrains the local Hilbert space on even (matter) sites of the bosonic system to $\{\ket{0}_{2\ell},\ket{1}_{2\ell}\}$, which represent the two local eigenstates of the Pauli operator $\hat{\sigma}^z_\ell$, and it restricts the local Hilbert space on odd (gauge) sites of Eq.~\eqref{eq:BHM} to $\{\ket{0}_{2\ell+1},\ket{2}_{2\ell+1}\}$, which represent the two local eigenstates of the spin-$1/2$ matrix $\hat{s}^z_{\ell,\ell+1}$. 
Thus, in this setup the even (\textit{shallow}) sites host the matter fields, while the odd (\textit{deep}) sites represent the links at which the gauge and electric fields reside; see Fig.~\ref{fig:concept}(b). 
Further, in this regime, the ``proto-Gauss's law'' generators $\hat{\mathcal{G}}_\ell$ become equivalent to those of the desired gauge theory, $\hat{G}_\ell$. 
The final term of $\hat{H}_\mathrm{diag}$ serves to protect the gauge symmetry against gauge-breaking terms \cite{Halimeh2020e,Lang2022SGP}, as we discuss further below at the end of Sec.~\ref{sec:QuenchDynamics_SS}.

The relation between the parameters of the Bose--Hubbard model~\eqref{eq:BHM} and those of the $\mathrm{U}(1)$ QLM~\eqref{eq:H} can be computed from degenerate perturbation theory, taking the hopping $\propto J$ as the small perturbation. The result is
\begin{subequations}
\begin{align}\label{eq:kappa}
    \kappa&=2\sqrt{2}J^2\bigg[\frac{\delta}{\delta^2-\Delta^2}+\frac{U-\delta}{(U-\delta)^2-\Delta^2}\bigg],\\
    \label{eq:mu}
    \mu&=\delta-\frac{U}{2}.
\end{align}
\end{subequations}
In addition, there will be second-order hopping to next-to-nearest-neighbor sites, which are suppressed by the tilt $\Delta$ by choosing $\Delta$ sufficiently larger than $J^2/ \delta$.

\begin{figure}[t!]
    \centering
    \includegraphics[width=\columnwidth]{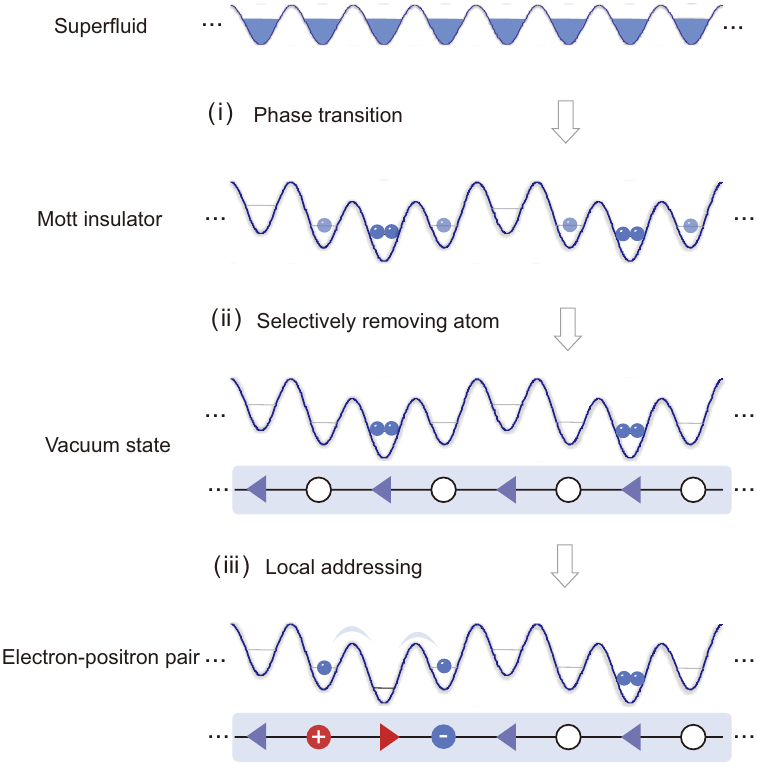}
    \caption{(Color online). Proposed experimental protocol for state initialization. 
    A single-species Bose--Einstein condensate is confined in one spatial dimension. An optical superlattice is ramped up to drive a phase transition from a superfluid to a Mott-insulator in small wedding-cake-like structures. After selectively removing atoms from every second well, the system is initialized in the ``vacuum state'' represented by $\ket{...,0,0,2,0,...}$. In scenarios where one is interested in the state with one electron-positron excitation in the vacuum, as $\ket{...0,2,0,0,1,0,1,0,0,2,0,...}$, local addressing can selectively lower the potential barriers near one doublon in order to allow for tunneling of the atoms to the matter site.  
}
    \label{fig:initilization}
\end{figure}

This proposed Bose--Hubbard quantum simulator uses only well-tested experimental resources and allows for a wide tunability of parameters.  
The hopping $J$ and on-site interaction $U$ terms are controlled primarily by tuning the depth of the main lattice with a periodicity of $\lambda/2$.
The energy offsets $\delta$ and $\chi$ in Eq.~\eqref{eq:BHM} can be generated by two additional optical lattices with double ($\lambda$) and quadruple wavelength ($2\lambda$) as compared to the main lattice, as shown in Fig.~\ref{fig:concept}(b). 
The lattices, with respective lattice depths $V_{1,2,3}$, are spatially overlapped along the $x$-axis to generate the superlattice potential $V(x)= V_1\cos^2(4\pi x/\lambda) - V_2\cos^2(\pi x/\lambda - \pi/4) + V_3\cos^2(\pi x/\lambda - \pi/8)$. Their relative phases are fixed according to Eq.~\eqref{eq:BHM} and can be stabilized with standard locking techniques. 
The energy offsets are given by $\delta= V_2$ and $\chi= V_3/2$ and can be easily tuned through the lattice depths.
Since the relevant regime is where $\chi$ is on the order of $\kappa$, i.e., of $J^2/\delta$, we have that $\chi/\delta\ll 1$. The addition of the tunable $\theta$-term thus does not incur any relevant experimental errors such as undesired resonant transitions in the superlattice.
Besides, the linear potential $\Delta$ can be created by the projection of gravity or a magnetic gradient field.
Based on the experimental setup in Ref.~\cite{Yang2020}, only a lattice with spacing $2\lambda$ needs to be added.
This lattice can be formed conveniently by interfering two $\lambda$-wavelength lasers with an intersection angle of $29$ degrees.

\subsection{Initial states and their preparation}\label{sec:InitialStates}
We are primarily interested in two initial states: a vacuum state $\ket{0,0,2,0}$ or $\ket{2,0,0,0}$, as defined on the two-link two-site unit cell of the corresponding $\mathrm{U}(1)$ QLM, which are the bosonic representations of the two doubly degenerate ground states of Eq.~\eqref{eq:H} at $\mu/\kappa\to\infty$ and $\theta=\pi$, and an electron-positron pair state, where in the center of the vacuum state the bosonic configuration $020$ is replaced with $101$. As indicated by its name, this state represents an electron-positron pair, which, as we will show, will be confined when the $\theta$-angle is tuned away from $\pi$. See Fig.~\ref{fig:initilization} for a depiction of these states and their preparation scheme, which we describe in the following.

The state initialization begins with a one-dimensional Mott insulator state represented by the site occupation as $\ket{...,0,0,2,0,...}$.
It can be obtained by adiabatically loading a Bose--Einstein condensate into the staggered superlattice.
When the average filling factor is $\bar{n}=0.75$, and the superlattice is set to $V_1\cos^2(4\pi x/\lambda) + 3U\cos^2(\pi x/\lambda - \pi/8)$, the mean occupation after the loading process is around $\ket{...,0,0.5,2,0.5,...}$, see Fig.~\ref{fig:initilization}. Here, the atoms undergo a quantum phase transition from a superfluid to a Mott insulator as the lattice depth $V_1$ is slowly ramped up.
Meanwhile, the atoms residing on sites $j\text{mod}4=3$ at the Mott state are cooled down by the neighboring superfluid reservoirs \cite{Yang:2020science}.
The atom configuration will be $\ket{...,0,0,2,0,...}$ after selectively removing the atoms residing on the even sites \cite{Yang:2017pra}.
In the following Section, we will benchmark a ramp protocol for driving Coleman's quantum phase transition from this initial state (Sec.~\ref{sec:numerics_ramp}) as well as abrupt quenches (Sec.~\ref{sec:QuenchDynamics_SS}). Although this initial state differs from the charge-proliferated state used in Ref.~\cite{Yang2020}, its preparation requires only demonstrated experimental capabilities.

To achieve the state with an electron-position pair in the vacuum, one can use a local addressing technique to enable local oscillations between the states $\ket{...,0,2,0,...}$ and $\ket{...,1,0,1,...}$, as seen in Ref.~\cite{Yang2020}, confined to a selected region.
This state will be a starting point for the quench dynamics in our numerical calculations in Sec.~\ref{sec:QuenchDynamics_PS}.

\section{Numerical Benchmarks}\label{sec:numerics}
We now present our main numerical results on the time evolution of our ramp and quench dynamics, obtained from $t$-DMRG \cite{Vidal2004,White2004,Daley2004} based on the Krylov-subspace method \cite{Schmitteckert2004,Feiguin2005,Ripoll2006,McCulloch2007}. In $t$-DMRG, the quantum many-body wave function is represented by matrix product states \cite{Uli_review,Paeckel_review} as facilitated by repeated truncations of small Schmidt coefficients. Upon appropriately tuning the \textit{fidelity threshold}, the accumulated error over evolution times in the numerical simulation can be controlled. This sets the desired accuracy over the calculation, the achievement of which will lead to a corresponding increase in the bond dimension of the calculated wave function. For the main results of our paper, we have used $L=32$ sites ($L_\mathrm{m}=16$ matter sites and $L_\mathrm{g}=16$ gauge links) with open boundary conditions, which we find converged with respect to system size for our purposes \cite{Yang2020,Halimeh2020d}. For our numerically most stringent calculations, we find excellent convergence for an on-site maximal boson occupation of $N_\mathrm{max}=2$ (indicative of good imposition of our constraint described in Sec.~\ref{sec:mapping}), a time-step of $dt=10^{-1}$ ms, and a fidelity threshold of $10^{-6}$ per time-step.

\subsection{Ramp}\label{sec:numerics_ramp}
At a topological $\theta$-angle of $\pi$ (i.e., $\chi=0$), the $\mathrm{U}(1)$ QLM Hamiltonian~\eqref{eq:H} is invariant under the parity transformation and charge conjugation. This CP symmetry is of a discrete $\mathbb{Z}_2$ nature, and can thus be spontaneously broken in one spatial dimension (at zero temperature), leading to the so-called Coleman's phase transition \cite{Coleman1976}. The phase transition can be captured by the order parameter $\mathcal{E}=\sum_{\ell=1}^{L_\mathrm{m}}(-1)^\ell\langle\hat{s}^z_{\ell,\ell+1}\rangle/L_\mathrm{m}$. Upon tuning the topological $\theta$-angle away from $\pi$, the last term in Eq.~\eqref{eq:H} will explicitly break CP symmetry, thus invalidating Coleman's phase transition. 

\begin{figure}[t!]
    \centering
    \includegraphics[width=0.48\columnwidth]{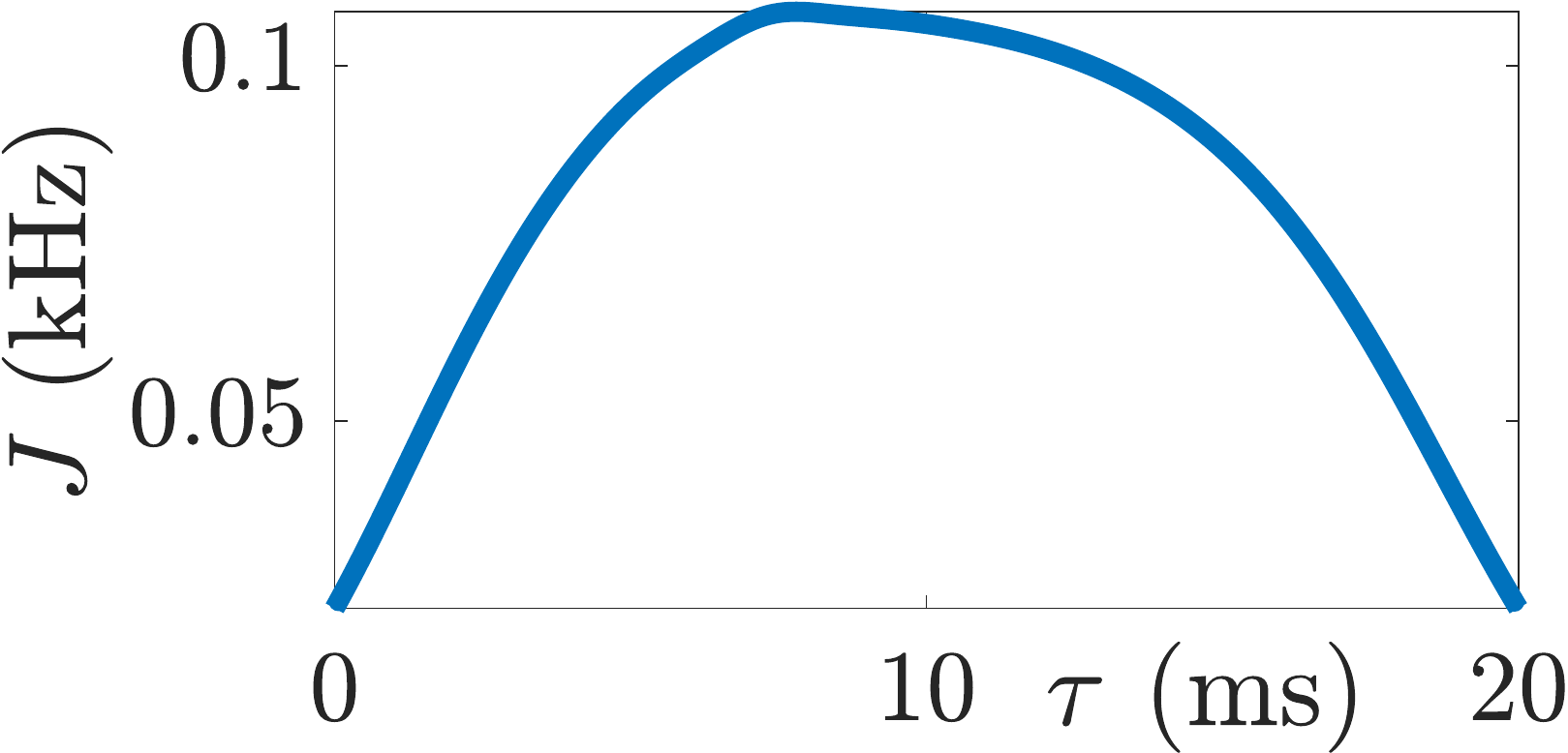}\quad\includegraphics[width=0.48\columnwidth]{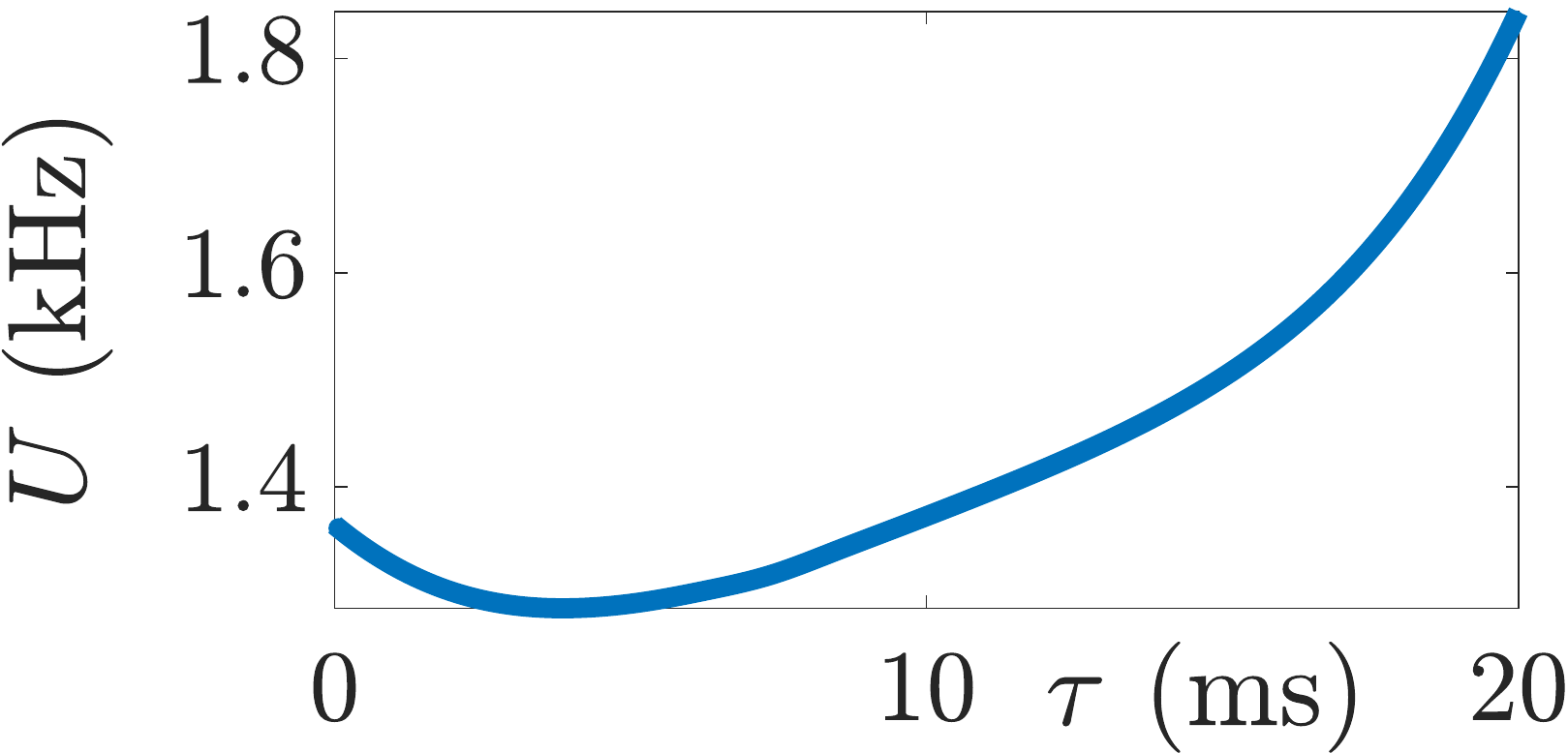}\\
	\vspace{1.1mm}
    \includegraphics[width=0.48\columnwidth]{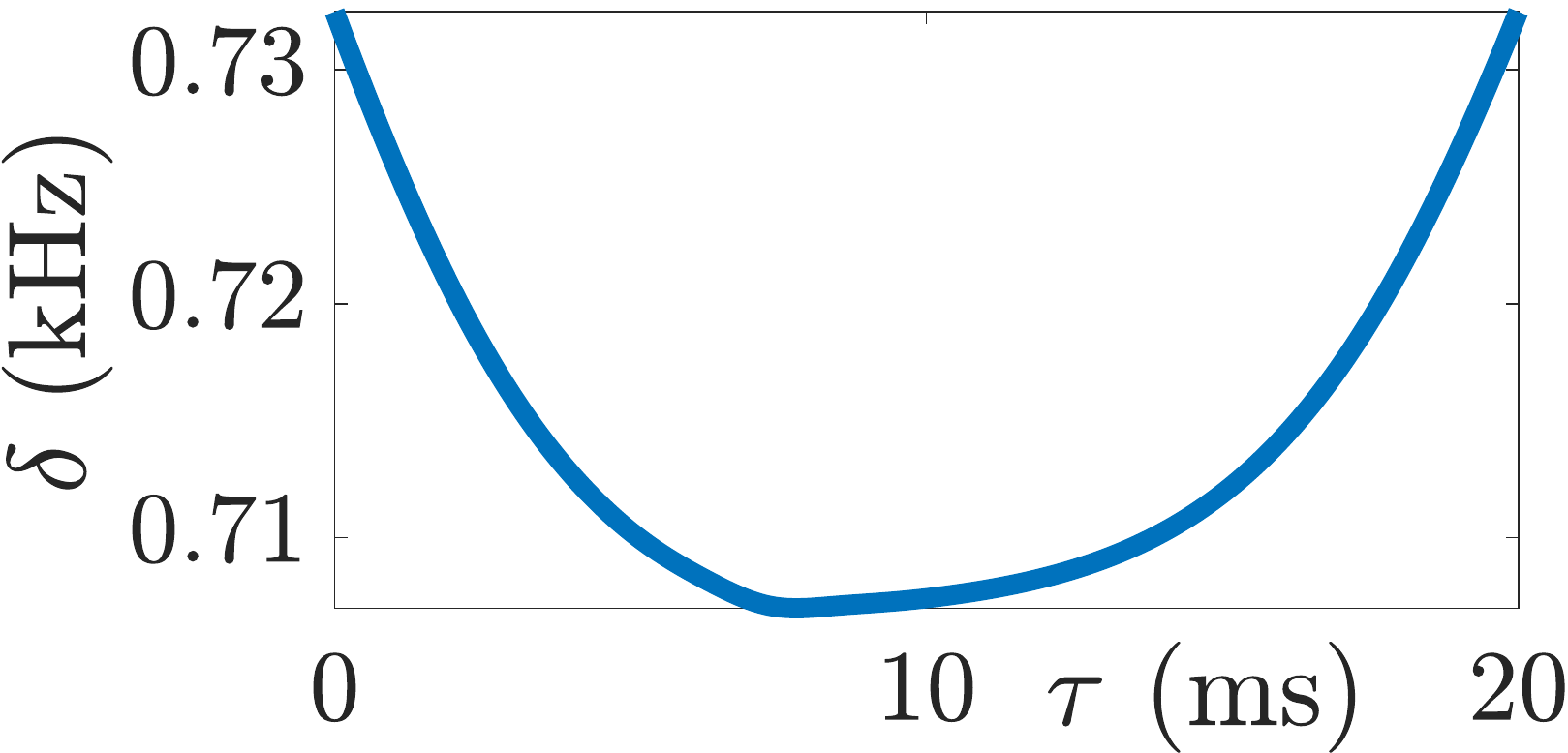}\quad\includegraphics[width=0.48\columnwidth]{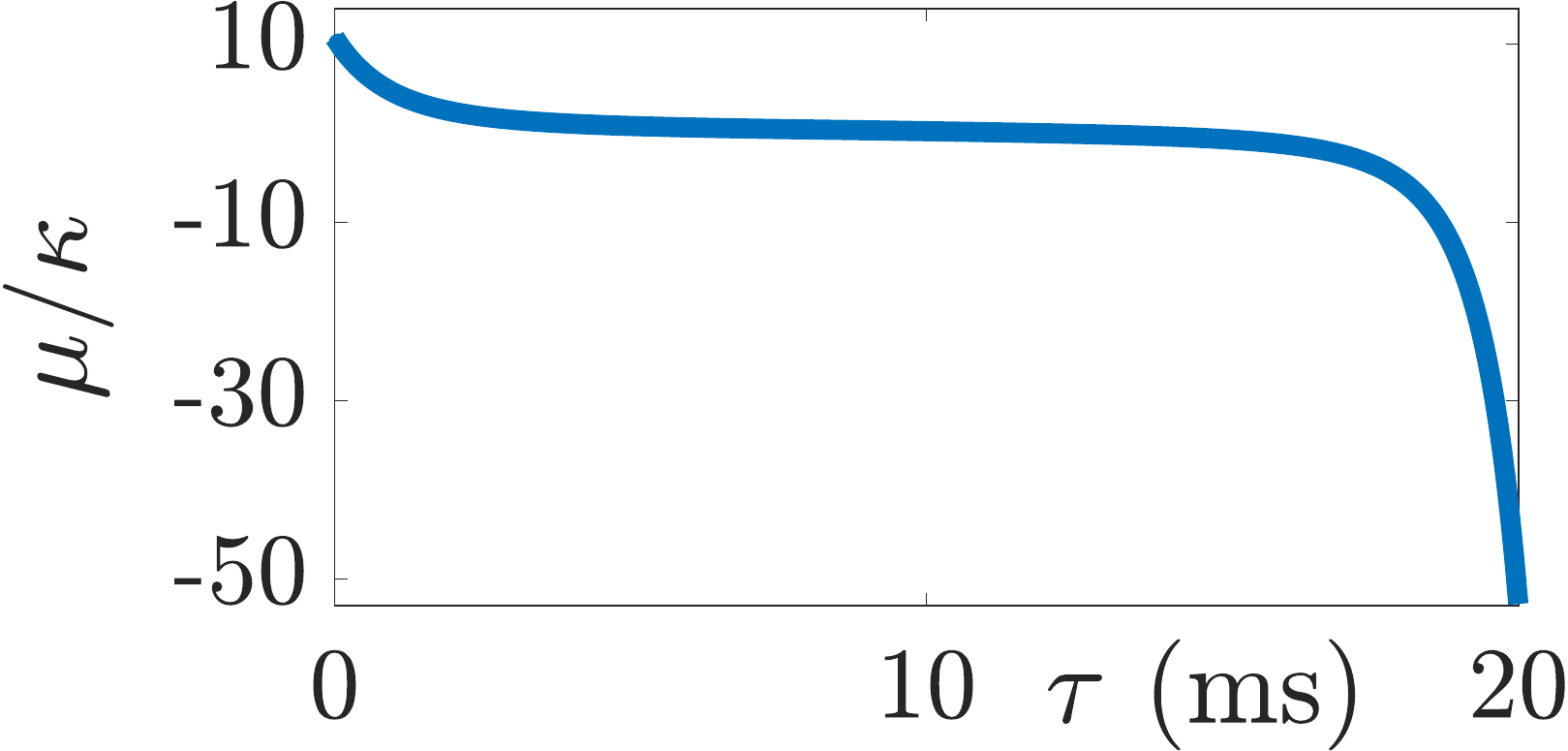}
    \caption{(Color online). The ramp protocol employed in this work, where $J$, $U$, and $\delta$ of Eq.~\eqref{eq:BHM} are tuned in such a way that the mass $\mu/\kappa$ of Eq.~\eqref{eq:H} is adiabatically ramped from a large positive value (where the ground state is one of two doubly degenerate $\mathbb{Z}_2$ symmetry-broken vacua in the deconfined regime) to a large negative value (where the ground state is the charge-proliferated state in the deconfined regime). Here, the total ramp time $\tau=20$ ms is much smaller than the coherence time of the experimental setup in Ref.~\cite{Yang2020}, making this protocol experimentally feasible.}
    \label{fig:ramp_protocol}
\end{figure}

To study this effect, we consider an adiabatic ramp protocol where we start in a $\mathbb{Z}_2$ symmetry-broken vacuum state at $\mu/\kappa\to\infty$ in Eq.~\eqref{eq:H} and slowly sweep the mass to $\mu/\kappa\to-\infty$ in the $\mathbb{Z}_2$-symmetric phase. In the bosonic model of Eq.~\eqref{eq:BHM}, this is equivalent to sweeping $J$, $U$, and $\delta$ as shown in Fig.~\ref{fig:ramp_protocol}, respectively. The resulting mass ramp of $\mu/\kappa$ in the $\mathrm{U}(1)$ QLM~\eqref{eq:H} that the proposed experiment would implement is depicted in the lower-right panel. We note that the reverse of this ramp has been performed experimentally, starting from the charge-proliferated state $\ket{0,1,0,1}$, to probe Coleman's phase transition~\cite{Yang2020} and for preparing far-from-equilibrium states to study thermalization dynamics in the wake of global quenches~\cite{Zhou2021}. 
In the bosonic picture relevant to the experimental implementation, the order parameter and chiral condensate are defined as
\begin{subequations}
\begin{align}\label{eq:OP_ramp}
&\mathcal{E}(\tau)=\frac{2}{L}\sum_{m=1}^{L/2}(-1)^m\bra{\psi(\tau)}\hat{n}^\text{d}_{2m-1}\ket{\psi(\tau)}\\\label{eq:CC_ramp}
&\mathcal{C}(\tau)=\frac{2}{L}\sum_{m=1}^{L/2}\bra{\psi(\tau)}\hat{n}_{2m}\ket{\psi(\tau)},
\end{align}
\end{subequations}
respectively. Here, $\tau$ denotes the time of the ramp, $\ket{\psi(\tau)}=\mathcal{T}e^{-i\int_0^\tau ds\,\hat{H}_\text{BH}(s)}\ket{\text{vac}}$, $\mathcal{T}$ is the time-ordering operator, $\hat{H}_\text{BH}(\tau)$ is Hamiltonian~\eqref{eq:BHM} with its parameters taking on the values specified in Fig.~\ref{fig:ramp_protocol} at ramp time $\tau$, and $\hat{n}^\text{d}_{j}=\hat{b}_j^\dagger\hat{b}_j^\dagger\hat{b}_j\hat{b}_j/2$ is the doublon-occupation operator at site $j$ of the optical superlattice, which represents the local electric flux when $j$ is odd.

\begin{figure}[t!]
    \centering
    \includegraphics[width=\columnwidth]{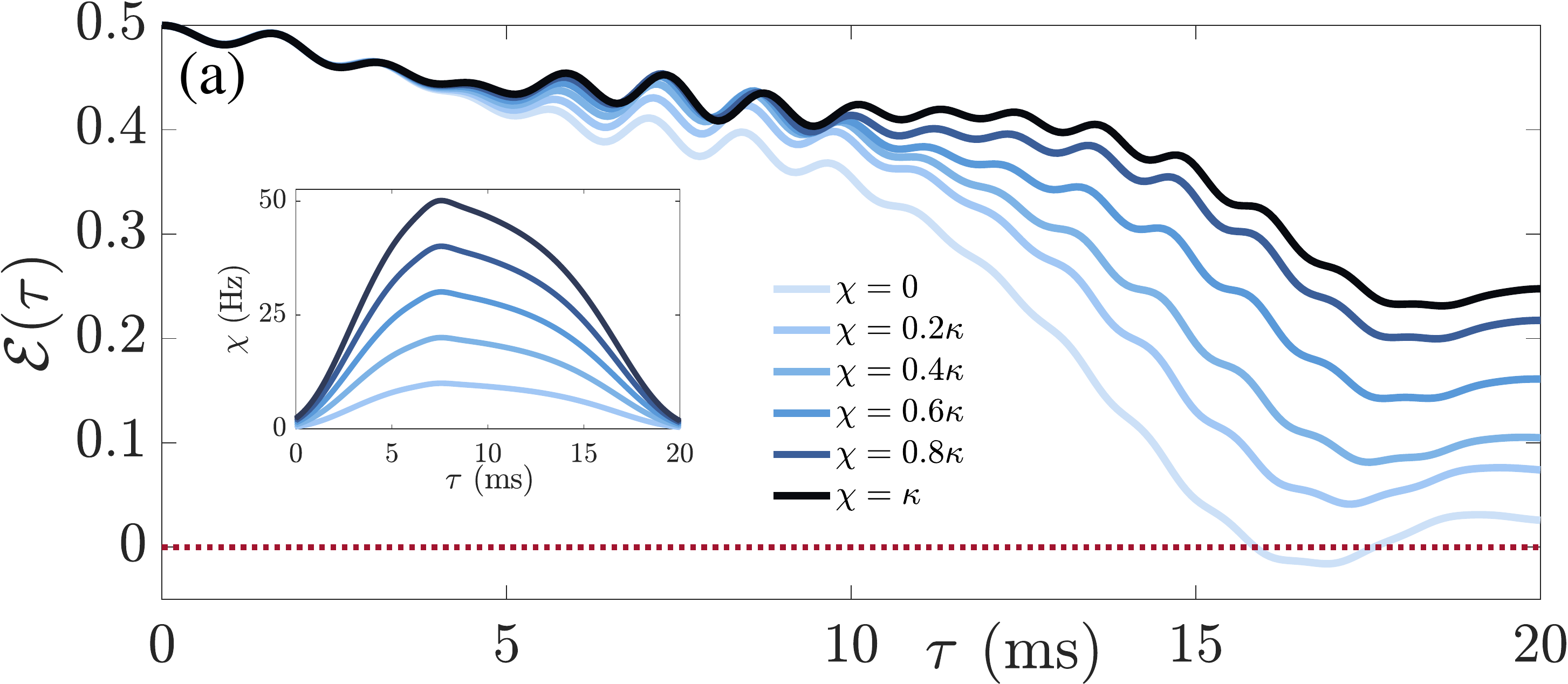}\\
    \vspace{1.1mm}
    \includegraphics[width=\columnwidth]{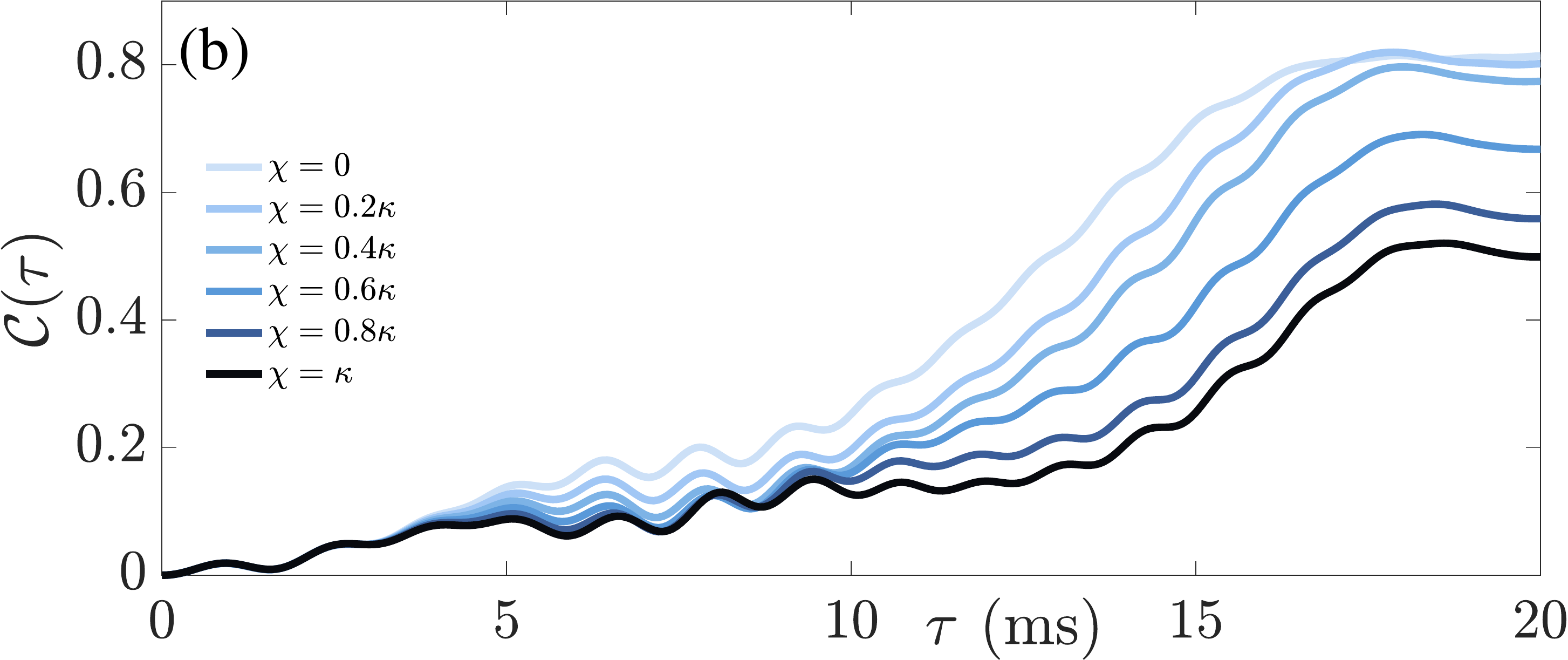}
    \caption{(Color online). (a) Time evolution of the electric flux $\mathcal{E}(\tau)$, defined in Eq.~\eqref{eq:OP_ramp}, throughout the ramp protocol of Fig.~\ref{fig:ramp_protocol}. The parameter $\chi$ is tuned in such a way that $\chi/\kappa$ is constant throughout the ramp (see inset), taking into account the corresponding variation of $\kappa$ during the ramp. The order parameter reaches zero only in the case of $\chi=0$ where Coleman's phase transition exists in the deconfined regime, while it is always finite throughout the entire duration in the ramp when $\chi\neq0$, since then the associated global $\mathbb{Z}_2$ symmetry is explicitly broken in this confined regime, lifting Coleman's phase transition. 
    (b) Time evolution of the chiral condensate $\mathcal{C}(\tau)$, defined in Eq.~\eqref{eq:CC_ramp}, throughout the ramp protocol of Fig.~\ref{fig:ramp_protocol}. The vacuum state is void of matter, so naturally $\mathcal{C}(0)=0$. In the deconfined regime ($\chi=0$) where Coleman's phase transition is present, an adiabatic ramp drives the system close to the charge-proliferated state within the $\mathbb{Z}_2$-symmetric phase, where it exhibits a chiral condensate close to unity. Ramping in the confined regime ($\chi\neq0$) leads to the final state having a lower value of the chiral condensate. This is because the $\theta$-angle term explicitly breaks the global $\mathbb{Z}_2$ symmetry, imparting a finite electric flux on the final state. Due to the effective stabilization of Gauss's law in our proposed quantum simulator, this in turn necessitates a smaller value of the chiral condensate.}
    \label{fig:vacuum_ramp}
\end{figure}

\begin{figure*}[t!]
    \centering
    \includegraphics[width=\columnwidth]{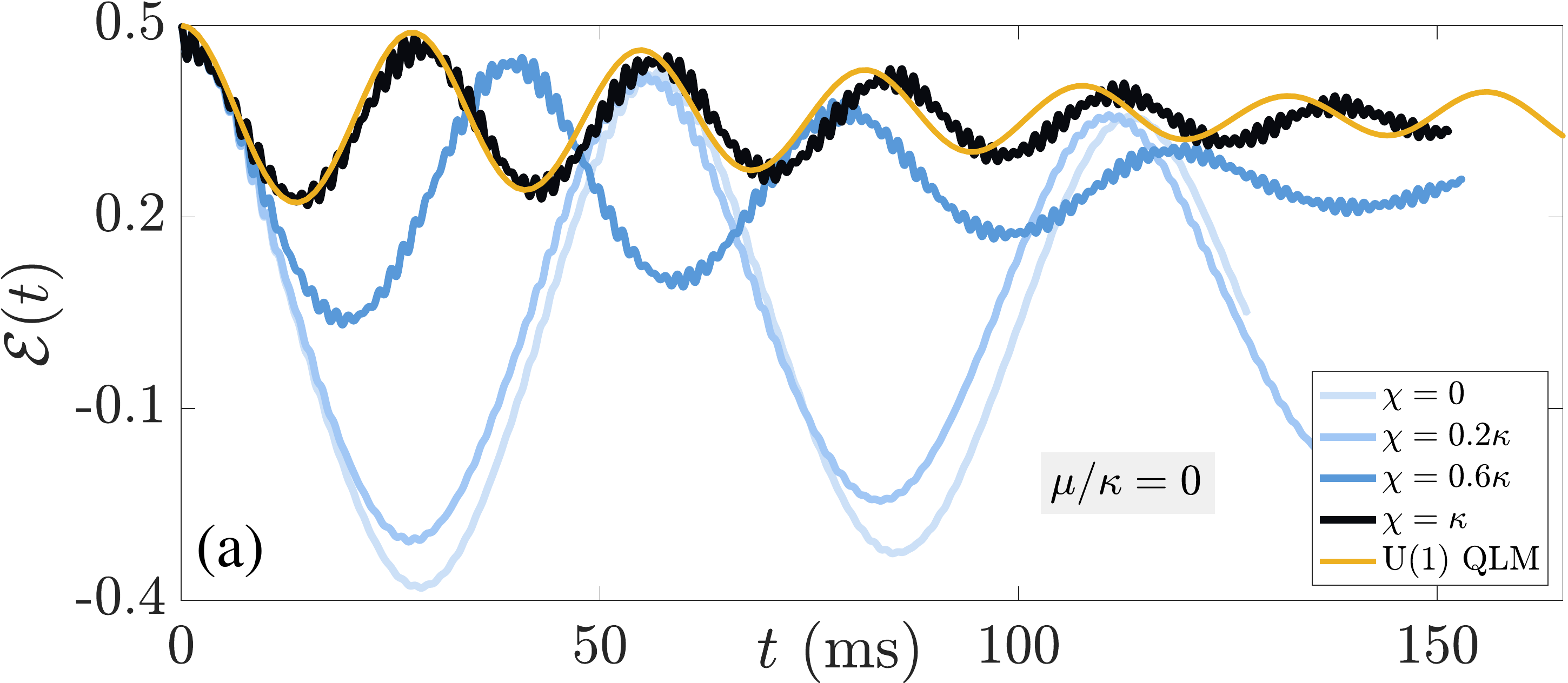}\quad\includegraphics[width=\columnwidth]{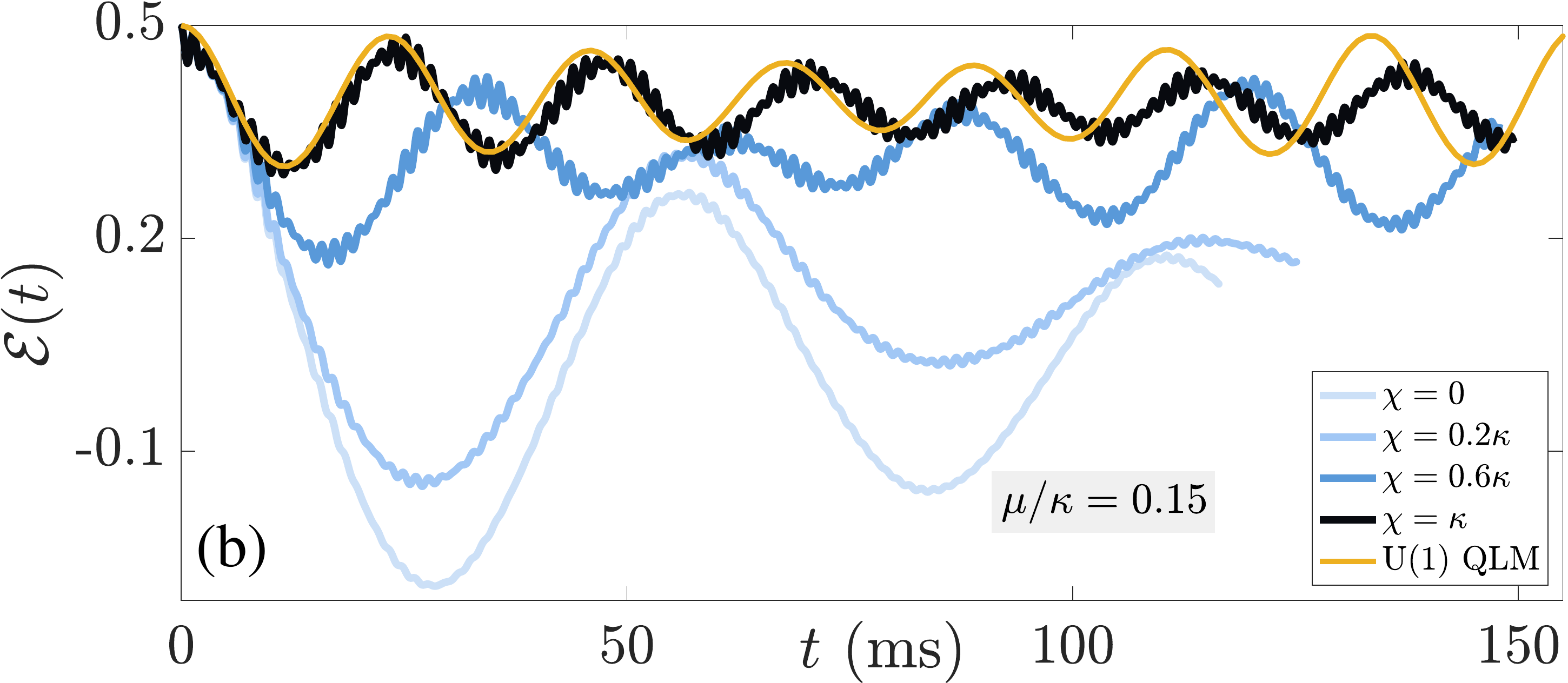}\\
    \vspace{1.1mm}
    \includegraphics[width=\columnwidth]{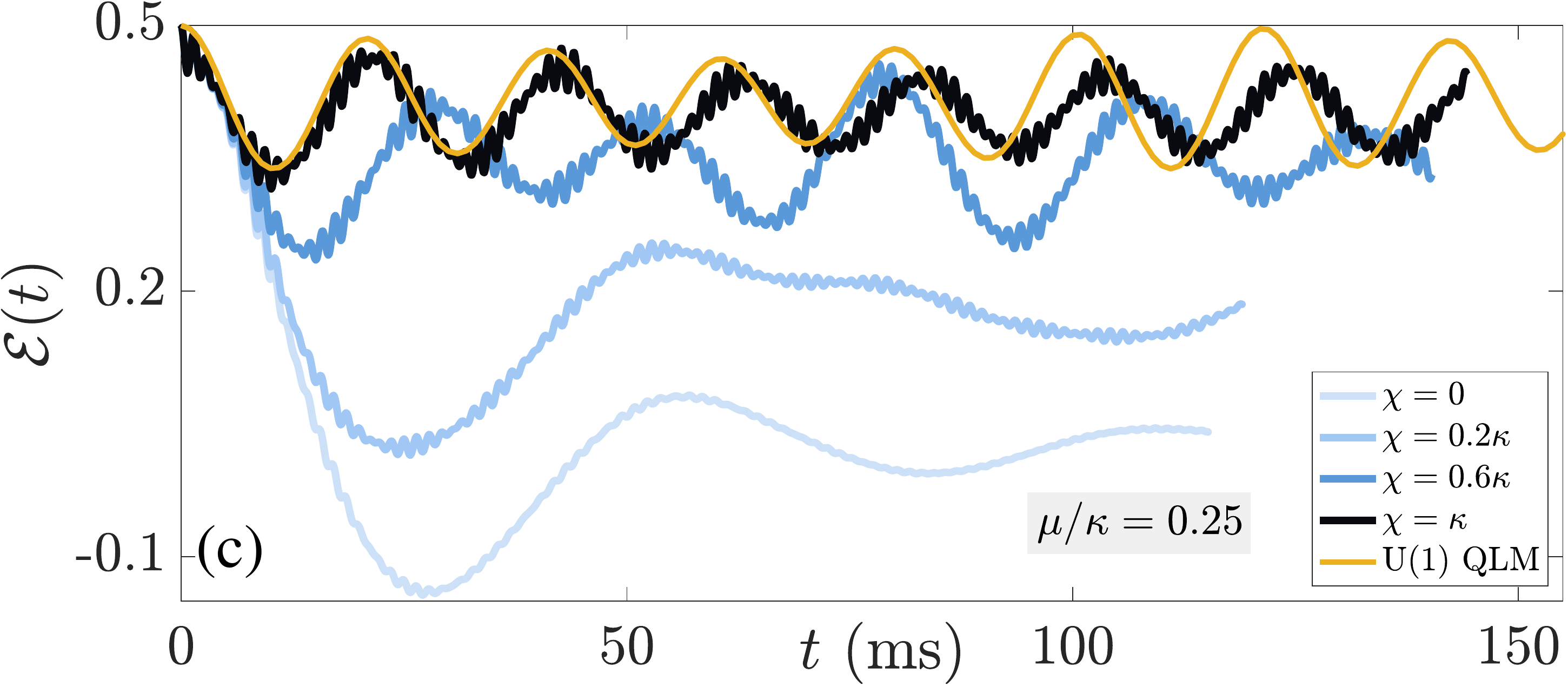}\quad\includegraphics[width=\columnwidth]{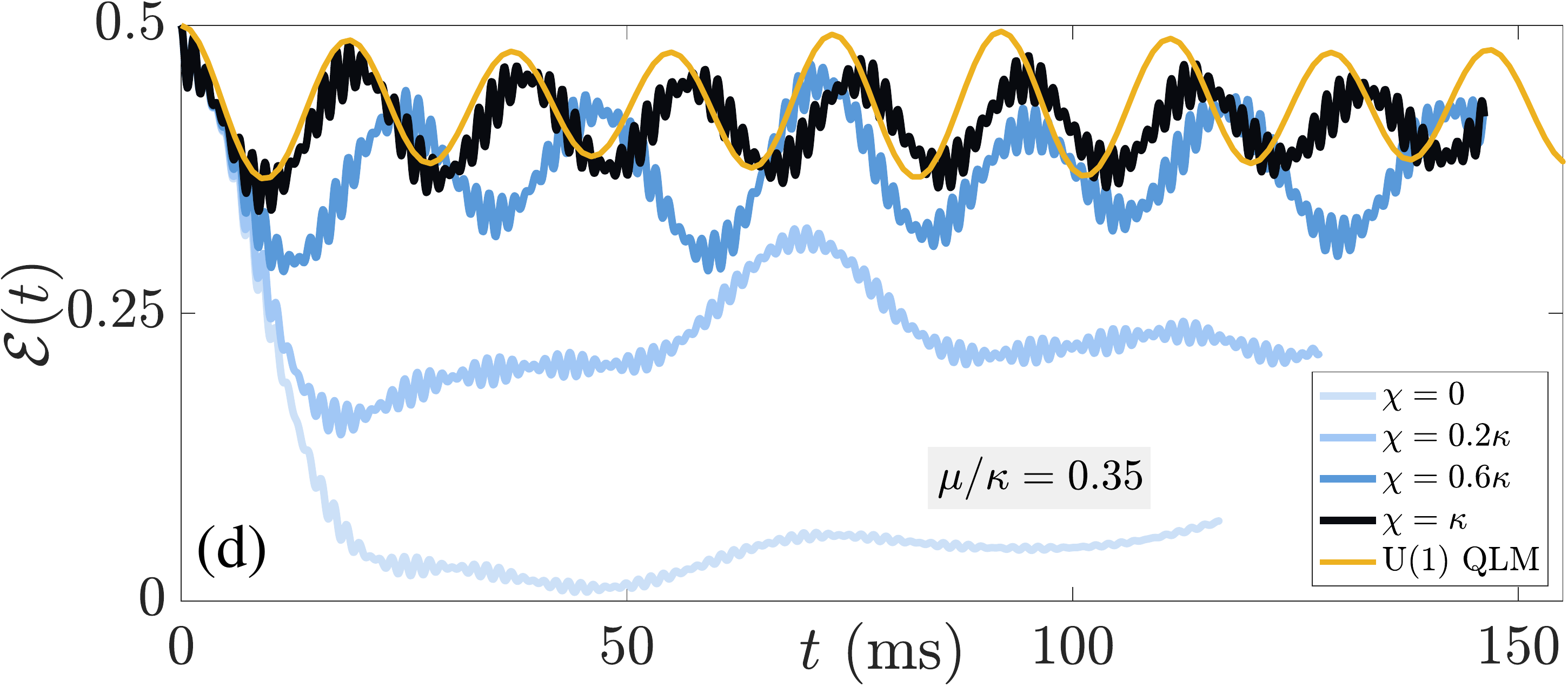}
    \caption{(Color online). $t$-DMRG calculations of the dynamics of the electric flux for a quench of the vacuum state with Hamiltonian~\eqref{eq:BHM} at mass (a) $\mu/\kappa=0$, (b) $\mu/\kappa=0.15$, (c) $\mu/\kappa=0.25$, and (d) $\mu/\kappa=0.35$, with the topological term at strength $\chi/\kappa=0,\,0.2,\,0.6,\,1$ (from light to dark blue). Solid yellow curve depicts the corresponding dynamics in the ideal $\mathrm{U}(1)$ QLM~\eqref{eq:H} for the same quench at $\chi/\kappa=1$, as obtained from ED. As $\chi$ increases, confinement forces the electric field to remain close to its initial value throughout all considered evolution times. For nonzero rest mass, the persistent oscillations are enhanced, exhibiting a larger frequency with increasing $\chi$. It is worth noting here that the massless quench at $\chi=0$ exhibits quantum many-body scarring, first observed in a Rydberg-atom setup~\cite{Bernien2017}. This scarring, which involves state transfer between the two degenerate vacua of QED, is destroyed in the strongly confined regime, where the wave function remains close to the initial state at all evolution times (see text).
    }
    \label{fig:SS_Nd}
\end{figure*}

We now prepare our initial state in the vacuum state $\ket{\text{vac}}=\ket{0,0,2,0}$, and perform the ramp protocol of Fig.~\ref{fig:ramp_protocol} at various values of $\chi/\kappa$. As $\kappa$ varies during the ramp, see Eq.~\eqref{eq:kappa}, $\chi$ is adjusted to compensate for the tuning of the lattice parameters and thus keep $\chi/\kappa$ constant; see inset of Fig.~\ref{fig:vacuum_ramp}(a). The dynamics of $\mathcal{E}(\tau)$ as calculated in $t$-DMRG is shown in the main panel of Fig.~\ref{fig:vacuum_ramp}(a). When $\chi=0$, the topological $\theta$-angle term vanishes, and Coleman's phase transition exists in Eq.~\eqref{eq:H}. This means that the ramp can lead the dynamics into a $\mathbb{Z}_2$-symmetric phase, which is indeed what we find in Fig.~\ref{fig:vacuum_ramp}(a) when $\chi=0$. Note that even though the critical point $(\mu/\kappa)_\mathrm{c}=0.3275$ is reached at $\tau\approx6$ ms, the order parameter first reaches zero at $\tau\approx15.87$ ms. This is expected because while ramping through the critical point, significant quantum fluctuations are generated that delay the onset of the $\mathbb{Z}_2$-symmetric phase in the ramp dynamics.

Upon switching the topological $\theta$-angle away from $\pi$ (i.e., $\chi\neq0$), the $\mathrm{U}(1)$ QLM in Eq.~\eqref{eq:H} no longer hosts Coleman's phase transition, because the topological $\theta$-angle term explicitly breaks the global $\mathbb{Z}_2$ symmetry. As such, one would expect $\mathcal{E}(\tau)$ to remain trivially finite throughout the entire ramp, relaxing at $\mu/\kappa\to\-\infty$ towards a value that increases with $\chi$. Our $t$-DMRG calculations confirm this picture, indicating that the larger $\chi$ is, the larger is the final value of $\mathcal{E}(\tau)$.

The chiral condensate dynamics during this ramp is shown in Fig.~\ref{fig:vacuum_ramp}(b). When $\chi=0$, we expect that the wave function goes from the initial vacuum state to something close to the charge-proliferated state in the $\mathbb{Z}_2$-symmetric phase, where then $\mathcal{C}(\tau)$ ought to be close to unity. Our $t$-DMRG calculations confirm this picture in this deconfined regime. For nonzero values of $\chi$, we see that the final value of $\mathcal{C}(\tau)$ in this ramp decreases the larger $\chi$ is. This can be understood by noting that Coleman's phase transition vanishes for nonzero $\chi$, and this explicit breaking of the $\mathbb{Z}_2$ symmetry will favor a finite $\mathcal{E}(\tau)$ at late times. Due to Gauss's law, this then leads to a decrease in the chiral condensate.

It is also worth noting that the ramp protocol we employ here may be adapted to probe the order and universality of possible underlying phase transitions in gauge theories. This is facilitated by adiabatically ramping the mass at fixed infinitesimally small values of $\chi$ and measuring the final value of $\mathcal{E}(\tau)$ to extract the corresponding critical exponents \cite{demidio2021}.

\subsection{Quench dynamics}\label{sec:QuenchDynamics}
We now turn to abrupt global quenches in our setup, which have recently been employed to experimentally probe thermalization dynamics in the $\mathrm{U}(1)$ QLM \cite{Zhou2021}. Such quench dynamics are essential in probing salient features in gauge-theory quantum simulations relevant to condensed matter physics such as, for example, quantum many-body scars \cite{Bernien2017,Moudgalya2018,Turner2018} and disorder-free localization \cite{Smith2017,Brenes2018}, and also those relevant to high-energy physics, such as thermalization of generic gauge theories \cite{Zhou2021,Berges_review}, string-breaking dynamics \cite{Hebenstreit2013,Surace2020}, and confinement \cite{Wang2021,Mildenberger2022}. In the following, we will focus on two main initial states: the vacuum and electron-positron pair states. In order to probe signatures of confinement, we numerically calculate the quench dynamics of the electric flux, the chiral condensate, and the von Neumann entanglement entropy,
\begin{subequations}
\begin{align}\label{eq:flux}
    &\mathcal{E}(t)=\frac{2}{L}\sum_{m=1}^{L/2}(-1)^m\bra{\psi(t)}\hat{n}^\text{d}_{2m-1}\ket{\psi(t)},\\\label{eq:CC}
    &\mathcal{C}(t)=\frac{2}{L}\sum_{m=1}^{L/2}\bra{\psi(t)}\hat{n}_{2m}\ket{\psi(t)},\\\label{eq:EE}
    &\mathcal{S}_j(t)=-\Tr{\hat{\rho}_{1\to j}(t)\ln \hat{\rho}_{1\to j}(t)},
\end{align}
\end{subequations}
respectively, where $\ket{\psi(t)}=e^{-i\hat{H}_\text{BH}t}\ket{\psi_0}$, $\ket{\psi_0}$ is the initial state, $\hat{H}_\text{BH}$ is the quench Hamiltonian~\eqref{eq:BHM}, and $\hat{\rho}_{1\to j}(t)$ is the reduced density matrix at evolution time $t$ of the subsystem formed on the lattice along the sites $1,\ldots,j$.

For all quenches considered in this paper, we have set $U=1368$ Hz, $J=58$ Hz, and $\Delta=57$ Hz. Then for a given value of $\mu/\kappa$, and employing Eqs.~\eqref{eq:kappa} and~\eqref{eq:mu}, we arrive at an implicit equation for $\delta$, which can then be solved using, for example, Newton's method. To leading approximation, the topological $\theta$-term strength $\chi$ can be set independently of the other parameters so long as $\lvert\delta\pm\Delta\rvert\gg\lvert\chi\rvert$ (see Appendix~\ref{app:degenPT} for details).

\subsubsection{Initial state: vacuum}\label{sec:QuenchDynamics_SS}
The vacuum state of the $\mathrm{U}(1)$ QLM is of particular interest in synthetic quantum matter experiments not just from a high-energy perspective, but also in terms of intriguing condensed matter features it can give rise to. For example, it has been shown in Ref.~\cite{Bernien2017} that a massless quench with $\chi=0$ of this vacuum state leads to the weak ergodicity breaking paradigm of quantum many-body scarring \cite{Serbyn2020,MoudgalyaReview,Moudgalya2018,Turner2018}. The vacuum state resides in a \textit{cold subspace} that is weakly connected to the rest of the Hilbert space of the quench Hamiltonian~\eqref{eq:H} at $\mu/\kappa=\chi/\kappa=0$, which leads to a significant delay of thermalization \cite{Turner2018}. The eigenstates of this cold subspace exhibit anomalously low entanglement entropy and are roughly equally spaced in energy across the entire spectrum. Furthermore, this scarring behavior manifests as persistent oscillations in the dynamics of local observables lasting well beyond relevant timescales, along with an anomalously low and slowly growing entanglement entropy \cite{ShiraishiMori,lin2018exact}.

Figure~\ref{fig:SS_Nd}(a) shows the resulting dynamics of the electric flux~\eqref{eq:flux} when starting in the vacuum state $\ket{0,0,2,0}$ and quenching with the Hamiltonian of Eq.~\eqref{eq:BHM} at $\mu/\kappa=0$. In agreement with known experimental results \cite{Bernien2017,Su2022}, when $\chi=0$ we see persistent oscillations around zero in the electric flux that last up to all accessible times. The dynamics in this case can be explained as ``state transfer'' \cite{Christandl2004} between the two doubly degenerate vacua of Eq.~\eqref{eq:H}. Upon tuning the topological $\theta$-angle away from $\pi$, we find that the oscillations remain, but the mean of $\mathcal{E}(t)$ is no longer around zero and instead takes on a finite value closer to $\mathcal{E}(0)=1/2$. Interestingly, we find that the frequency of oscillations increases with confinement, in agreement with numerical results on confined dynamics in quantum spin chains with long-range interactions~\cite{Halimeh2017,Liu2019,Halimeh2020quasiparticle} and the quantum Ising model with both transverse and longitudinal fields \cite{Kormos2017}. For the largest deviation of the $\theta$-angle from $\pi$ that we investigate ($\chi=\kappa)$, we find strong confinement of the electric field, where it is always $\mathcal{E}(t)\gtrsim0.22$ over all accessible evolution times in $t$-DMRG. This indicates that the time-evolved wave function remains very close to the initial vacuum state throughout all accessible evolution times, and does not approach the second vacuum. In other words, confinement prohibits state transfer between the two vacua, and therefore destroys quantum many-body scarring in the $\mathrm{U}(1)$ QLM.

\begin{figure*}[t!]
    \centering
    \includegraphics[width=\columnwidth]{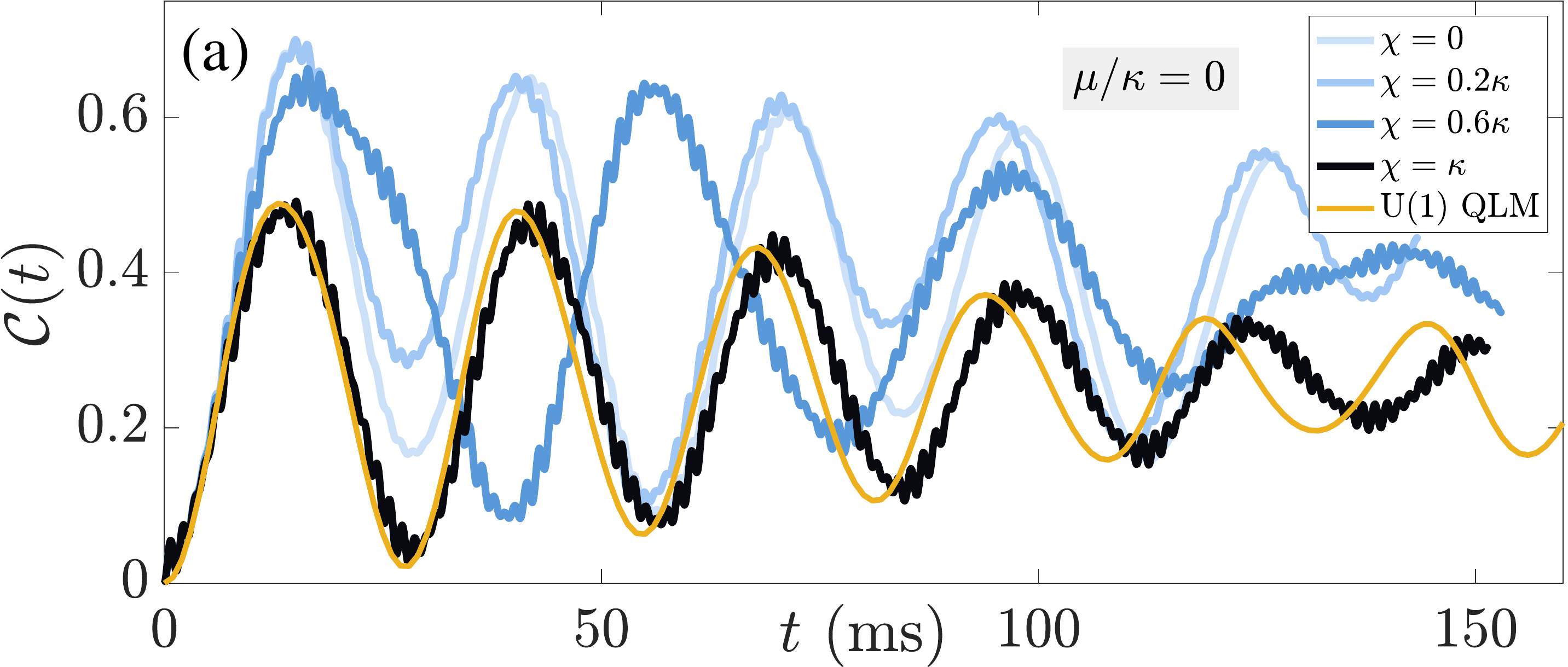}\quad\includegraphics[width=\columnwidth]{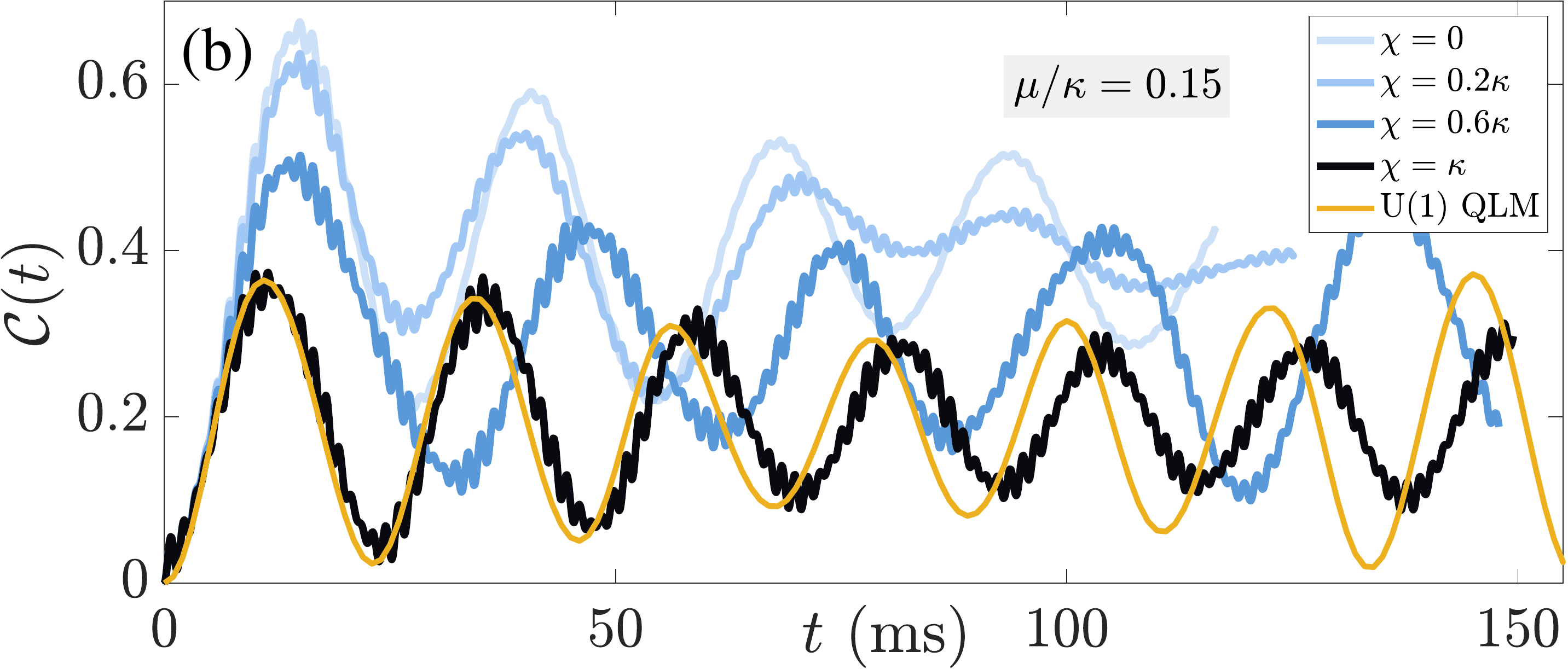}\\
    \vspace{1.1mm}
    \includegraphics[width=\columnwidth]{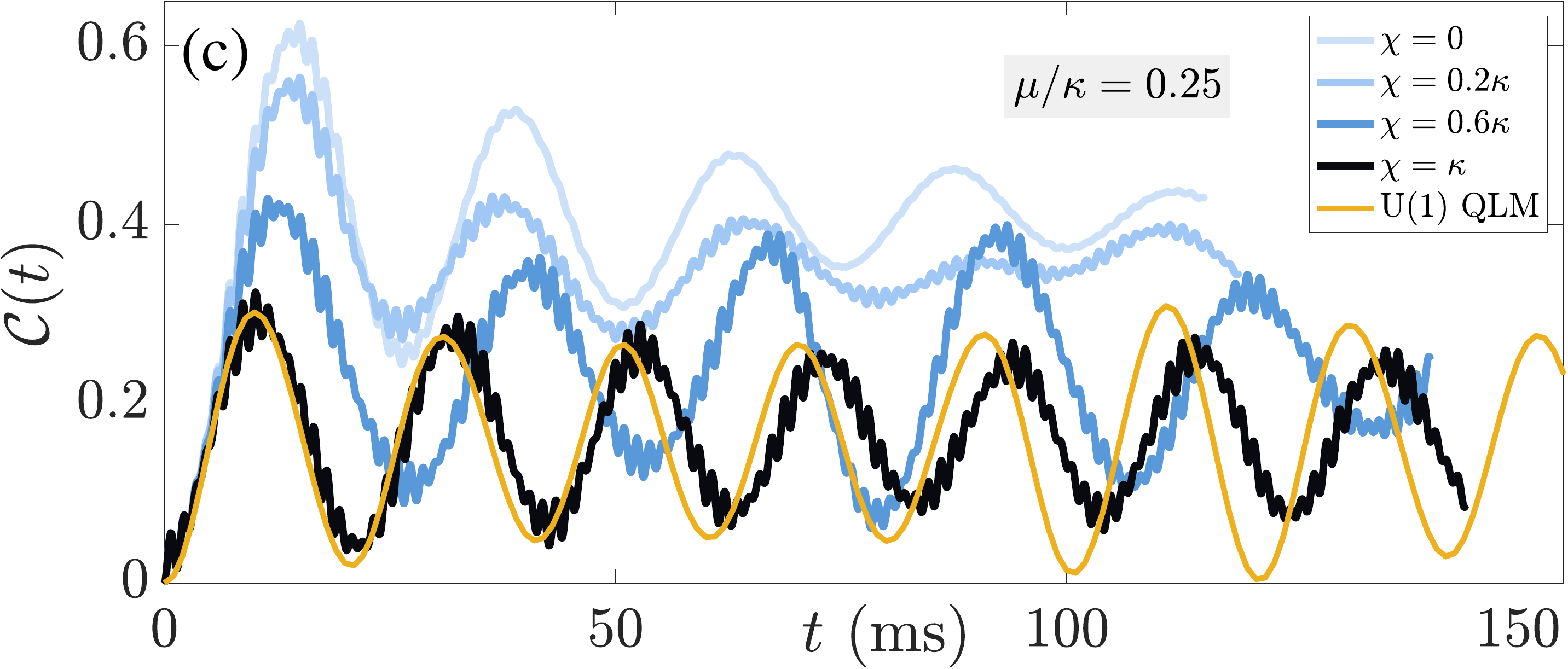}\quad\includegraphics[width=\columnwidth]{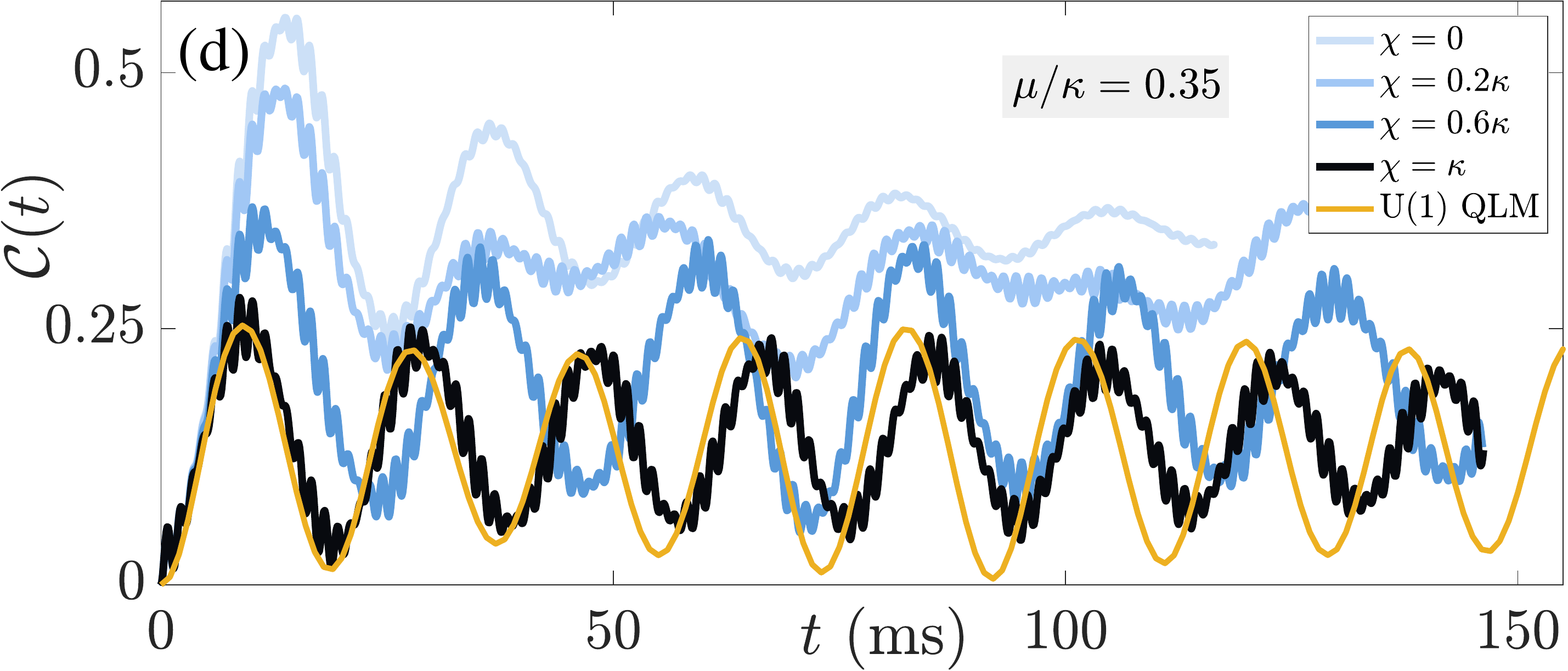}
    \caption{(Color online). Same as Fig.~\ref{fig:SS_Nd} but for the chiral condensate~\eqref{eq:CC}. The dynamics is qualitatively different between the deconfined and confined regimes. Whereas for $\chi/\kappa=0$ we find dynamics deviating significantly from the initial value of the chiral condensate, as $\chi/\kappa$ is increased, the dynamics exhibits confined behavior with $\mathcal{C}(t)$ oscillating much closer to $0$. For massive quenches, we find that the frequency of oscillations increases with $\chi$, but this is not so clear for the case of $\mu/\kappa=0$. This may be related to scars, in that for this quench with $\chi=0$, the electric flux has half the frequency of the chiral condensate due to the state transfer between the two doubly degenerate vacua during the dynamics; see text and Fig.~\ref{fig:SS_Nd}(a). In the confined regime, however, the chiral condensate and the electric flux both exhibit the same frequency.}
    \label{fig:SS_CC}
\end{figure*}

Using exact diagonalization (ED) calculations, we benchmark the case of a massless quench at $\chi=\kappa$ with the corresponding quench in the ideal $\mathrm{U}(1)$ QLM of Eq.~\eqref{eq:H} at $\mu/\kappa=0$ and $\chi=\kappa$. We find excellent quantitative agreement at short times, and very good qualitative agreement over all times. It is not surprising that the quantitative agreement at late times is not as good as at early times, because in our mapping we obtain Eq.~\eqref{eq:BHM} up to leading order in perturbation theory, but subleading orders that break gauge invariance will become less innocuous at later times. As we will elucidate later, this leads to a renormalized gauge theory that nevertheless hosts the same $\mathrm{U}(1)$ gauge symmetry of the ideal $\mathrm{U}(1)$ QLM, and which persists over all relevant timescales.

In Fig.~\ref{fig:SS_Nd}(b,c), we show $t$-DMRG results for quenches of the vacuum state $\ket{0,0,2,0}$ at other values of the mass $\mu/\kappa=0.15,\,0.25$, which like the massless case are across Coleman's phase transition when $\chi=0$. We again find that the order parameter oscillates around zero in the deconfined regime ($\chi=0$), albeit these oscillations are not persistent since the dynamics is not scarred for $\mu/\kappa\neq0$ \cite{Turner2018}. In the confined regime ($\chi\neq0$), we again see that $\mathcal{E}(t)$ oscillates around a positive finite mean closer to its initial value. For large $\chi$, the oscillations in $\mathcal{E}(t)$ become more persistent---in that they do not exhibit much decay---and possess a higher frequency. Again, the overall qualitative agreement with the corresponding case of the ideal $\mathrm{U}(1)$ QLM is very good, with excellent quantitative agreement at early times.

We now consider a quench at $\mu/\kappa=0.35$, which does not cross the critical point at $\chi=0$. The corresponding quench dynamics of the electric flux are shown in Fig.~\ref{fig:SS_Nd}(d). As typical of quenches close to the critical point, we find in the case of $\chi=0$ that the order parameter approaches zero neither crossing it nor displaying violent dynamics throughout all accessible evolution times in $t$-DMRG. There is no evidence of oscillations similar to those for quenches across the critical point. Repeating this quench at $\chi\neq0$ indicates a qualitative change in the behavior of $\mathcal{E}(t)$. Even for a small value of $\chi=0.2\kappa$, $\mathcal{E}(t)$ settles to a value much larger than $0$ over all accessible evolution times. For larger $\chi$ values that we consider, the electric flux exhibits oscillations that become faster with increasing $\chi$. The qualitative difference for this quench between the deconfined and confined case is quite remarkable, because the system goes from exhibiting asymptotic decay towards zero in its dynamics to persistent oscillations around a mean value much closer to its initial value of $1/2$. The ED results for the corresponding dynamics in the $\mathrm{U}(1)$ QLM show very good qualitative agreement with the $t$-DMRG results for the dynamics in the bosonic mapping~\eqref{eq:BHM} in the demonstrated case of $\chi=\kappa$. The quantitative comparison at early times is also excellent.

\begin{figure*}[t!]
    \centering
    \includegraphics[width=\columnwidth]{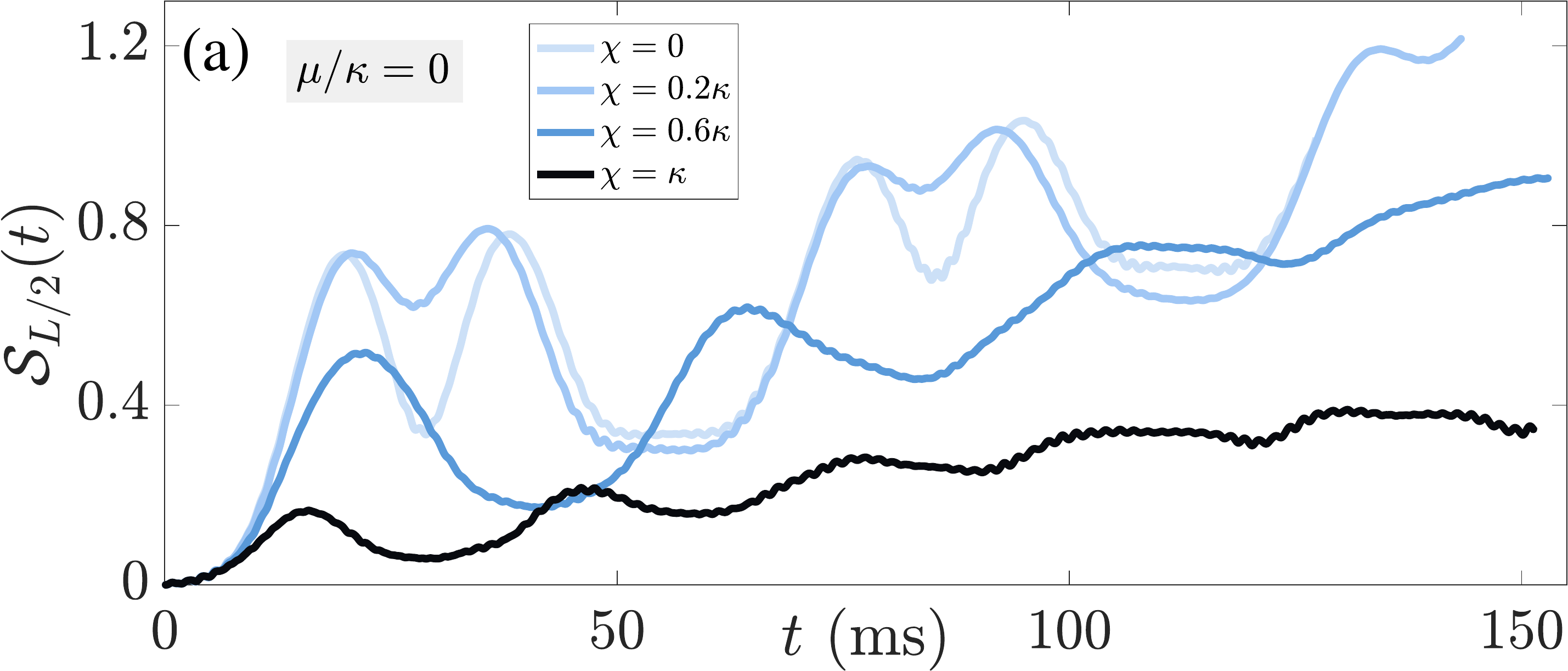}\quad\includegraphics[width=\columnwidth]{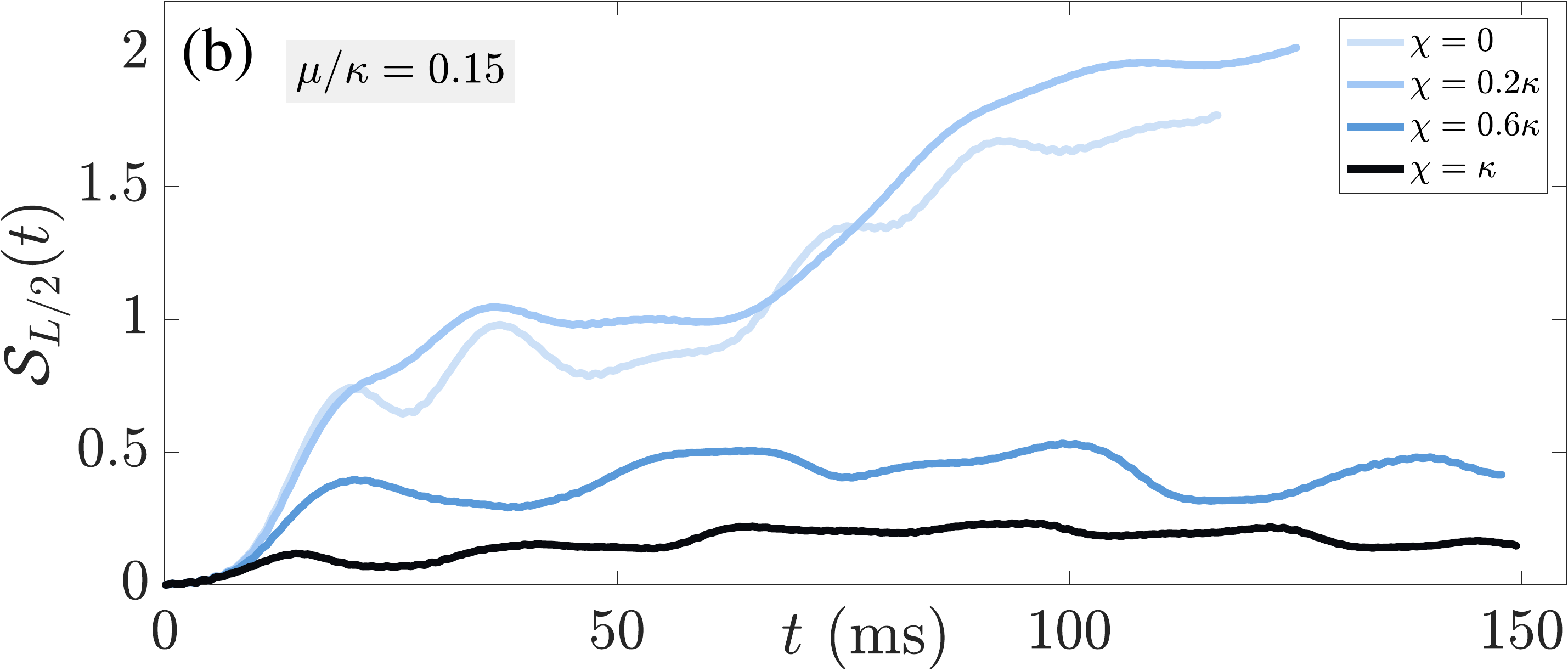}\\
    \vspace{1.1mm}
    \includegraphics[width=\columnwidth]{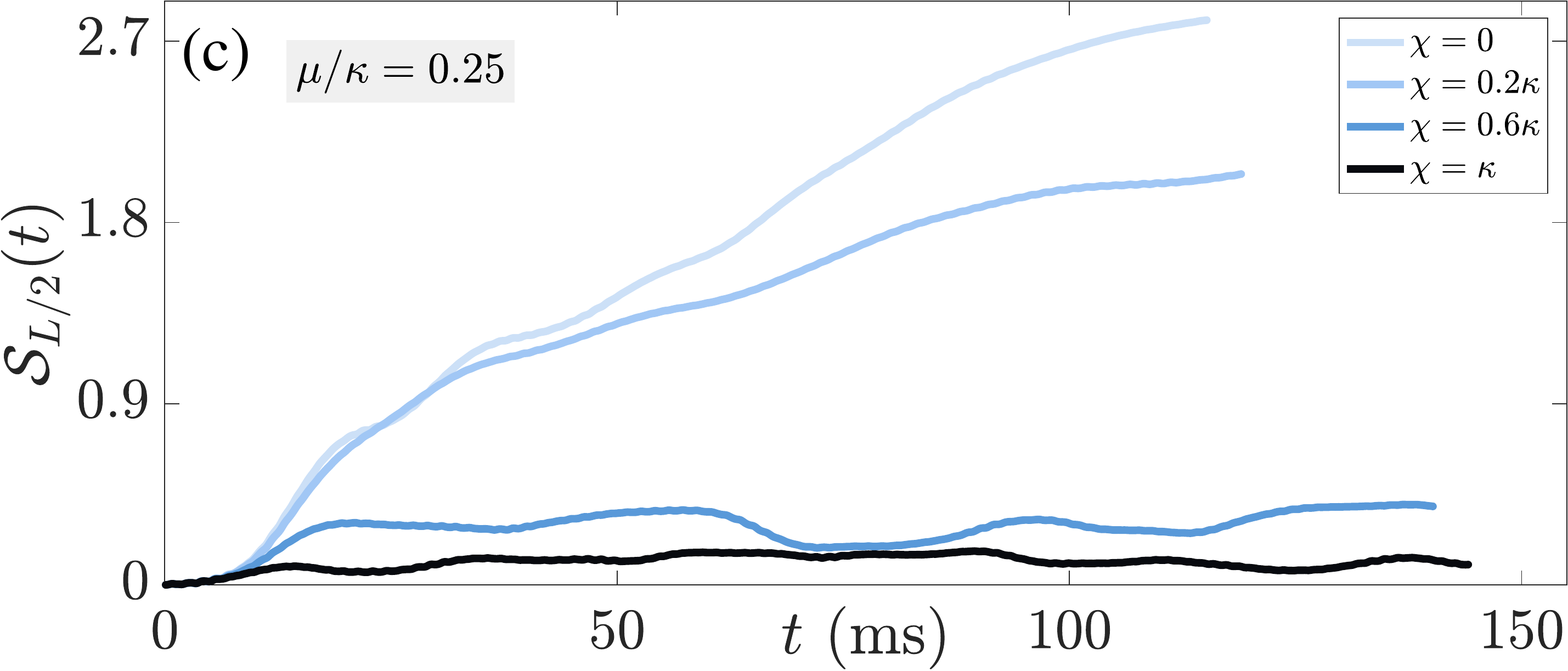}\quad\includegraphics[width=\columnwidth]{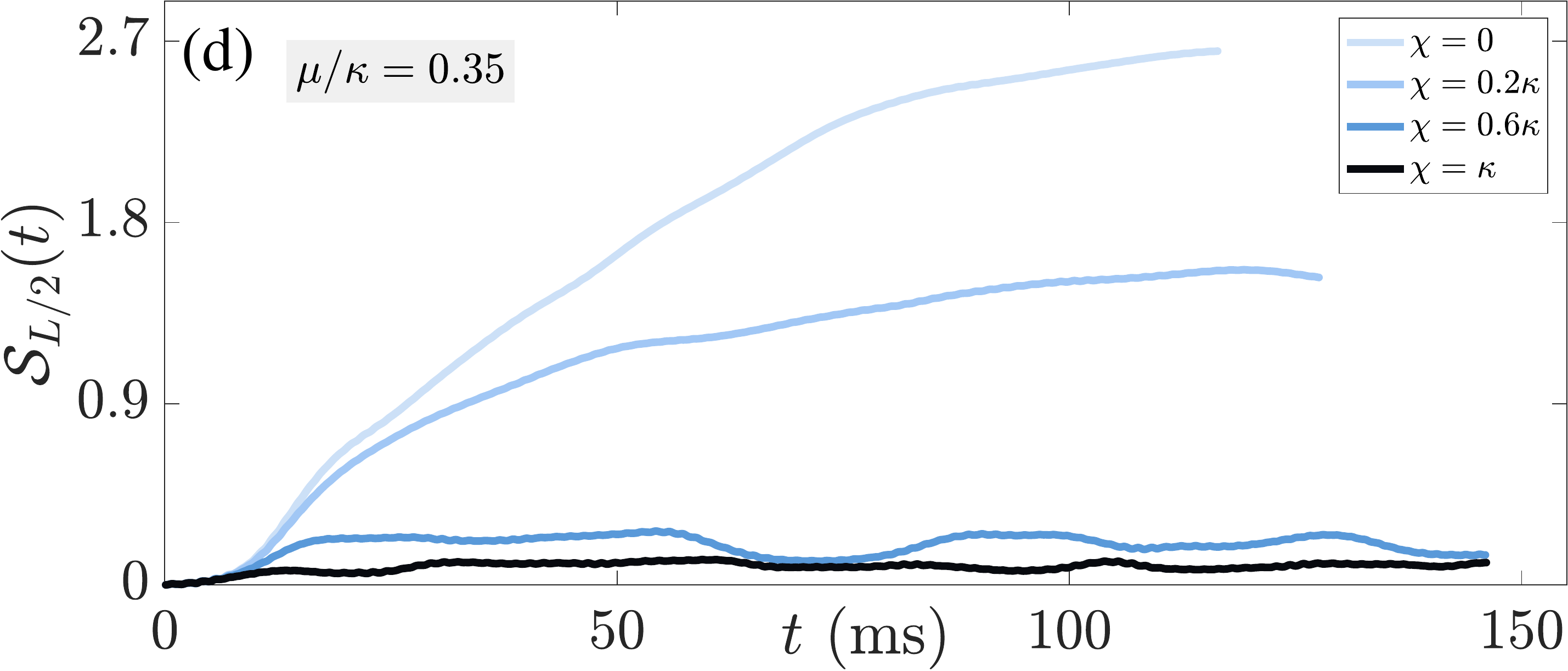}
    \caption{(Color online). Same as Fig.~\ref{fig:SS_Nd} but for the bipartite entanglement entropy, which is Eq.~\eqref{eq:EE} for $j=L/2$. (a) In the case of $\chi/\kappa=0$, this quench of the vacuum state gives rise to scarred dynamics, which manifest in an anomalously low and slowly growing $\mathcal{S}_{L/2}(t)$. Nevertheless, upon increasing $\chi$, the bipartite entanglement entropy is suppressed even further, indicating confinement that constrains the dynamics in the quench Hamiltonian's Hilbert space more than scarring does. (b-d) For quenches at a finite mass $\mu/\kappa>0$, the confined regime shows much more suppressed growth of the bipartite entanglement entropy, where at $\chi=\kappa$ there is almost no growth at late times within the timescales of our numerical calculations.}
    \label{fig:SS_EE}
\end{figure*}

These oscillations we observe in Fig.~\ref{fig:SS_Nd} at larger values of $\chi$ are a hallmark of confinement \cite{Kormos2017}, where the time-evolved wave function is always coming back very close to the initial state as soon as it begins to deviate away from it in its dynamics. A feature worth noting in Fig.~\ref{fig:SS_Nd} is that for $\chi=\kappa$, the oscillations in $\mathcal{E}(t)$ exhibit a larger frequency the larger $\mu/\kappa$ is. In other words, a larger mass makes the confinement more drastic, which has also been numerically found in Ref.~\cite{Surace2020}, for example.

The qualitative picture described in Fig.~\ref{fig:SS_Nd} remains largely intact when considering the corresponding dynamics of the chiral condensate~\eqref{eq:CC}, shown in Fig.~\ref{fig:SS_CC}. Since the vacuum is void of matter, $\mathcal{C}(0)=0$. For $\mu/\kappa=0$, shown in Fig.~\ref{fig:SS_CC}(a), we again see for $\chi=0$ clear signatures of scarring with the chiral condensate exhibiting persistent oscillations up to all accessible evolution times in our $t$-DMRG calculations. Note how in the case of $\chi=0$, the chiral condensate has double the frequency of the order parameter, which we have studied in Fig.~\ref{fig:SS_Nd}(a). This is because the state transfer occurring between the two degenerate vacua due to quantum many-body scarring, which takes half a cycle in the order parameter, will necessitate that meanwhile the chiral condensate complete a full cycle, since it has to be at a local minimum when the wave function is at either vacuum. Upon tuning $\chi$ to larger values, the chiral condensate shows signatures of confinement, remaining closer to its initial value and exhibiting persistent oscillations. Focusing on the case of $\chi=\kappa$, we find that both the chiral condensate and the electric flux share the same frequency, which further confirms that scarring is no longer present and there is therefore no resulting state transfer that leads to a factor of two difference in the frequencies of these observables. Excellent qualitative agreement is displayed in the case of $\chi=\kappa$ with ED results for the corresponding dynamics in the ideal $\mathrm{U}(1)$ QLM.

The transition from deconfined to confined dynamics is even more striking for quenches at a finite mass $\mu/\kappa\neq0$, shown in Fig.~\ref{fig:SS_CC}(b-d). In all cases, we find that the larger $\chi$ is, the more confined is the dynamics of $\mathcal{C}(t)$, with the latter remaining closer to its initial value of zero. Unlike in the zero-mass case where the frequency does not appear to change with $\chi$, for $\mu/\kappa>0$ there is a clear increase in the frequency of $\mathcal{C}(t)$ with increasing $\chi$. In all cases, we find greater persistence in oscillations when $\chi$ is large, particularly at larger $\mu/\kappa$. As in the zero-mass quench, the dynamics shows good qualitative agreement with the corresponding dynamics in the ideal $\mathrm{U}(1)$ QLM at $\chi=\kappa$, as obtained from ED calculations.

\begin{figure*}[t!]
    \centering
    \includegraphics[width=\columnwidth]{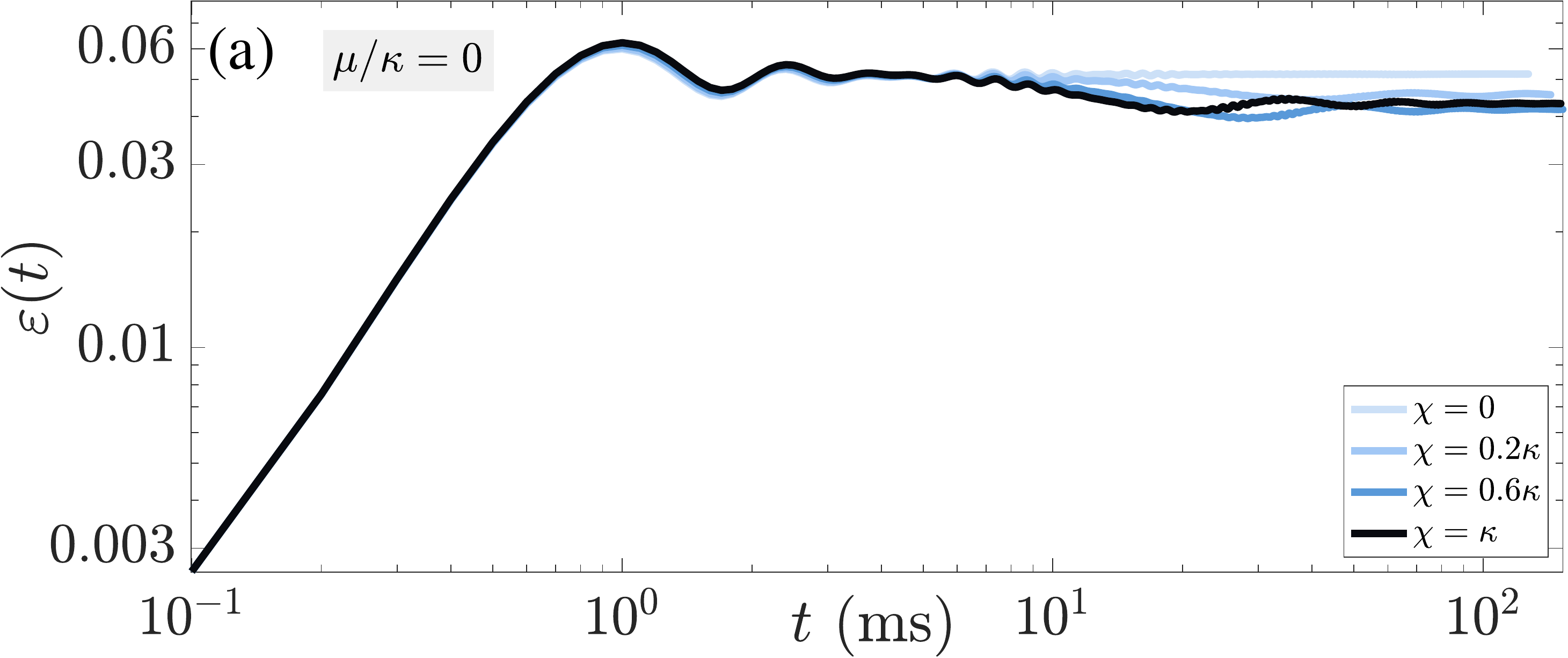}\quad\includegraphics[width=\columnwidth]{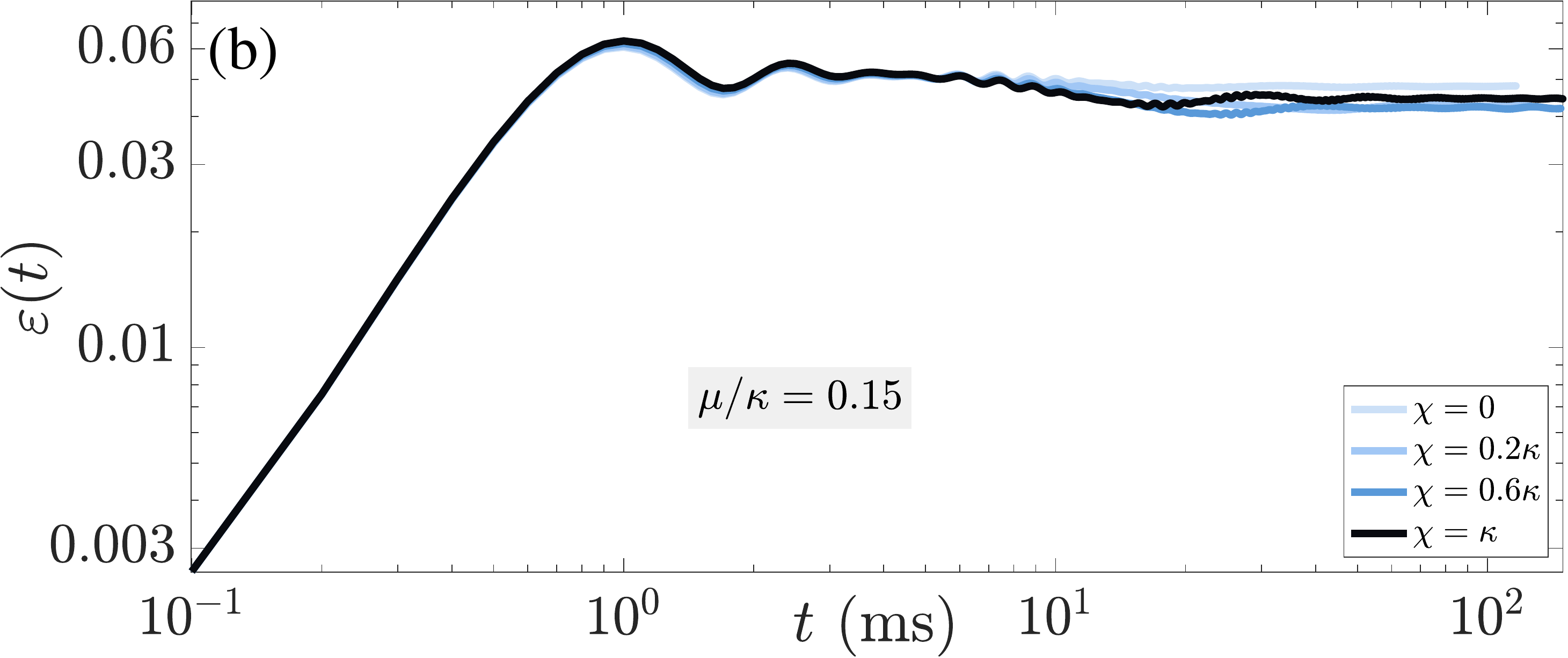}\\
    \vspace{1.1mm}
    \includegraphics[width=\columnwidth]{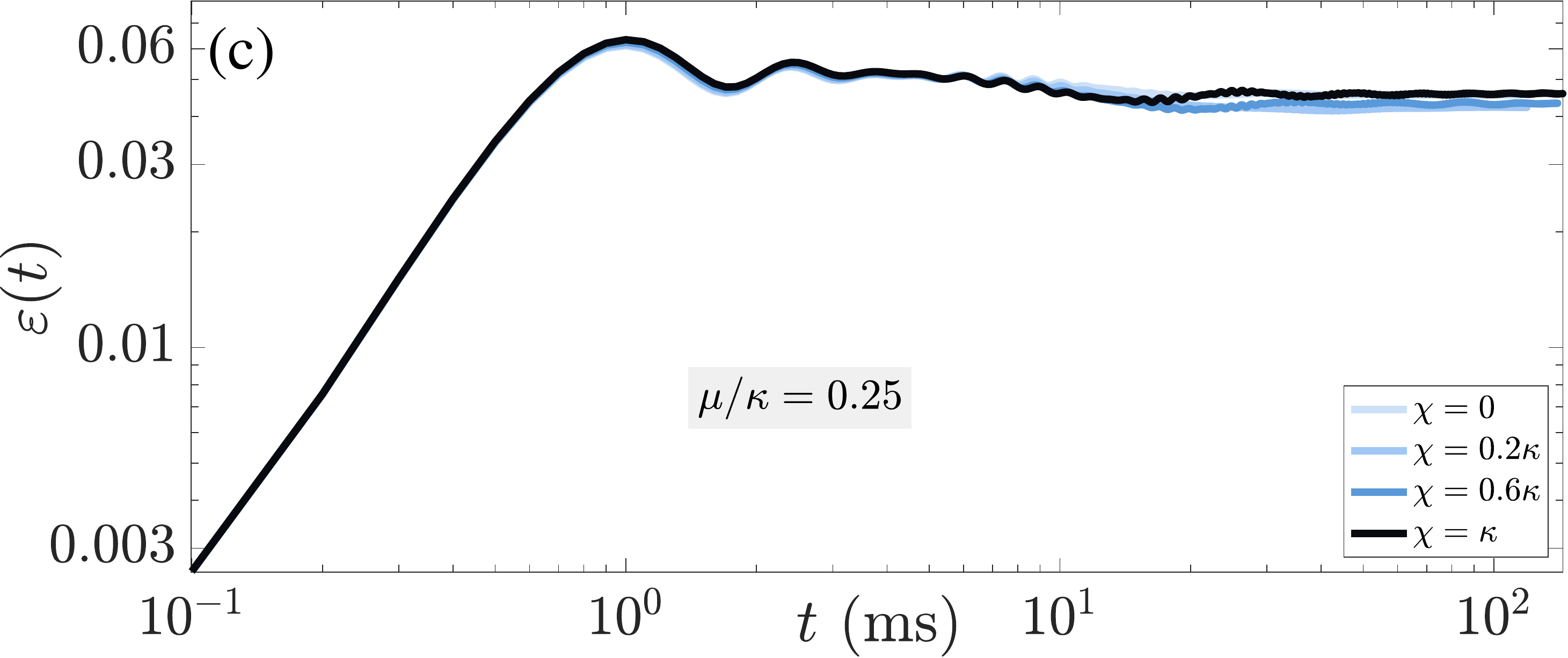}\quad\includegraphics[width=\columnwidth]{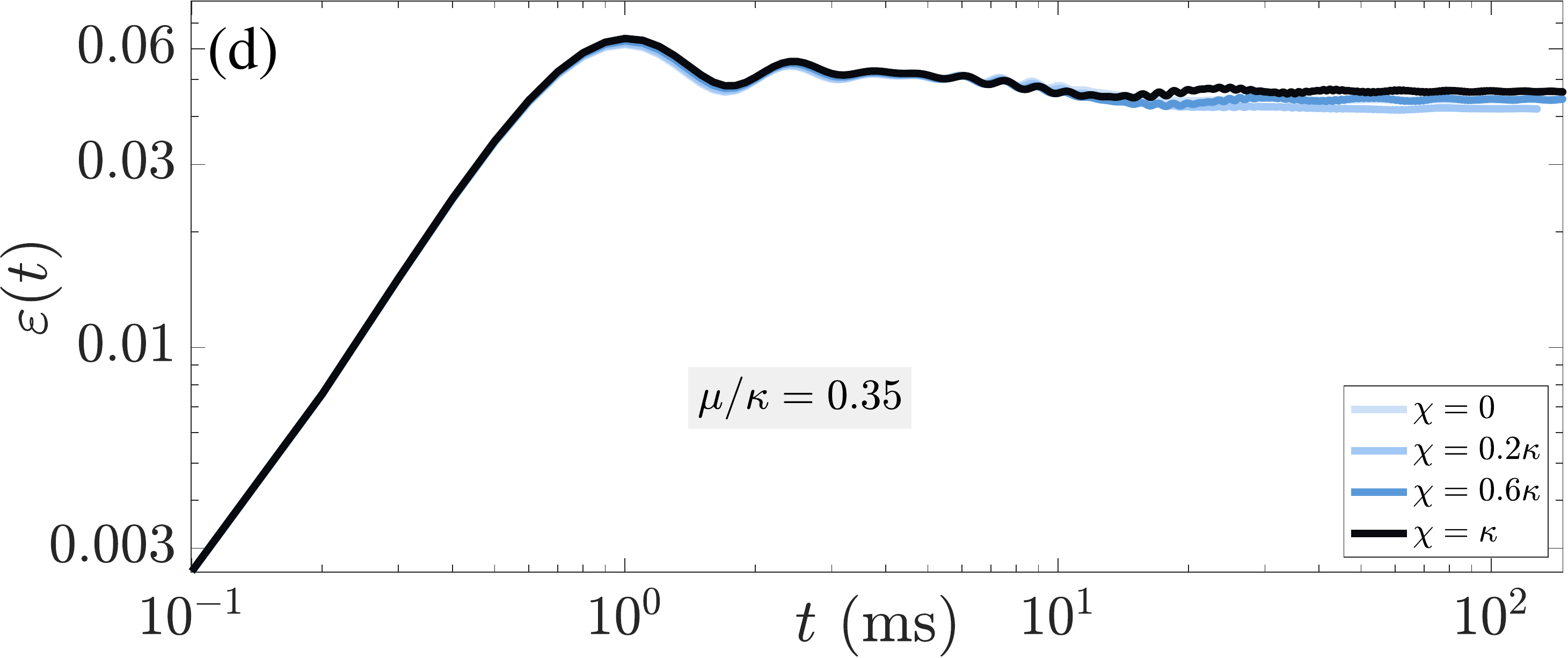}
    \caption{(Color online). Same as Fig.~\ref{fig:SS_Nd} but for the gauge violation~\eqref{eq:viol}. In the effective implementation using a single bosonic species, gauge invariance is enforced through energy penalties, leading to small violations of the exact gauge symmetry generated by the operators defined in Eq.~\eqref{eq:Gj}. Nevertheless, the gauge violations remain small and controlled throughout the entire considered evolution times, leading to a renormalized gauge theory with the same $\mathrm{U}(1)$ gauge symmetry as the ideal model~\eqref{eq:H} \cite{Halimeh2020e}. The exact form of the energy penalty and how it arises in our mapping is discussed in Sec.~\ref{sec:experimentalsetup}, and it is closely related to the recently introduced concept of Stark gauge protection \cite{Lang2022SGP}.}
    \label{fig:SS_viol}
\end{figure*}

Confinement can greatly constrain the spread of the wave function in the Hilbert space of the quench Hamiltonian. A good measure of this spread is the bipartite entanglement entropy, Eq.~\eqref{eq:EE} for $j=L/2$, the dynamics of which we have calculated in $t$-DMRG for the above quenches, shown in Fig.~\ref{fig:SS_EE}. In the case of the zero-mass quench, even the deconfined regime ($\chi=0$) shows an anomalously low and slowly growing bipartite entanglement entropy in Fig.~\ref{fig:SS_EE}(a)---in contrast with the deconfined case for nonzero-mass quenches in Fig.~\ref{fig:SS_EE}(b-d). Upon increasing $\chi$ for a quench at any mass, $\mathcal{S}_{L/2}(t)$ is further suppressed and grows significantly slower, indicating constrained dynamics in a confined subspace of the quench-Hamiltonian Hilbert space. In the massless quench, the bipartite entanglement entropy is suppressed even further in the confined regime than in the deconfined one where scarring is present. This shows that whereas scarring involves constrained dynamics in a cold subspace of the quench Hamiltonian's Hilbert space, confinement further constrains the dynamics within this subspace. For quenches at a nonzero $\mu/\kappa$ with $\chi\gtrsim0.6\kappa$, confinement is so striking that the bipartite entanglement entropy seems to no longer grow beyond short times.

\begin{figure*}[t!]
    \centering
    \includegraphics[width=0.35\columnwidth]{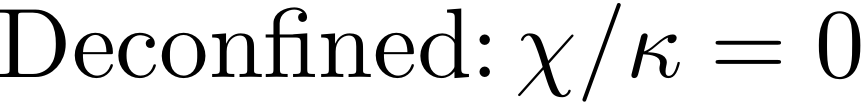}\quad\quad\quad\quad\quad\quad\quad\quad\quad\quad\quad\quad\quad\quad\quad\quad\quad\quad\quad\quad\includegraphics[width=0.3\columnwidth]{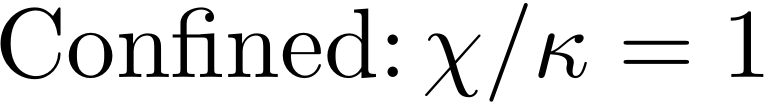}\\
    \vspace{2.1mm}
    \includegraphics[width=\columnwidth]{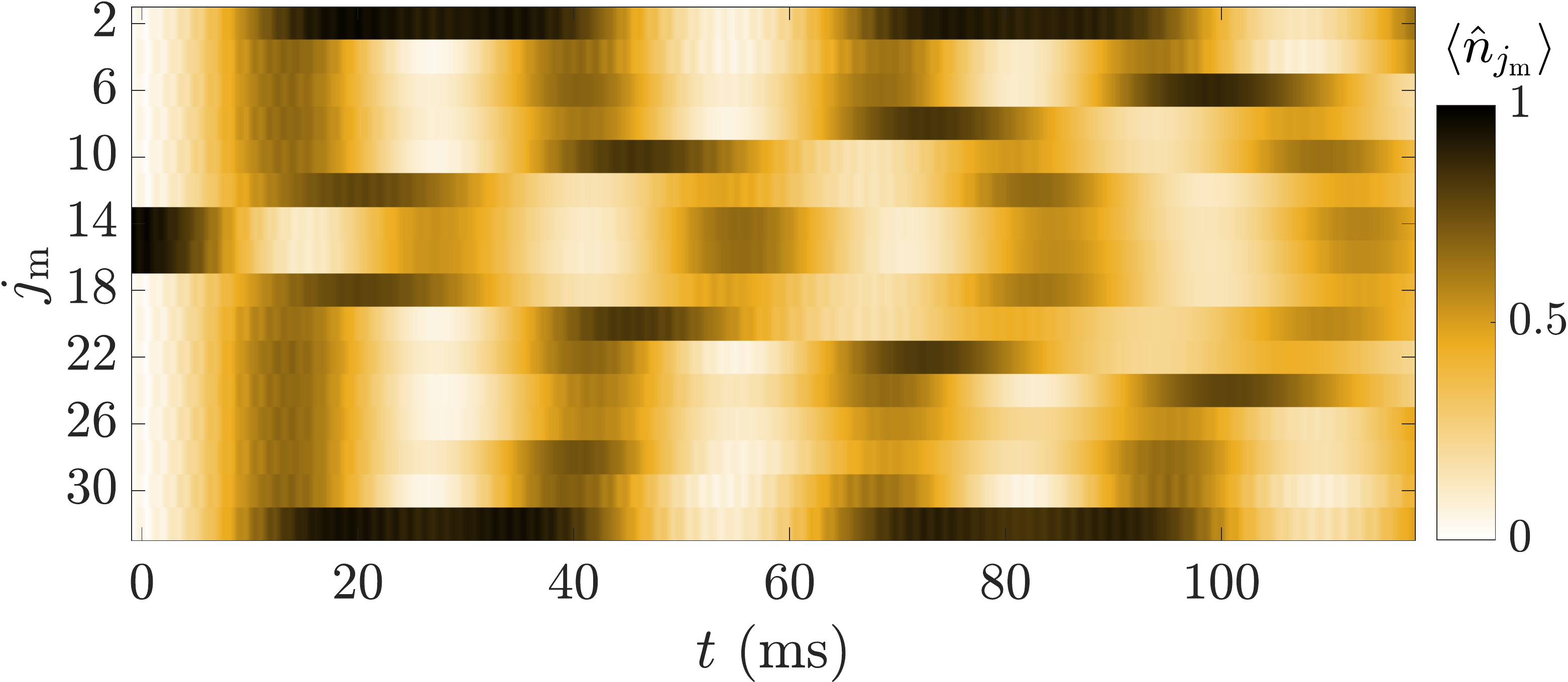}\quad\includegraphics[width=\columnwidth]{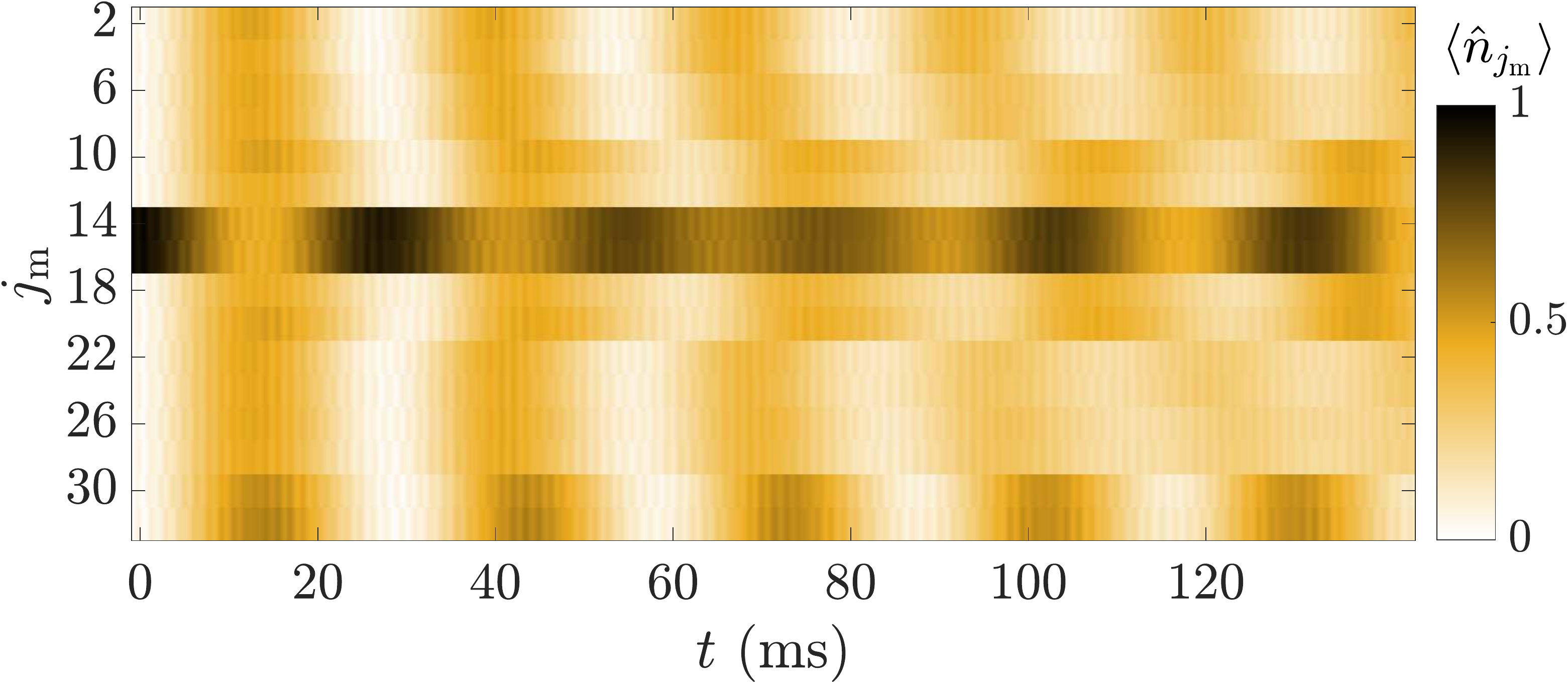}\\
    \vspace{1.1mm}
    \includegraphics[width=\columnwidth]{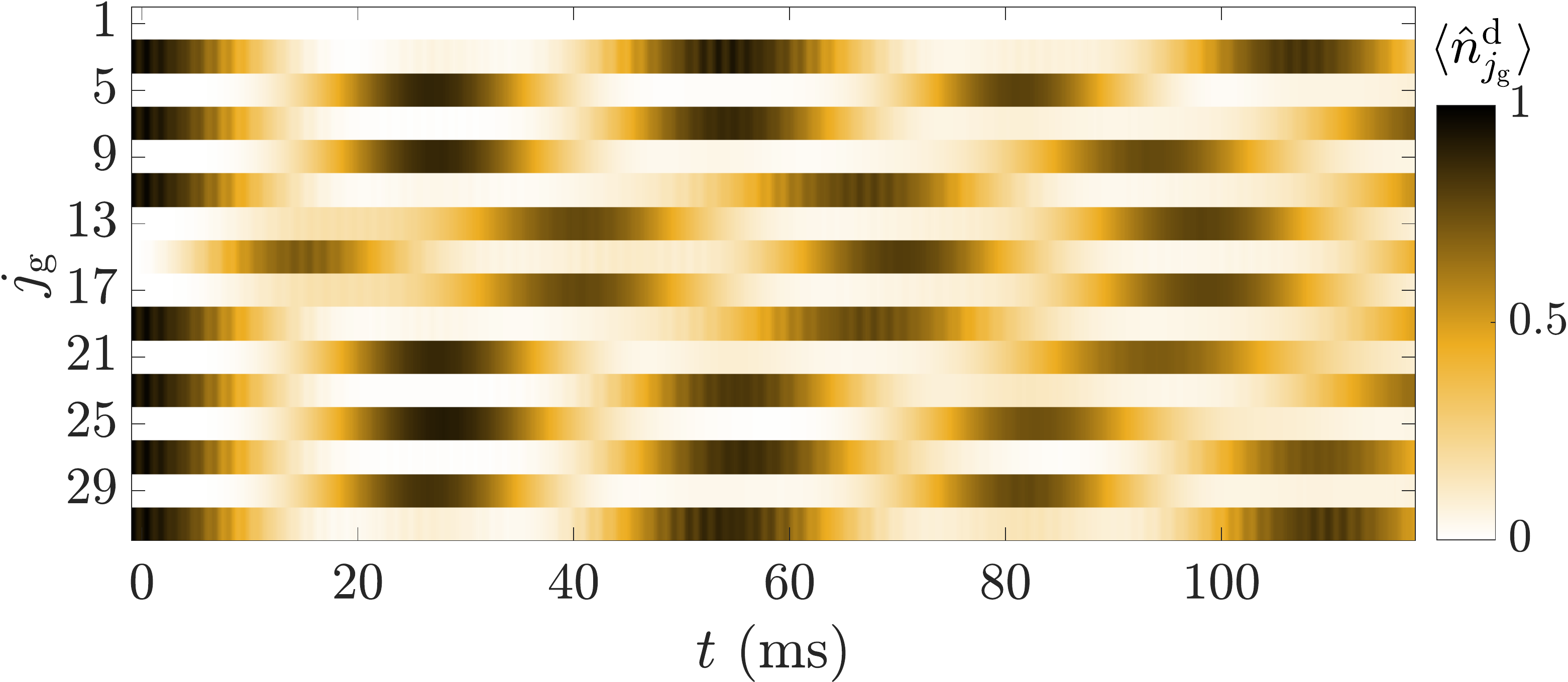}\quad\includegraphics[width=\columnwidth]{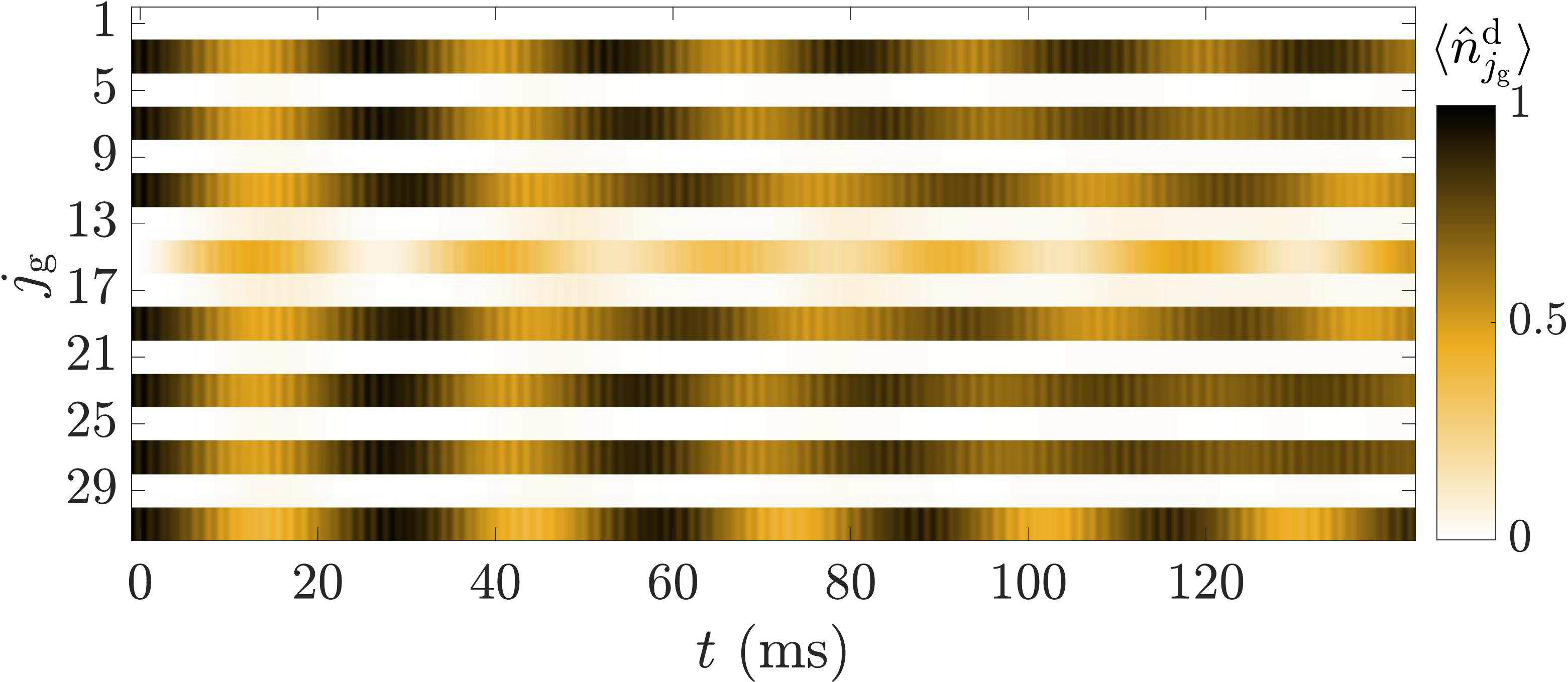}\\
    \vspace{1.1mm}
    \includegraphics[width=\columnwidth]{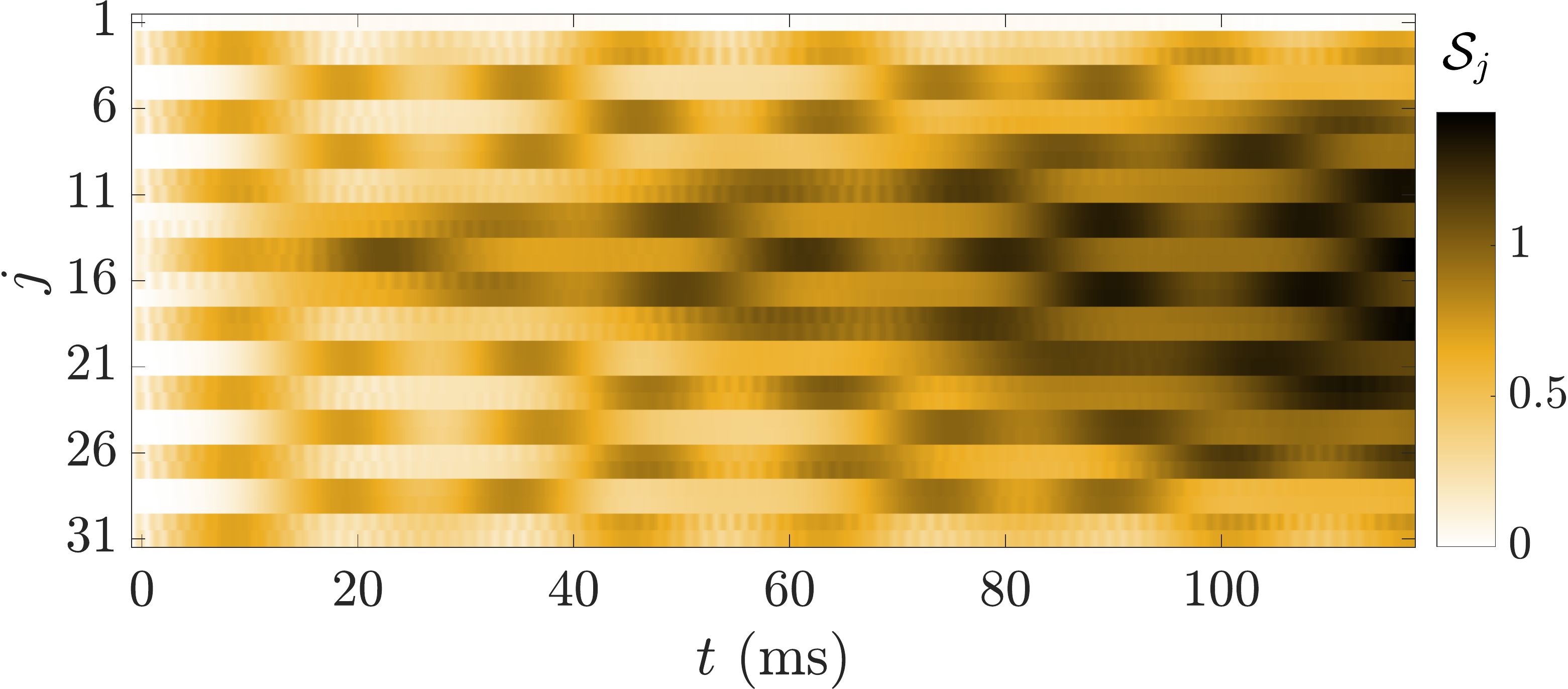}\quad\includegraphics[width=\columnwidth]{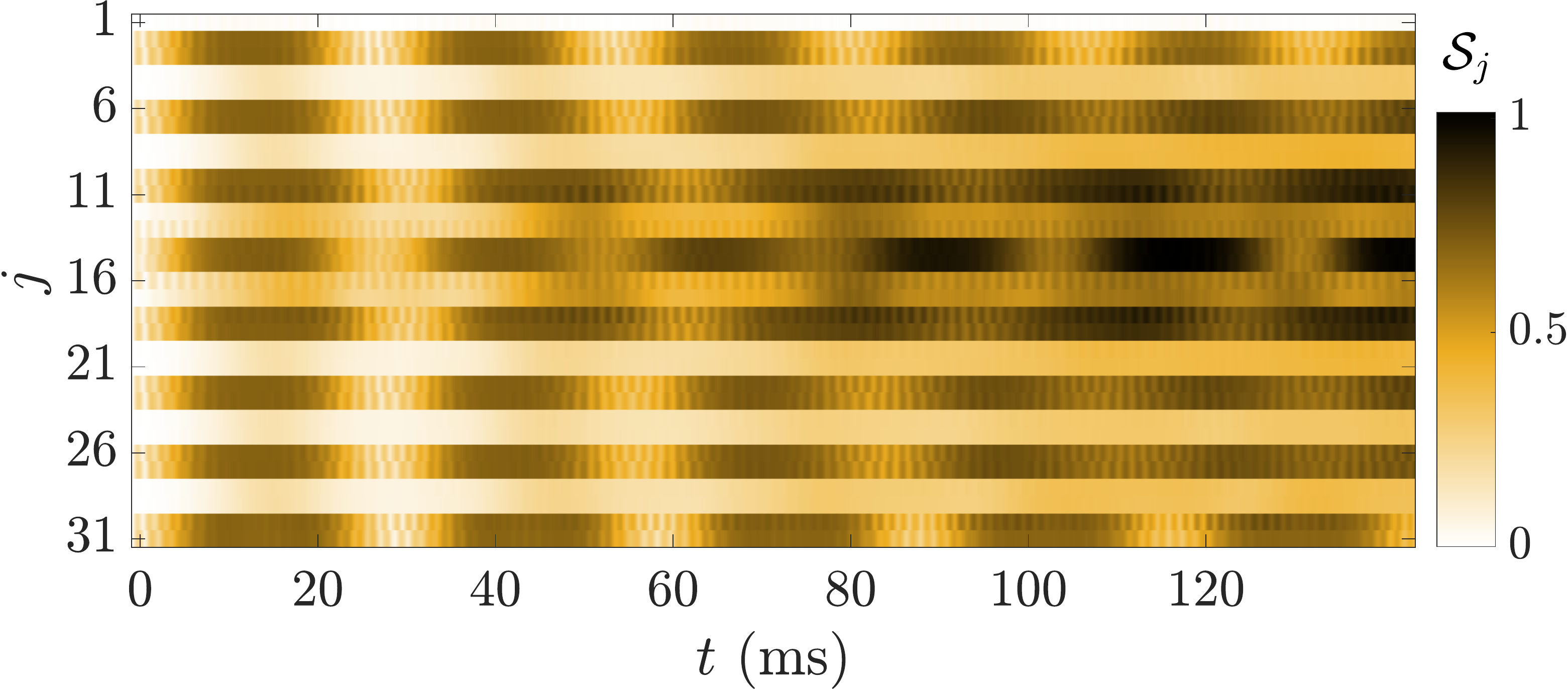}
    \caption{(Color online). Dynamics of the electron-positron pair state (see Fig.~\ref{fig:initilization}) in the wake of a quench by Hamiltonian~\eqref{eq:BHM} at $\mu/\kappa=0$, with the strength of the topological $\theta$-term set to $\chi/\kappa=0$ (left-column panels) or $\chi/\kappa=1$ (right-column panels). Top panels show the quench dynamics of the matter occupation on the even sites of the optical superlattice (where matter fields reside), representing the local chiral condensate. In the deconfined case ($\chi/\kappa=0$), we find clear ballistic dynamics, which can be interpreted as the electron and positron propagating away from each other linearly in time at no energy cost. The dynamics fundamentally changes when $\chi/\kappa=1$, where the electron-positron pair is confined, and there is virtually no dynamics. This qualitative picture is confirmed in the dynamics of the site-resolved doublon occupation (middle panels) on odd sites of the superlattice, which represents the local electric flux. Whereas the flipped electric flux between the electron and positron can no longer be distinguished at long times from the electric fluxes at other sites in the deconfined case, when $\chi=\kappa$ confinement stabilizes this flux up to all accessible evolution times. As shown in the bond entanglement entropy in the bottom panels, in the deconfined regime $\mathcal{S}_j(t)$ spreads qualitatively much faster than in the confined regime.}
    \label{fig:PS_mbykappa0}
\end{figure*}

\begin{figure*}[t!]
    \centering
    \includegraphics[width=0.35\columnwidth]{{cap_left}.pdf}\quad\quad\quad\quad\quad\quad\quad\quad\quad\quad\quad\quad\quad\quad\quad\quad\quad\quad\quad\quad\includegraphics[width=0.31\columnwidth]{{cap_right}.pdf}\\
    \vspace{2.5mm}
    \includegraphics[width=\columnwidth]{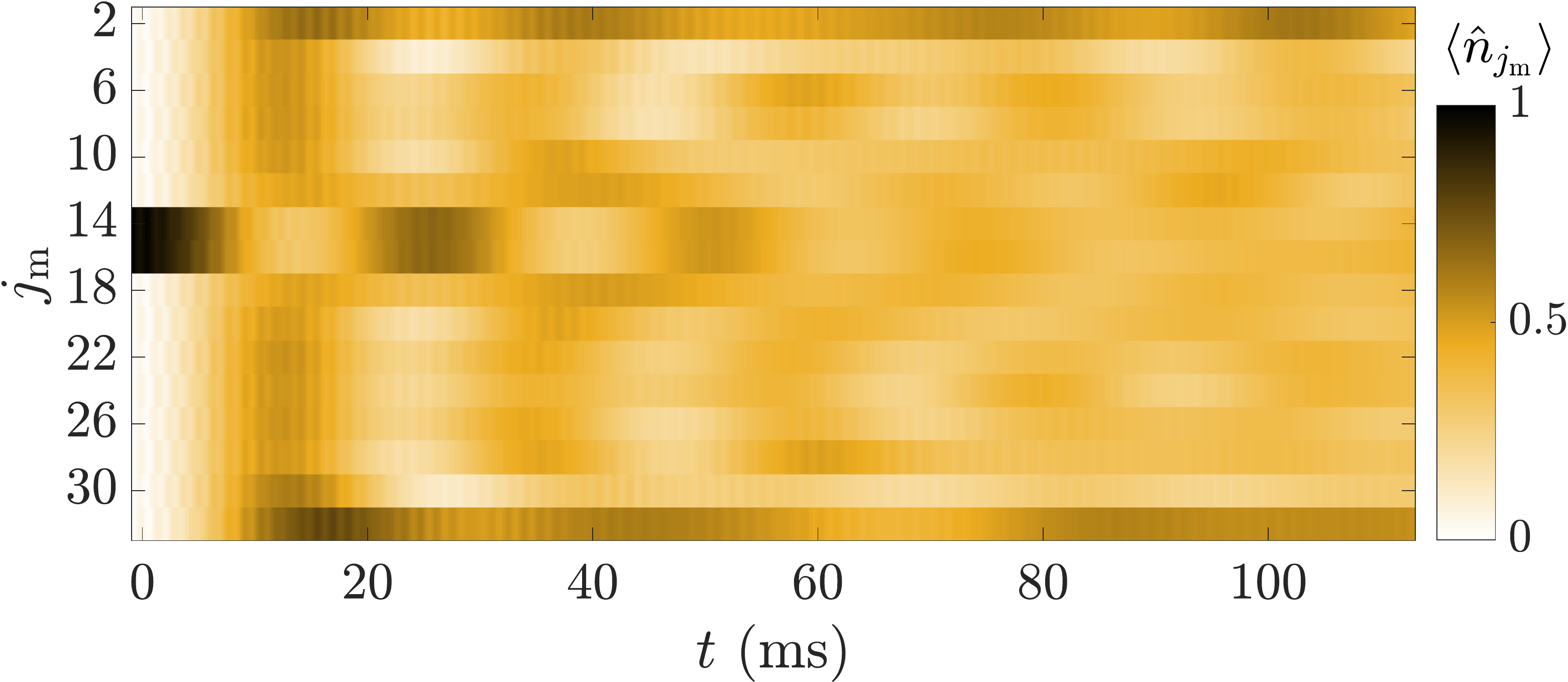}\quad\includegraphics[width=\columnwidth]{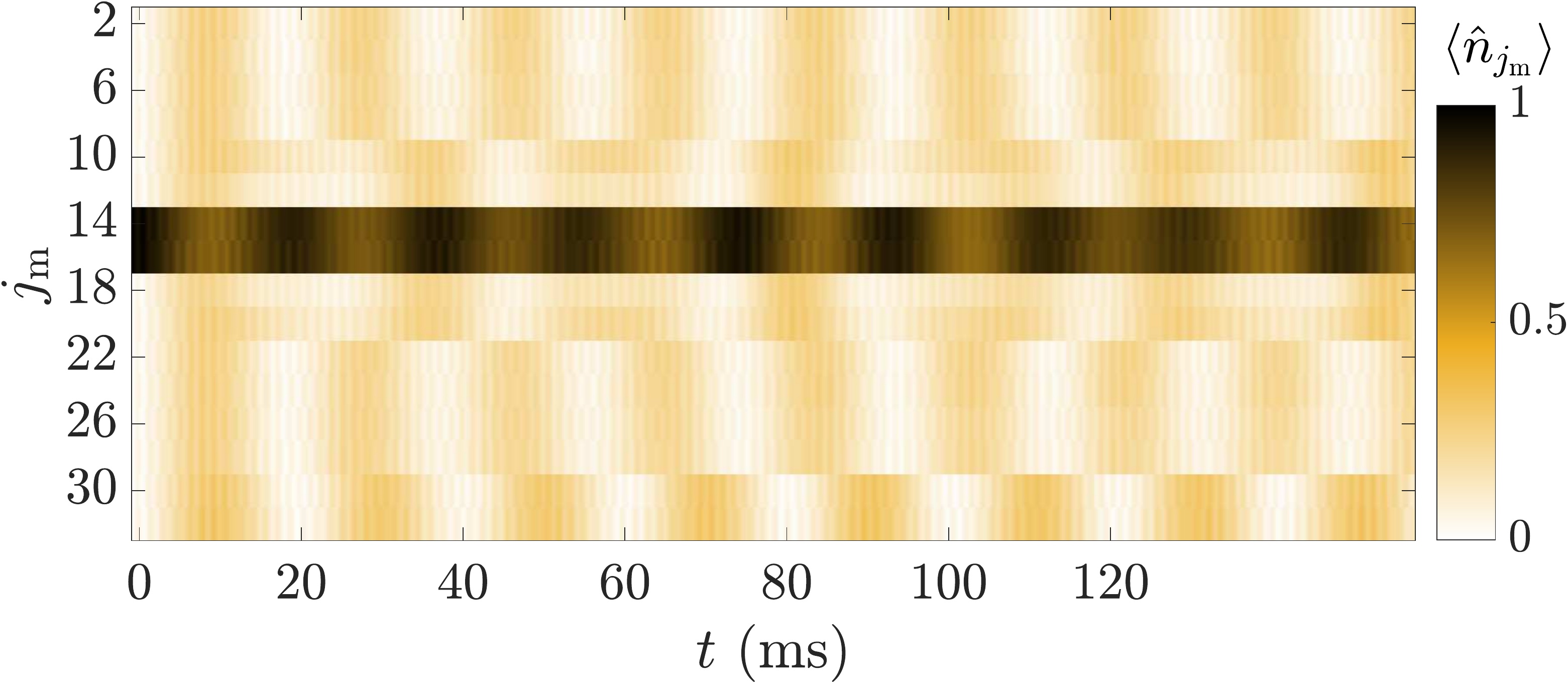}\\
    \vspace{1.1mm}
    \includegraphics[width=\columnwidth]{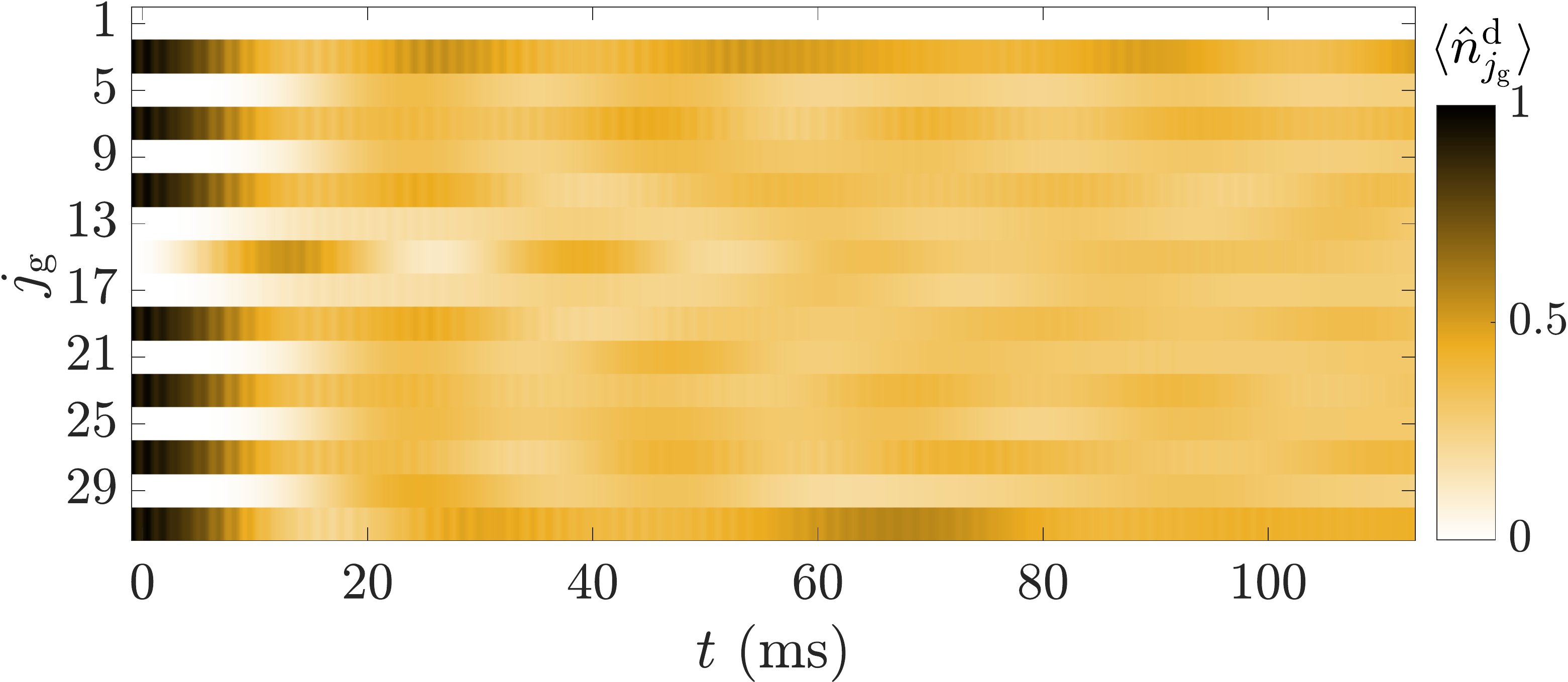}\quad\includegraphics[width=\columnwidth]{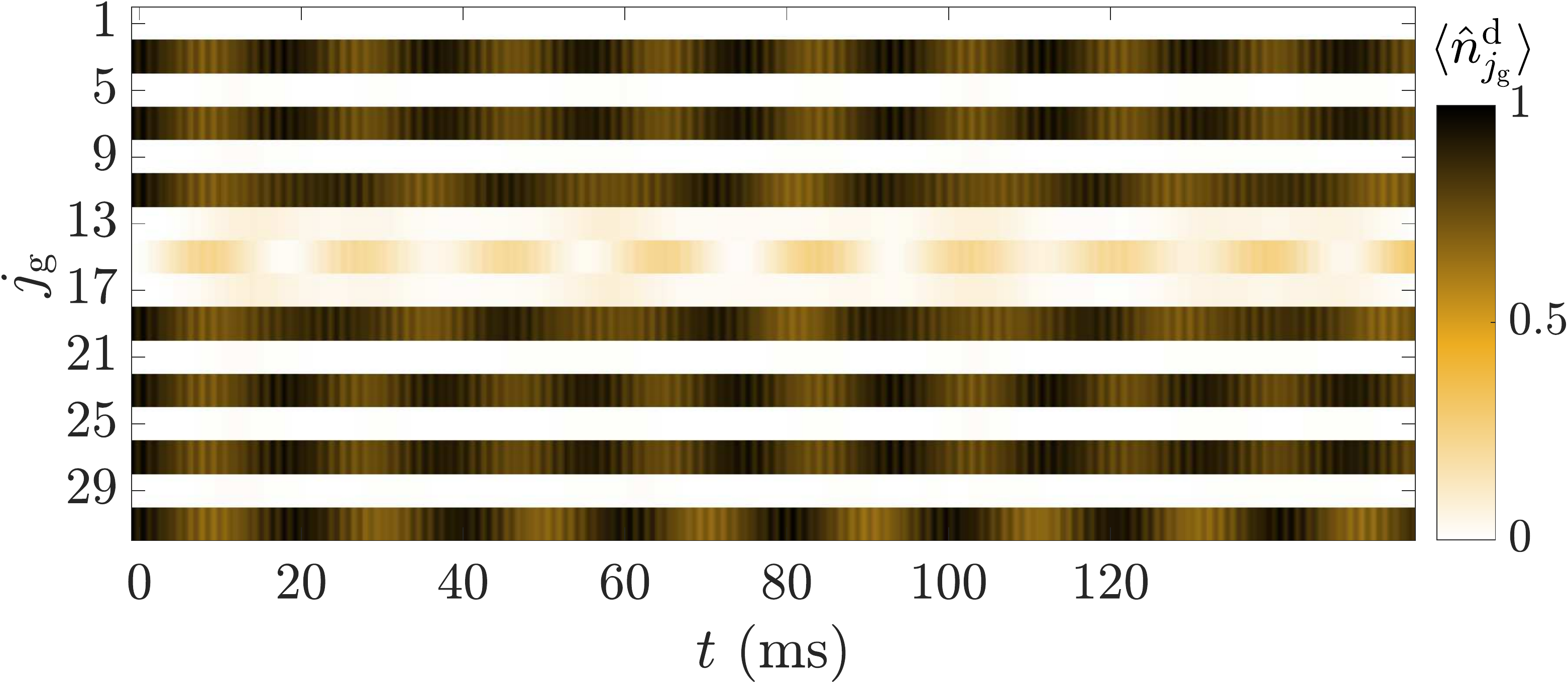}\\
    \vspace{1.1mm}
    \includegraphics[width=\columnwidth]{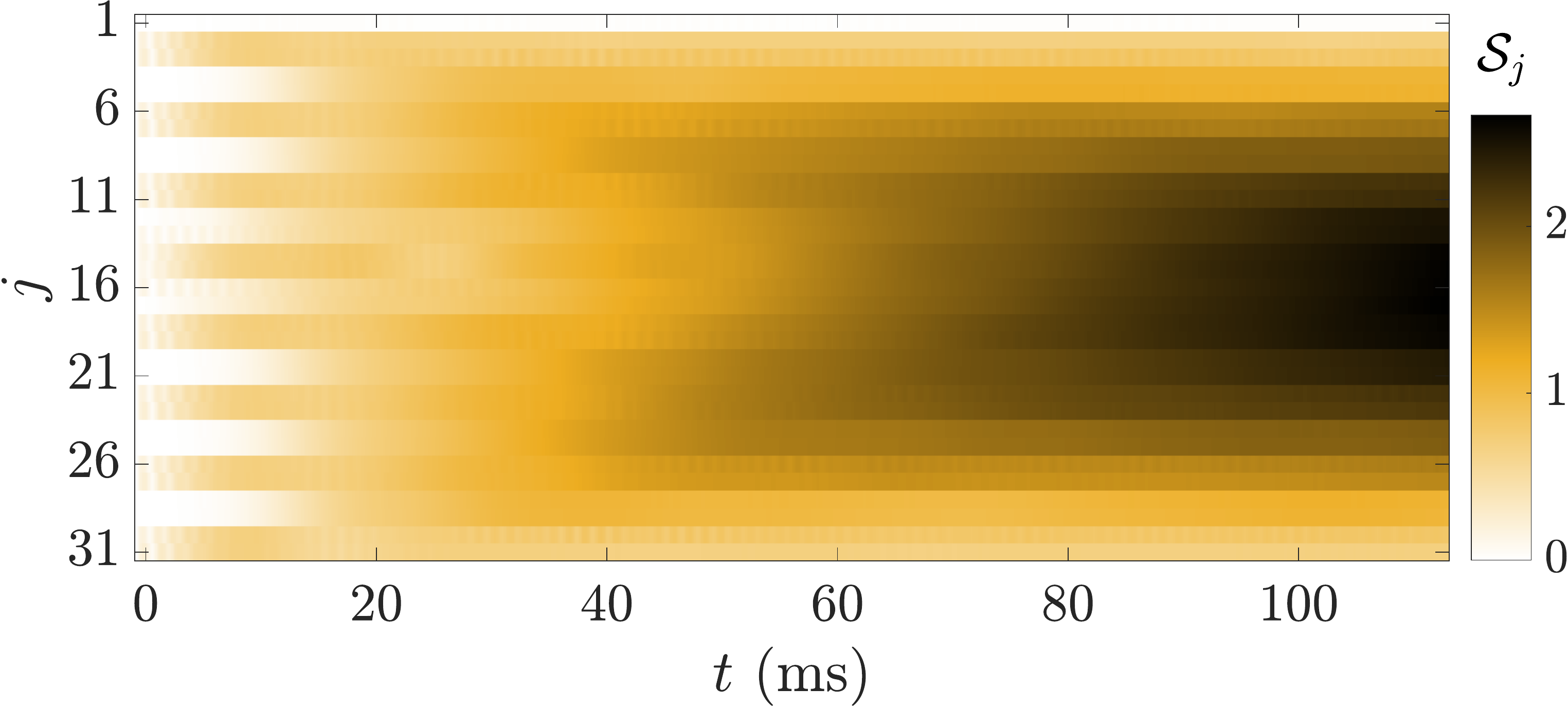}\quad\includegraphics[width=\columnwidth]{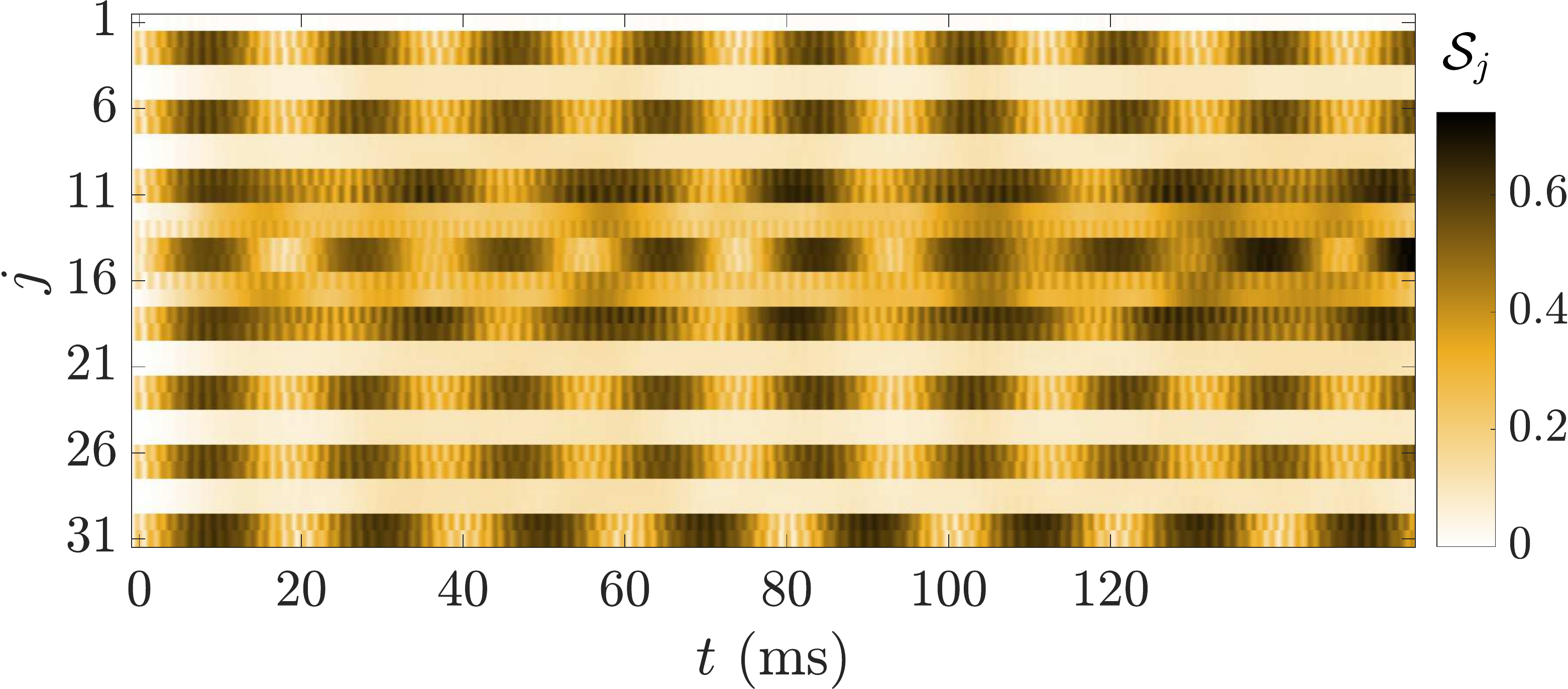}
    \caption{(Color online). Time evolution of the site-resolved matter occupation (top panels), which resides on even sites of the superlattice and represents the local chiral condensate, the site-resolved doublon occupation (middle panels) on odd sites of the superlattice, which represents the local electric flux, and the site-resolved bond entanglement entropy (bottom panels) in the wake of a global quench at $\mu/\kappa=0.35$ starting in the electron-positron pair state, within either the deconfined regime ($\chi/\kappa=0$, left-column panels) or its confined counterpart ($\chi/\kappa=1$, right-column panels). In agreement with results in Sec.~\ref{sec:QuenchDynamics_SS}, we find that a finite mass enhances the confinement for a given value of $\chi/\kappa$. In the deconfined regime, a finite mass leads to quench dynamics that is markedly faster than the ballistic behavior seen in the corresponding case of a massless quench (see Fig.~\ref{fig:PS_mbykappa0}, left-column panels). However, at $\chi=\kappa$, the confinement is much stronger for the quench at $\mu/\kappa=0.35$ than that at $\mu/\kappa=0$.}
    \label{fig:PS_mbykappa0.35}
\end{figure*}

Even though our numerical benchmarks with ED results on the ideal $\mathrm{U}(1)$ QLM have shown good qualitative agreement, indicating that the bosonic mapping of Eq.~\eqref{eq:BHM} is an adequate renormalized $\mathrm{U}(1)$ gauge theory, it is interesting to look at the dynamics of the gauge violation in the considered quenches. Considering local constraints specified at a matter (even) site $j_\mathrm{m}$, then only three local configurations at this matter site and its neighboring gauge sites in the bosonic system satisfy Gauss's law: $\ket{0,0,2}$, $\ket{2,0,0}$ that locally correspond to vacuum, and $\ket{0,1,0}$ that locally corresponds to the charge-proliferated state. The local projector onto the vacua local configurations is then $\hat{\mathcal{P}}_{j_\mathrm{m}}^{\ket{002}+\ket{200}}$, and that onto the charge-proliferated local configuration is $\hat{\mathcal{P}}_{j_\mathrm{m}}^{\ket{010}}$. In the ideal $\mathrm{U}(1)$ QLM, these are the only allowed local configurations. However, as explained in Sec.~\ref{sec:mapping}, our mapping is perturbatively valid in the limit of $U\approx2\delta\gg J$, but subleading terms from perturbation theory violate Gauss's law. Here, we define the gauge violation as
\begin{align}\nonumber
    \varepsilon(t)=1-\frac{2}{Lt}\int_0^tds\sum_{m=1}^{L/2}\Big[&\bra{\psi(s)}\hat{\mathcal{P}}_{2m}^{\ket{002}+\ket{200}}\ket{\psi(s)}\\\label{eq:viol}
    +&\bra{\psi(s)}\hat{\mathcal{P}}_{2m}^{\ket{010}}\ket{\psi(s)}\Big].
\end{align}
The dynamics of the gauge violation~\eqref{eq:viol} is shown in Fig.~\ref{fig:SS_viol} for the above considered quenches. Regardless of the values of $\mu/\kappa$ and $\chi$, we find very good stability of gauge invariance throughout the whole duration of the experiment, where the gauge violation settles into a steady state with finite value after a small increase at short times. In fact, it seems that at larger values of $\chi$ the value of the gauge-violation plateau is slightly lower. Nevertheless, the overall picture of a gauge violation that is constant at intermediate to long evolution times, and which is significantly below $10\%$, indicates a very reliable implementation that quantum-simulates faithful gauge-theory dynamics.

This stability arises from an effective \textit{linear gauge protection} term
\begin{align}\label{eq:LinPro}
    V\hat{H}_G=\sum_\ell c_\ell\hat{G}_\ell,
\end{align}
where $\hat{G}_\ell$ is the generator of Gauss's law, given in Eq.~\eqref{eq:Gj}, and $c_\ell$ are a sequence of numbers depending on the matter site $\ell$ and the parameters of Eq.~\eqref{eq:BHM}. 
If the $c_\ell$ are chosen appropriately, violations of Gauss's law are energetically penalized, which in ideal situations can stabilize the gauge symmetry up to exponentially long times \cite{Halimeh2020a}. 
In the present case, the energy penalty is provided by on-site interaction strength $U$, the tilt $\Delta$, and the staggered potential $\delta$. As discussed in Sec.~\ref{sec:experimentalsetup}, with coefficients explicitly read $c_\ell = (-1)^\ell[\Delta \ell + (U-\delta + \Delta/2)]$. 
As comparison with Ref.~\cite{Halimeh2020e} shows, the addition of the $\theta$-angle does not modify the protection with respect to the recent experiments of Refs.~\cite{Yang2020,Zhou2021}, where approximately gauge-invariant adiabatic and quench dynamics have already been shown. The success of this protection scheme with the above coefficients, that include a staggered term as well as a tilt, can be further corroborated based on the concept of \textit{Stark gauge protection} \cite{Lang2022SGP}, from which it can be shown that gauge symmetry is stabilized up to essentially all relevant timescales.

\subsubsection{Initial state: electron-positron pair}\label{sec:QuenchDynamics_PS}
A very pertinent state to consider when investigating confinement is that of an electron-positron pair on top of vacuum \cite{Hebenstreit2013simulating}. In our bosonic mapping, this is equivalent to having all sites empty except for the middle two matter sites, with each hosting a single boson. With the impressive advancements of quantum gas microscopes \cite{Bakr2009}, it is now possible to probe the dynamics of electron-positron pairs in modern quantum simulations. We will now quench this system for different values of $\mu/\kappa$ and $\chi/\kappa$ and study the ensuing dynamics of the electron-positron pair.

The first quench is at $\mu/\kappa=0$, shown in Fig.~\ref{fig:PS_mbykappa0} for the deconfined case with $\chi=0$ (left column) and for the confined case with $\chi=\kappa$ (right column). The top row shows strikingly different dynamics for the matter fields---defined as the bosonic occupation on even sites of the superlattice---between the deconfined and confined regimes. Whereas for $\chi=0$ the electron and positron propagate ballistically away from each other indicative of deconfined dynamics, for $\chi=\kappa$ they are strongly confined up to all accessible evolution times in $t$-DMRG.

This picture is confirmed by the corresponding dynamics of the electric fluxes on gauge links, represented in the bosonic model by the doublon-occupation $\hat{n}^\mathrm{d}_{j_\mathrm{g}}$ on the odd sites of the superlattice. Whereas for $\chi=0$ there is evidence of ballistic propagation in the associated flux in between the electron-positron pair, for $\chi=\kappa$ it displays strongly confined dynamics, with the local flux remaining in roughly the same configuration throughout all accessible evolution times. This is to be expected given that the associated electron-positron pair is also confined.

To further validate this conclusion, we look at the von Neumann entanglement entropy at each bond in the lattice. There is a clear ballistic spread in the entanglement entropy in the deconfined dynamics with $\chi=0$. In contrast, when $\chi=\kappa$, we find that the entanglement entropy growth is strongly suppressed, which is typical of confined dynamics.

Let us now repeat the same quenches but for a mass $\mu/\kappa=0.35$. The corresponding quench dynamics are shown in Fig.~\ref{fig:PS_mbykappa0.35} for $\chi=0$ (left column) and $\chi=\kappa$ (right column). The contrast here between the deconfined and confined regimes is even more striking than in the case of the zero-mass quenches in Fig.~\ref{fig:PS_mbykappa0}. In the deconfined regime ($\chi=0$, left column), the initial presence of the electron-particle pair is quickly washed out in both the site-resolved matter occupation and electric flux. However, for $\chi=\kappa$ (right column), the dynamics is strongly confined, and the matter and electric-flux configurations remain virtually unchanged throughout the dynamics. This is also reflected in the entanglement entropy dynamics shown in the bottom row of panels in Fig.~\ref{fig:PS_mbykappa0.35}. Whereas for $\chi=0$ (left) the dynamics of $\mathcal{S}_j(t)$ indicates a very fast spread throughout the Hilbert space of the quench Hamiltonian typical of deconfinement, for $\chi=\kappa$ it shows very slow spreading indicative of strong confinement.

\section{Conclusions and Outlook}\label{sec:conc}
We have presented an experimental proposal for realizing the spin-$1/2$ quantum link formulation of $1+1$D quantum electrodynamics on a cold-atom quantum simulator including a tunable topological $\theta$-angle. The setup is composed of a tilted Bose--Hubbard superlattice with three periodicities, that not only allow for the stabilization of gauge invariance through the entire duration of the experiment, but also give rise to the topological $\theta$-term in the effective model derived in leading order perturbation theory. We discuss how an effective \textit{Stark gauge protection} term \cite{Lang2022SGP} emerges that allows for a renormalized gauge theory with the same local gauge symmetry as the ideal model.

Using the time-dependent density matrix renormalization group method, we have calculated the time evolution of the vacuum state during an adiabatic ramp to probe the effect of a modification of the $\theta$-angle term on Coleman's phase transition. Our results suggest that our proposed experiment can probe how Coleman's phase transition disappears once the topological $\theta$-angle is tuned away from $\pi$.

We further calculated the far-from-equilibrium dynamics of the vacuum state for massless and massive quenches in both the deconfined ($\chi=0$) and confined ($\chi\neq0$) regimes. The qualitative difference between both regimes is striking, especially at large fermionic rest mass. For a massless quench of the vacuum, the weak ergodicity breaking paradigm of quantum many-body scars emerges in the form of persistent oscillations in local observables lasting well beyond relevant timescales within the deconfined regime. However, at finite values of the $\theta$-term strength, scarring is undermined by confinement, and state transfer between the two degenerate vacua, a staple of scarring in this system, is prohibited as the time-evolved wave function stays close to the initial vacuum state over all times. Numerical benchmarks comparing the dynamics of the bosonic model with those of the ideal target gauge theory show very good qualitative agreement throughout the accessible timescales, with excellent quantitative agreement at short times.

A paradigmatic ``Gedanken'' state for the investigation of confinement in particle physics is that of an electron-positron pair in a vacuum background. We have presented numerical simulations of a massless as well as a massive global quench on this initial state at various values of the $\theta$-term strength. Whereas in the deconfined regime the dynamics exhibits ballistic propagation in local observables, and a fast spread in the bond entanglement entropy, the confined regime shows constrained dynamics, where the electron-positron pair remains confined up to all accessible evolution times. For the massive quench, the qualitative difference between the deconfined and confined regimes is even more striking, with much stronger confinement in the latter.

We focused here on signatures of tuning the topological $\theta$-angle as are observable in cold-atom experiments such as Ref.~\cite{Yang2020,Zhou2021} without significant additional overhead. Other phenomena that could be interesting to observe in future experiments include the extraction of the meson spectrum that leads to confinement, dynamical quantum phase transitions following a quench of the topological $\theta$-angle \cite{Zache2019}, or how different values of the $\theta$-angle modify thermalization in a gauge theory \cite{Berges_review,Zhou2021}. 

\textit{Note added.} For a related work, see Ref.~\cite{Cheng2022tunable}, posted to the arXiv on the same day.

\begin{acknowledgments}
We acknowledge fruitful discussions with Robert Ott, Guo-Xian Su, Zhao-Yu Zhou, Hui Sun, Zhen-Sheng Yuan, and Jian-Wei Pan. J.C.H.~acknowledges funding from the European Research Council (ERC) under the European Union’s Horizon 2020 research and innovation programm (Grant Agreement no 948141) — ERC Starting Grant SimUcQuam, and by the Deutsche Forschungsgemeinschaft (DFG, German Research Foundation) under Germany's Excellence Strategy -- EXC-2111 -- 390814868. I.P.M.~acknowledges support from the Australian Research Council (ARC) Discovery Project Grants No.~DP190101515 and DP200103760. This work is part of and supported by Provincia Autonoma di Trento, the ERC Starting Grant StrEnQTh (project ID 804305), the Google Research Scholar Award ProGauge, and Q@TN — Quantum Science and Technology in Trento.
\end{acknowledgments}

\appendix
\section{Lattice Schwinger model and its quantum link formulation}\label{app:LSM}
The lattice Schwinger model is described by the Hamiltonian \cite{Kogut1975,Byrnes2002,Buyens2014,Buyens2016,Banuls2017}
\begin{align}\nonumber
    \hat{H}=&\,-\frac{\kappa}{2a}\sum_{j=1}^{L_\mathrm{m}-1}\big(\hat{\psi}^\dagger_\ell\hat{U}_{\ell,\ell+1}\hat{\psi}_{\ell+1}+\text{H.c.}\big)\\\label{eq:LSM}
    &+\mu\sum_{\ell=1}^{L_\mathrm{m}}(-1)^\ell\hat{\psi}_\ell^\dagger\hat{\psi}_\ell+\frac{a}{2}\sum_{\ell=1}^{L_\mathrm{m}-1}\big(\hat{E}_{\ell,\ell+1}+E_\text{bg}\big)^2,
\end{align}
where matter on site $\ell$ is described by Kogut--Susskind (staggered) fermions of annihilation and creation operators $\hat{\psi}_\ell,\hat{\psi}_\ell^\dagger$, with mass $\mu$. We set the lattice spacing $a=1$ throughout our work. Equation~\eqref{eq:LSM} adopts the Wilsonian lattice formulation where the gauge (electric) field on the link between sites $\ell$ and $\ell+1$ is described by the infinite-dimensional operator $\hat{U}_{\ell,\ell+1}$ ($\hat{E}_{\ell,\ell+1}$), satisfying the commutation relations
\begin{subequations}\label{eq:comm}
\begin{align}\label{eq:comm1}
    \big[\hat{E}_{\ell,\ell+1},\hat{U}_{r,r+1}\big]&=g\delta_{\ell,r}\hat{U}_{\ell,\ell+1},\\\label{eq:comm2}
    \big[\hat{U}_{\ell,\ell+1},\hat{U}_{r,r+1}^\dagger\big]&=0.
\end{align}
\end{subequations}
The lattice Schwinger model hosts a $\mathrm{U}(1)$ gauge symmetry with generator
\begin{align}
    \hat{G}_\ell=\hat{E}_{\ell,\ell+1}-\hat{E}_{\ell-1,\ell}-g\bigg[\hat{\psi}_\ell^\dagger\hat{\psi}_\ell+\frac{(-1)^\ell-1}{2}\bigg].
\end{align}
The topological $\theta$-angle is incorporated here through the background field  $E_\text{bg}=g\theta/(2\pi)$ \cite{Tong_LectureNotes}. 

We now perform the Jordan--Wigner transformation
\begin{subequations}
\begin{align}
    &\hat{\psi}^\dagger_\ell=\exp\bigg[i\pi\sum_{r<\ell}\frac{\hat{\sigma}^z_r+\mathds{1}}{2}\bigg]\hat{\sigma}^+_\ell,\\
    &\hat{\psi}_\ell=\exp\bigg[-i\pi\sum_{r<\ell}\frac{\hat{\sigma}^z_r+\mathds{1}}{2}\bigg]\hat{\sigma}^-_\ell,\\
    &\hat{\psi}^\dagger_\ell\hat{\psi}_\ell=\frac{\hat{\sigma}^z_\ell+\mathds{1}}{2}\,.
\end{align}
\end{subequations}
Moreover, we adopt the quantum link formulation, where
\begin{subequations}
\begin{align}
    \hat{U}_{\ell,\ell+1}&\to\frac{\hat{s}^+_{\ell,\ell+1}}{\sqrt{S(S+1)}},\\
    \hat{E}_{\ell,\ell+1}&\to g\hat{s}^z_{\ell,\ell+1}.
\end{align}
\end{subequations}
This transforms the commutation relations~\eqref{eq:comm} as
\begin{subequations}\label{eq:commp}
\begin{align}\nonumber
    \big[\hat{E}_{\ell,\ell+1},\hat{U}_{r,r+1}\big]&\to\frac{g}{\sqrt{S(S+1)}}\big[\hat{s}^z_{\ell,\ell+1},\hat{s}^+_{r,r+1}\big]\\\label{eq:commp1}
    &=g\delta_{\ell,r}\frac{\hat{s}^+_{\ell,\ell+1}}{{\sqrt{S(S+1)}}},\\\nonumber
    \big[\hat{U}_{\ell,\ell+1},\hat{U}_{r,r+1}^\dagger\big]&\to\frac{1}{S(S+1)}\big[\hat{s}^+_{\ell,\ell+1},\hat{s}^-_{r,r+1}\big]\\\label{eq:commp2}
    &=\frac{2}{S(S+1)}\delta_{\ell,r}\hat{s}^z_{\ell,\ell+1}.
\end{align}
\end{subequations}
Equation~\eqref{eq:commp1} reproduces the canonical commutation relation given by Eq.~\eqref{eq:comm1} for any $S$, while Eq.~\eqref{eq:commp2} reduces to Eq.~\eqref{eq:comm2} in the limit of $S\to\infty$. 
The above transformations render Eq.~\eqref{eq:LSM}, up to an inconsequential energetic constant, in the form
\begin{align}\nonumber
    \hat{H}=&-\kappa_S\sum_{\ell=1}^{L_\mathrm{m}-1}\big(\hat{\sigma}^+_\ell\hat{s}^+_{\ell,\ell+1}\hat{\sigma}^-_{\ell+1}+\text{H.c.}\big)+\frac{\mu}{2}\sum_{\ell=1}^{L_\mathrm{m}}(-1)^\ell\hat{\sigma}^z_\ell\\\label{eq:QLM1}
    &+\frac{g^2}{2}\sum_{\ell=1}^{L_\mathrm{m}-1}\bigg[\big(\hat{s}^z_{\ell,\ell+1}\big)^2+\frac{\theta-\pi}{\pi}\hat{s}^z_{\ell,\ell+1}\bigg],
\end{align}
with $\kappa_S=\kappa/\big[2\sqrt{S(S+1)}\big]$. 
For half-integer $S$, the QLM formulation naturally incorporates a $\theta$-angle of $\pi$ \cite{Surace2020,Zache2021achieving}. To account for that as we work with $S=1/2$, we have shifted the last term by $\pi$. 
In this formulation, Gauss's law takes the form
\begin{align}
    \hat{G}_\ell=\hat{s}^z_{\ell,\ell+1}-\hat{s}^z_{\ell-1,\ell}-\frac{\hat{\sigma}^z_\ell+(-1)^\ell}{2}.
\end{align}

For experimental purposes, $S=1/2$ is the most feasible choice, which we shall use henceforth. 
In the case of $S=1/2$, $\big(\hat{s}^z_{\ell,\ell+1}\big)^2=\mathds{1}$, and neglecting this irrelevant energetic constant further simplifies our Hamiltonian to
\begin{align}\nonumber
    \hat{H}=&-\frac{\kappa}{\sqrt{3}}\sum_{\ell=1}^{L_\mathrm{m}-1}\big(\hat{\sigma}^+_\ell\hat{s}^+_{\ell,\ell+1}\hat{\sigma}^-_{\ell+1}+\text{H.c.}\big)\\\label{eq:QLM2}
    &+\frac{\mu}{2}\sum_{\ell=1}^{L_\mathrm{m}}(-1)^\ell\hat{\sigma}^z_\ell+\frac{g^2(\theta-\pi)}{2\pi}\sum_{\ell=1}^{L_\mathrm{m}-1}\hat{s}^z_{\ell,\ell+1}.
\end{align}
We further employ the particle-hole transformation \cite{Hauke2013,Yang2016}
\begin{subequations}
\begin{align}
    \hat{\sigma}^{z(y)}_\ell&\to(-1)^\ell\hat{\sigma}^{z(y)}_\ell,\\
    \hat{s}^{z(y)}_{\ell,\ell+1}&\to(-1)^{\ell+1}\hat{s}^{z(y)}_{\ell,\ell+1},
\end{align}
\end{subequations}
which leaves our $\mathrm{U}(1)$ QLM Hamiltonian in the form
\begin{align}\nonumber
    \hat{H}=&-\frac{\kappa}{\sqrt{3}}\sum_{\ell=1}^{L_\mathrm{m}-1}\big(\hat{\sigma}^-_\ell\hat{s}^+_{\ell,\ell+1}\hat{\sigma}^-_{\ell+1}+\text{H.c.}\big)\\\label{eq:QLM3}
    &+\frac{\mu}{2}\sum_{\ell=1}^{L_\mathrm{m}}\hat{\sigma}^z_\ell-\frac{g^2(\theta-\pi)}{2\pi}\sum_{\ell=1}^{L_\mathrm{m}-1}(-1)^\ell\hat{s}^z_{\ell,\ell+1}.
\end{align}
Up to a renormalized tunneling coefficient $\kappa$, this is the Hamiltonian of Eq.~\eqref{eq:H} studied in our work. Furthermore, the generator of the $\mathrm{U}(1)$ gauge symmetry now reads
\begin{align}
    \hat{G}_\ell=(-1)^{\ell+1}\bigg[\hat{s}^z_{\ell,\ell+1}+\hat{s}^z_{\ell-1,\ell}+\frac{\hat{\sigma}^z_\ell+\mathds{1}}{2}\bigg].
\end{align}

\section{Further details on mapping between quantum link model and Bose--Hubbard model}\label{app:degenPT}
In this section, we provide further details on the mapping of Eq.~\eqref{eq:H} onto a single-species bosonic model that can be implemented using ultracold atoms. We first identify the two local bosonic states $\ket{0}_\ell$ and $\ket{1}_\ell$, associated with the bosonic ladder operators $\hat{a}_\ell$ and $\hat{a}_\ell^\dagger$, on matter site $\ell$ with the two eigenstates of the local Pauli operator:
\begin{subequations}\label{eq:sigma_mapping}
\begin{align}
    &\hat{\sigma}^-_\ell=\hat{\mathcal{P}}_\ell\hat{a}_\ell\hat{\mathcal{P}}_\ell,\\
    &\hat{\sigma}^+_\ell=\hat{\mathcal{P}}_\ell\hat{a}_\ell^\dagger\hat{\mathcal{P}}_\ell,\\
    &\hat{\sigma}^z_\ell=\hat{\mathcal{P}}_\ell\big(2\hat{a}_\ell^\dagger\hat{a}_\ell-1\big)\hat{\mathcal{P}}_\ell,
\end{align}
\end{subequations}
where $\hat{\mathcal{P}}_\ell$ is the projector onto the local subspace $\mathcal{H}_\ell=\text{span}\big\{\ket{0}_\ell,\ket{1}_\ell\big\}$. Restricting to this subspace, the bosonic commutation relation $\big[\hat{a}_\ell,\hat{a}_\ell^\dagger\big]=1$ reproduces the spin commutation relations $\big[\hat{\sigma}^+_\ell,\hat{\sigma}^-_\ell\big]=\hat{\sigma}^z_\ell$ and $\big[\hat{\sigma}^z_\ell,\hat{\sigma}^\pm_\ell\big]=\pm2\hat{\sigma}^\pm_\ell$.

In a similar fashion, we identify the local bosonic states $\ket{0}_{\ell,\ell+1}$ and $\ket{2}_{\ell,\ell+1}$, associated with the bosonic ladder operators $\hat{a}_{\ell,\ell+1}$ and $\hat{a}^\dagger_{\ell,\ell+1}$, on the link between matter sites $\ell$ and $\ell+1$ with the two eigenstates of the local spin-$1/2$ operator:
\begin{subequations}\label{eq:s_mapping}
\begin{align}
    \hat{s}^-_{\ell,\ell+1}&=\frac{1}{\sqrt{2}}\hat{\mathcal{P}}_{\ell,\ell+1}\big(\hat{a}_{\ell,\ell+1}\big)^2\hat{\mathcal{P}}_{\ell,\ell+1},\\
    \hat{s}^+_{\ell,\ell+1}&=\frac{1}{\sqrt{2}}\hat{\mathcal{P}}_{\ell,\ell+1}\big(\hat{a}_{\ell,\ell+1}^\dagger\big)^2\hat{\mathcal{P}}_{\ell,\ell+1},\\
    \hat{s}^z_{\ell,\ell+1}&=\frac{1}{2}\hat{\mathcal{P}}_{\ell,\ell+1}\big(\hat{a}_{\ell,\ell+1}^\dagger\hat{a}_{\ell,\ell+1}-1\big)\hat{\mathcal{P}}_{\ell,\ell+1},
\end{align}
\end{subequations}
where $\hat{\mathcal{P}}_{\ell,\ell+1}$ is the projector onto the local subspace $\mathcal{H}_{\ell,\ell+1}=\text{span}\big\{\ket{0}_{\ell,\ell+1},\ket{2}_{\ell,\ell+1}\big\}$. Restricting to this subspace, the bosonic commutation relation $\big[\hat{a}_{\ell,\ell+1},\hat{a}_{\ell,\ell+1}^\dagger\big]=1$ reproduces the spin commutation relations $\big[\hat{s}^+_{\ell,\ell+1},\hat{s}^-_{\ell,\ell+1}\big]=2\hat{s}^z_{\ell,\ell+1}$ and $\big[\hat{s}^z_{\ell,\ell+1},\hat{s}^\pm_{\ell,\ell+1}\big]=\pm\hat{s}^\pm_{\ell,\ell+1}$.

Inserting Eqs.~\eqref{eq:sigma_mapping} and~\eqref{eq:s_mapping} into Eq.~\eqref{eq:H}, and neglecting inconsequential constant energetic terms, renders Eq.~\eqref{eq:H} in the form
\begin{align}\nonumber
    \label{eq:HQLMboson}
    \hat{H}=\hat{\mathcal{P}}\sum_\ell\bigg\{&-\frac{\kappa}{2\sqrt{2}}\Big[\hat{a}_\ell\big(\hat{a}_{\ell,\ell+1}^\dagger\big)^2\hat{a}_{\ell+1}+\text{H.c.}\Big]+\mu\hat{a}_\ell^\dagger\hat{a}_\ell\\
    &-\frac{g^2(\theta-\pi)}{4\pi}(-1)^\ell\hat{a}_{\ell,\ell+1}^\dagger\hat{a}_{\ell,\ell+1}\bigg\}\hat{\mathcal{P}},
\end{align}
where $\hat{\mathcal{P}}=\prod_\ell\hat{\mathcal{P}}_\ell\hat{\mathcal{P}}_{\ell,\ell+1}$.

To map this model to an effective Hamiltonian derived from the Bose--Hubbard model, one can follow degenerate perturbation theory as outlined in Ref.~\cite{Yang2020}. Using $U$ and $\delta$ as large energy scales, the hopping term $\propto J$ in the Bose--Hubbard model, Eq.~\eqref{eq:BHM}, becomes a perturbation to the diagonal terms $\hat{H}_\mathrm{diag}$ as collected in Eq.~\eqref{eq:Hdiag}.
Focusing on the target subsector of the Bose--Hubbard model consisting of bosonic occupations  $\{\ket{0}_{2\ell},\ket{1}_{2\ell}\}$ on even (matter) optical lattice sites $j=2\ell$ and $\{\ket{0}_{2\ell+1},\ket{2}_{2\ell+1}\}$ on odd (gauge) sites $j=2\ell+1$, and states that fulfil the Gauss's law, second-order degenerate perturbation theory yields the effective Hamiltonian \eqref{eq:HQLMboson}, 
where
\begin{align}\nonumber
    \kappa=\sqrt{2}J^2\bigg(&\frac{1}{\delta+\Delta\mp\chi}+\frac{1}{U-\delta+\Delta\pm\chi}\\
    &+\frac{1}{\delta-\Delta\mp\chi}+\frac{1}{U-\delta-\Delta\pm\chi}\bigg),
\end{align}
with the alternating sign of $\chi$ in this expression occurring between odd and even sites. The rest mass of fermions is given by $\mu=\delta-U/2$.
In the experiment, the large energy scale $\delta$ is on the order of $700$ Hz, while the largest value of $\chi$ we use here is always below $30$ Hz, ensuring that $\lvert\delta\pm\Delta\rvert\gg\lvert\chi\rvert$ and, since $U\approx2\delta$, also that $\lvert U-\delta\pm\Delta\rvert\gg\lvert\chi\rvert$. This permits us to neglect $\chi$ in the expression for $\kappa$, leading to Eq.~\eqref{eq:kappa} used in the main text.

\section{Ramping the charge-proliferated state}\label{app:InverseRamp}
\begin{figure}[t!]
    \centering
    \includegraphics[width=\columnwidth]{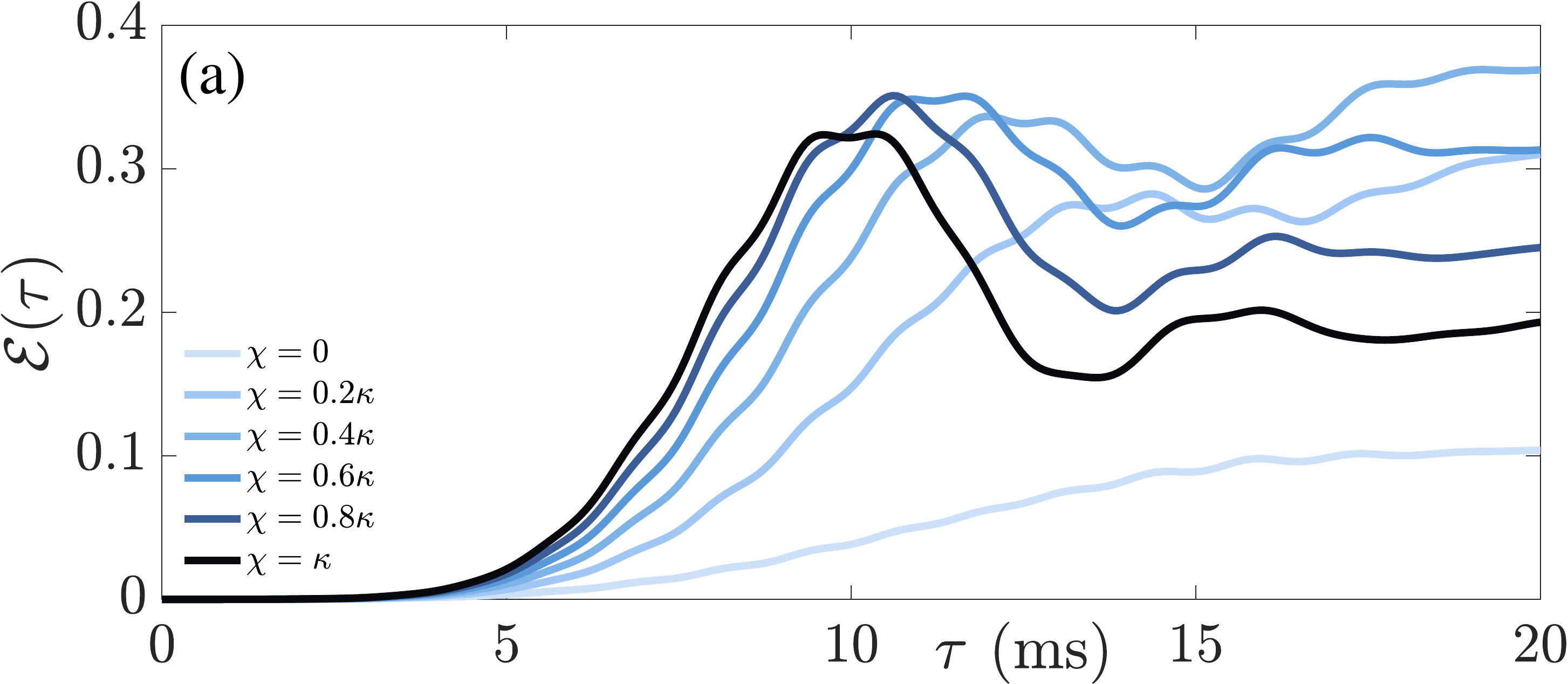}\\
    \vspace{1.1mm}
    \includegraphics[width=\columnwidth]{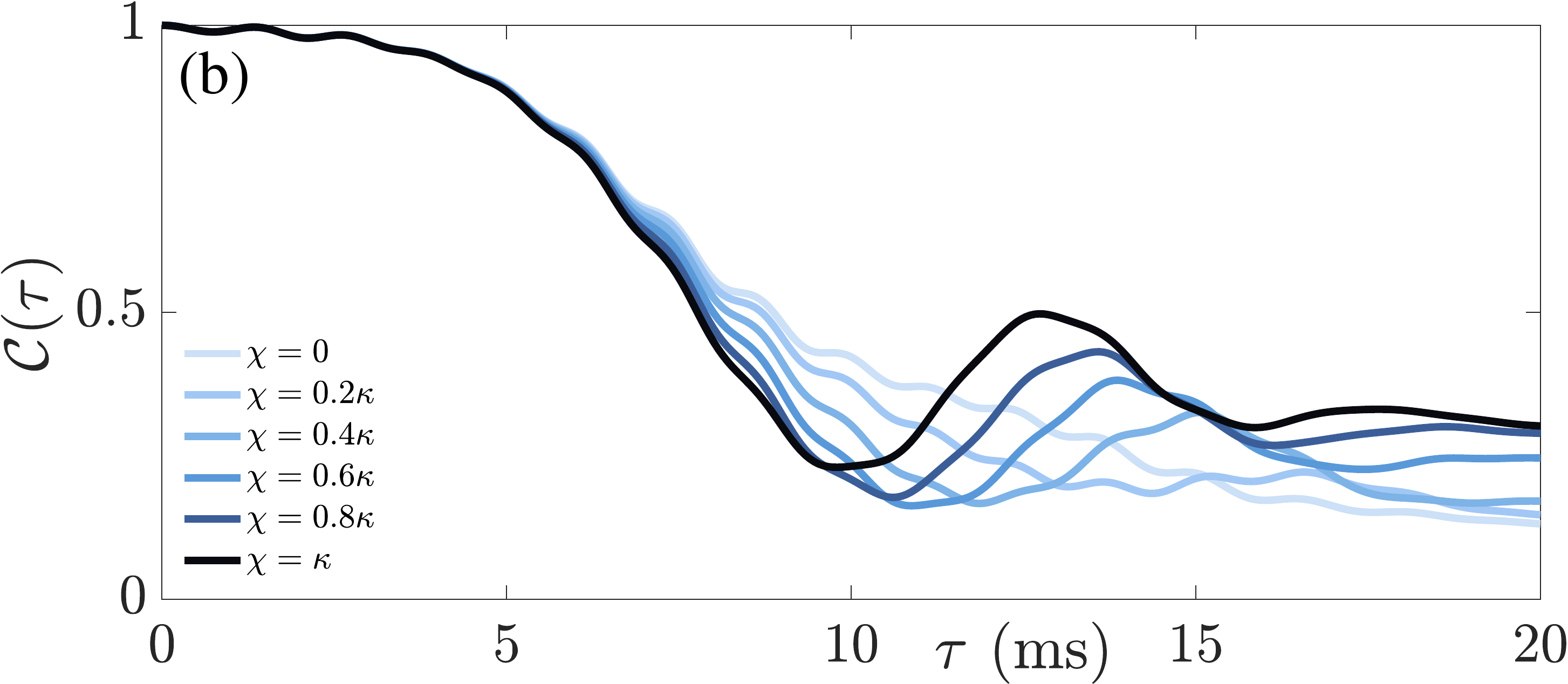}
    \caption{(Color online). Starting in the charge-proliferated state $\ket{\ldots,0,1,0,1,0,1,0,\ldots}$, we perform the inverse of the adiabatic ramp protocol depicted in Fig.~\ref{fig:ramp_protocol} for various values of $\chi/\kappa$ (see legends). (a) The ensuing dynamics of the electric flux. The explicit symmetry breaking introduced by the $\theta$-term allows the electric flux to go to larger values at moderate values of $\chi$, since now the system is no longer mostly in a superposition of vacua at late times as in the case of the deconfined regime at $\chi=0$. At larger values of $\chi$, the physics is completely changed, and the monotonic behavior of the electric flux is no longer there. (b) The ensuing dynamics of the chiral condensate confirms this picture (see text).}
    \label{fig:polarized_ramp}
\end{figure}
As mentioned in the main text, the experiments of Refs.~\cite{Yang2020,Zhou2021} started in the charge-proliferated state $\ket{\ldots,0,1,0,1,0,1,0,\ldots}$, which represents the ground state of Eq.~\eqref{eq:H} at $\mu/\kappa\to-\infty$, and employed the inverse of the ramp protocol described in Fig.~\ref{fig:ramp_protocol}. Let us now consider the same adiabatic ramp as Refs.~\cite{Yang2020,Zhou2021} but with the additional $\theta$-term of strength $\chi$. The corresponding dynamics of the electric flux and chiral condensate are shown in Fig.~\ref{fig:polarized_ramp}(a,b), respectively. We find perfect quantitative agreement in the chiral-condensate dynamics for the case of $\chi=0$ with the corresponding result in Ref.~\cite{Zhou2021}.

Focusing first on the case of $\chi=0$, we find that the electric flux starts at zero, as the state at $\tau=0$ is the charge-proliferated state, and then goes to $0.1$ at the end of the ramp. Even though the state is now in the $\mathbb{Z}_2$ symmetry-broken phase, it is roughly a superposition between the two vacua of the $\mathrm{U}(1)$ QLM. Indeed, the charge-proliferated initial state is $\mathbb{Z}_2$-symmetric as is $\hat{H}_\text{BH}$, and so numerically this symmetry is only slightly broken at the edges of the chain (we employ open boundary conditions for experimental feasibility). This finite-size effect is the main reason why $\mathcal{E}(\tau)$ is not exactly zero throughout the whole ramp. A small value of the $\theta$-term strength, $\chi=0.2\kappa$, helps bias the system toward one of the two vacua, rendering $\mathcal{E}(\tau)$ larger at the end of the ramp. However, when $\chi$ is quite large, the dynamics of the electric flux is no longer monotonic, exhibiting fast growth at early times, but then drops to a lower value that further decreases with larger $\chi$. This reversal in the growth of the electric flux also occurs earlier with increasing $\chi$.

Turning to the chiral condensate in Fig.~\ref{fig:polarized_ramp}(b), the behavior is as expected for $\chi=0$. The chiral condensate is maximal at unity in the charge-proliferated state at $\tau=0$, and then steadily decreases during the ramp, approaching close to zero at its end ($\tau=20$ ms), where the system is deep in the $\mathbb{Z}_2$ symmetry-broken phase. As $\chi$ gets larger, this monotonic decay is fundamentally altered, in congruence with the corresponding case in the electric flux. Again, the reversal of the monotonic behavior occurs earlier with increasing $\chi$, with the chiral condensate finishing the ramp at a finite value that increases with $\chi$.

The qualitatively different behavior in the ramp dynamics of the charge-proliferated state at larger values of $\chi$ indicate that the physics is fundamentally different under strong confinement. 

\bibliography{biblio}

\begin{thebibliography}{97}%
\makeatletter
\providecommand \@ifxundefined [1]{%
 \@ifx{#1\undefined}
}%
\providecommand \@ifnum [1]{%
 \ifnum #1\expandafter \@firstoftwo
 \else \expandafter \@secondoftwo
 \fi
}%
\providecommand \@ifx [1]{%
 \ifx #1\expandafter \@firstoftwo
 \else \expandafter \@secondoftwo
 \fi
}%
\providecommand \natexlab [1]{#1}%
\providecommand \enquote  [1]{``#1''}%
\providecommand \bibnamefont  [1]{#1}%
\providecommand \bibfnamefont [1]{#1}%
\providecommand \citenamefont [1]{#1}%
\providecommand \href@noop [0]{\@secondoftwo}%
\providecommand \href [0]{\begingroup \@sanitize@url \@href}%
\providecommand \@href[1]{\@@startlink{#1}\@@href}%
\providecommand \@@href[1]{\endgroup#1\@@endlink}%
\providecommand \@sanitize@url [0]{\catcode `\\12\catcode `\$12\catcode
  `\&12\catcode `\#12\catcode `\^12\catcode `\_12\catcode `\%12\relax}%
\providecommand \@@startlink[1]{}%
\providecommand \@@endlink[0]{}%
\providecommand \url  [0]{\begingroup\@sanitize@url \@url }%
\providecommand \@url [1]{\endgroup\@href {#1}{\urlprefix }}%
\providecommand \urlprefix  [0]{URL }%
\providecommand \Eprint [0]{\href }%
\providecommand \doibase [0]{http://dx.doi.org/}%
\providecommand \selectlanguage [0]{\@gobble}%
\providecommand \bibinfo  [0]{\@secondoftwo}%
\providecommand \bibfield  [0]{\@secondoftwo}%
\providecommand \translation [1]{[#1]}%
\providecommand \BibitemOpen [0]{}%
\providecommand \bibitemStop [0]{}%
\providecommand \bibitemNoStop [0]{.\EOS\space}%
\providecommand \EOS [0]{\spacefactor3000\relax}%
\providecommand \BibitemShut  [1]{\csname bibitem#1\endcsname}%
\let\auto@bib@innerbib\@empty
\bibitem [{\citenamefont {Bloch}\ \emph {et~al.}(2008)\citenamefont {Bloch},
  \citenamefont {Dalibard},\ and\ \citenamefont {Zwerger}}]{Bloch2008}%
  \BibitemOpen
  \bibfield  {author} {\bibinfo {author} {\bibfnamefont {Immanuel}\
  \bibnamefont {Bloch}}, \bibinfo {author} {\bibfnamefont {Jean}\ \bibnamefont
  {Dalibard}}, \ and\ \bibinfo {author} {\bibfnamefont {Wilhelm}\ \bibnamefont
  {Zwerger}},\ }\bibfield  {title} {\enquote {\bibinfo {title} {Many-body
  physics with ultracold gases},}\ }\href {\doibase 10.1103/RevModPhys.80.885}
  {\bibfield  {journal} {\bibinfo  {journal} {Rev. Mod. Phys.}\ }\textbf
  {\bibinfo {volume} {80}},\ \bibinfo {pages} {885--964} (\bibinfo {year}
  {2008})}\BibitemShut {NoStop}%
\bibitem [{\citenamefont {Georgescu}\ \emph {et~al.}(2014)\citenamefont
  {Georgescu}, \citenamefont {Ashhab},\ and\ \citenamefont
  {Nori}}]{Georgescu_review}%
  \BibitemOpen
  \bibfield  {author} {\bibinfo {author} {\bibfnamefont {I.~M.}\ \bibnamefont
  {Georgescu}}, \bibinfo {author} {\bibfnamefont {S.}~\bibnamefont {Ashhab}}, \
  and\ \bibinfo {author} {\bibfnamefont {Franco}\ \bibnamefont {Nori}},\
  }\bibfield  {title} {\enquote {\bibinfo {title} {Quantum simulation},}\
  }\href {\doibase 10.1103/RevModPhys.86.153} {\bibfield  {journal} {\bibinfo
  {journal} {Rev. Mod. Phys.}\ }\textbf {\bibinfo {volume} {86}},\ \bibinfo
  {pages} {153--185} (\bibinfo {year} {2014})}\BibitemShut {NoStop}%
\bibitem [{\citenamefont {Alexeev}\ \emph {et~al.}(2021)\citenamefont
  {Alexeev}, \citenamefont {Bacon}, \citenamefont {Brown}, \citenamefont
  {Calderbank}, \citenamefont {Carr}, \citenamefont {Chong}, \citenamefont
  {DeMarco}, \citenamefont {Englund}, \citenamefont {Farhi}, \citenamefont
  {Fefferman}, \citenamefont {Gorshkov}, \citenamefont {Houck}, \citenamefont
  {Kim}, \citenamefont {Kimmel}, \citenamefont {Lange}, \citenamefont {Lloyd},
  \citenamefont {Lukin}, \citenamefont {Maslov}, \citenamefont {Maunz},
  \citenamefont {Monroe}, \citenamefont {Preskill}, \citenamefont {Roetteler},
  \citenamefont {Savage},\ and\ \citenamefont {Thompson}}]{Alexeev_review}%
  \BibitemOpen
  \bibfield  {author} {\bibinfo {author} {\bibfnamefont {Yuri}\ \bibnamefont
  {Alexeev}}, \bibinfo {author} {\bibfnamefont {Dave}\ \bibnamefont {Bacon}},
  \bibinfo {author} {\bibfnamefont {Kenneth~R.}\ \bibnamefont {Brown}},
  \bibinfo {author} {\bibfnamefont {Robert}\ \bibnamefont {Calderbank}},
  \bibinfo {author} {\bibfnamefont {Lincoln~D.}\ \bibnamefont {Carr}}, \bibinfo
  {author} {\bibfnamefont {Frederic~T.}\ \bibnamefont {Chong}}, \bibinfo
  {author} {\bibfnamefont {Brian}\ \bibnamefont {DeMarco}}, \bibinfo {author}
  {\bibfnamefont {Dirk}\ \bibnamefont {Englund}}, \bibinfo {author}
  {\bibfnamefont {Edward}\ \bibnamefont {Farhi}}, \bibinfo {author}
  {\bibfnamefont {Bill}\ \bibnamefont {Fefferman}}, \bibinfo {author}
  {\bibfnamefont {Alexey~V.}\ \bibnamefont {Gorshkov}}, \bibinfo {author}
  {\bibfnamefont {Andrew}\ \bibnamefont {Houck}}, \bibinfo {author}
  {\bibfnamefont {Jungsang}\ \bibnamefont {Kim}}, \bibinfo {author}
  {\bibfnamefont {Shelby}\ \bibnamefont {Kimmel}}, \bibinfo {author}
  {\bibfnamefont {Michael}\ \bibnamefont {Lange}}, \bibinfo {author}
  {\bibfnamefont {Seth}\ \bibnamefont {Lloyd}}, \bibinfo {author}
  {\bibfnamefont {Mikhail~D.}\ \bibnamefont {Lukin}}, \bibinfo {author}
  {\bibfnamefont {Dmitri}\ \bibnamefont {Maslov}}, \bibinfo {author}
  {\bibfnamefont {Peter}\ \bibnamefont {Maunz}}, \bibinfo {author}
  {\bibfnamefont {Christopher}\ \bibnamefont {Monroe}}, \bibinfo {author}
  {\bibfnamefont {John}\ \bibnamefont {Preskill}}, \bibinfo {author}
  {\bibfnamefont {Martin}\ \bibnamefont {Roetteler}}, \bibinfo {author}
  {\bibfnamefont {Martin~J.}\ \bibnamefont {Savage}}, \ and\ \bibinfo {author}
  {\bibfnamefont {Jeff}\ \bibnamefont {Thompson}},\ }\bibfield  {title}
  {\enquote {\bibinfo {title} {Quantum computer systems for scientific
  discovery},}\ }\href {\doibase 10.1103/PRXQuantum.2.017001} {\bibfield
  {journal} {\bibinfo  {journal} {PRX Quantum}\ }\textbf {\bibinfo {volume}
  {2}},\ \bibinfo {pages} {017001} (\bibinfo {year} {2021})}\BibitemShut
  {NoStop}%
\bibitem [{\citenamefont {Klco}\ \emph {et~al.}(2021)\citenamefont {Klco},
  \citenamefont {Roggero},\ and\ \citenamefont {Savage}}]{klco2021standard}%
  \BibitemOpen
  \bibfield  {author} {\bibinfo {author} {\bibfnamefont {Natalie}\ \bibnamefont
  {Klco}}, \bibinfo {author} {\bibfnamefont {Alessandro}\ \bibnamefont
  {Roggero}}, \ and\ \bibinfo {author} {\bibfnamefont {Martin~J.}\ \bibnamefont
  {Savage}},\ }\bibfield  {title} {\enquote {\bibinfo {title} {Standard model
  physics and the digital quantum revolution: Thoughts about the interface},}\
  }\href@noop {} {\bibfield  {journal} {\bibinfo  {journal} {arXiv preprint}\ }
  (\bibinfo {year} {2021})},\ \Eprint {http://arxiv.org/abs/2107.04769}
  {arXiv:2107.04769 [quant-ph]} \BibitemShut {NoStop}%
\bibitem [{\citenamefont {Ba{\~n}uls}\ \emph {et~al.}(2020)\citenamefont
  {Ba{\~n}uls}, \citenamefont {Blatt}, \citenamefont {Catani}, \citenamefont
  {Celi}, \citenamefont {Cirac}, \citenamefont {Dalmonte}, \citenamefont
  {Fallani}, \citenamefont {Jansen}, \citenamefont {Lewenstein}, \citenamefont
  {Montangero}, \citenamefont {Muschik}, \citenamefont {Reznik}, \citenamefont
  {Rico}, \citenamefont {Tagliacozzo}, \citenamefont {Van~Acoleyen},
  \citenamefont {Verstraete}, \citenamefont {Wiese}, \citenamefont {Wingate},
  \citenamefont {Zakrzewski},\ and\ \citenamefont {Zoller}}]{Pasquans_review}%
  \BibitemOpen
  \bibfield  {author} {\bibinfo {author} {\bibfnamefont {Mari~Carmen}\
  \bibnamefont {Ba{\~n}uls}}, \bibinfo {author} {\bibfnamefont {Rainer}\
  \bibnamefont {Blatt}}, \bibinfo {author} {\bibfnamefont {Jacopo}\
  \bibnamefont {Catani}}, \bibinfo {author} {\bibfnamefont {Alessio}\
  \bibnamefont {Celi}}, \bibinfo {author} {\bibfnamefont {Juan~Ignacio}\
  \bibnamefont {Cirac}}, \bibinfo {author} {\bibfnamefont {Marcello}\
  \bibnamefont {Dalmonte}}, \bibinfo {author} {\bibfnamefont {Leonardo}\
  \bibnamefont {Fallani}}, \bibinfo {author} {\bibfnamefont {Karl}\
  \bibnamefont {Jansen}}, \bibinfo {author} {\bibfnamefont {Maciej}\
  \bibnamefont {Lewenstein}}, \bibinfo {author} {\bibfnamefont {Simone}\
  \bibnamefont {Montangero}}, \bibinfo {author} {\bibfnamefont {Christine~A.}\
  \bibnamefont {Muschik}}, \bibinfo {author} {\bibfnamefont {Benni}\
  \bibnamefont {Reznik}}, \bibinfo {author} {\bibfnamefont {Enrique}\
  \bibnamefont {Rico}}, \bibinfo {author} {\bibfnamefont {Luca}\ \bibnamefont
  {Tagliacozzo}}, \bibinfo {author} {\bibfnamefont {Karel}\ \bibnamefont
  {Van~Acoleyen}}, \bibinfo {author} {\bibfnamefont {Frank}\ \bibnamefont
  {Verstraete}}, \bibinfo {author} {\bibfnamefont {Uwe-Jens}\ \bibnamefont
  {Wiese}}, \bibinfo {author} {\bibfnamefont {Matthew}\ \bibnamefont
  {Wingate}}, \bibinfo {author} {\bibfnamefont {Jakub}\ \bibnamefont
  {Zakrzewski}}, \ and\ \bibinfo {author} {\bibfnamefont {Peter}\ \bibnamefont
  {Zoller}},\ }\bibfield  {title} {\enquote {\bibinfo {title} {Simulating
  lattice gauge theories within quantum technologies},}\ }\href {\doibase
  10.1140/epjd/e2020-100571-8} {\bibfield  {journal} {\bibinfo  {journal} {The
  European Physical Journal D}\ }\textbf {\bibinfo {volume} {74}},\ \bibinfo
  {pages} {165} (\bibinfo {year} {2020})}\BibitemShut {NoStop}%
\bibitem [{\citenamefont {Dalmonte}\ and\ \citenamefont
  {Montangero}(2016)}]{Dalmonte_review}%
  \BibitemOpen
  \bibfield  {author} {\bibinfo {author} {\bibfnamefont {M.}~\bibnamefont
  {Dalmonte}}\ and\ \bibinfo {author} {\bibfnamefont {S.}~\bibnamefont
  {Montangero}},\ }\bibfield  {title} {\enquote {\bibinfo {title} {Lattice
  gauge theory simulations in the quantum information era},}\ }\href {\doibase
  10.1080/00107514.2016.1151199} {\bibfield  {journal} {\bibinfo  {journal}
  {Contemporary Physics}\ }\textbf {\bibinfo {volume} {57}},\ \bibinfo {pages}
  {388--412} (\bibinfo {year} {2016})},\ \Eprint
  {http://arxiv.org/abs/https://doi.org/10.1080/00107514.2016.1151199}
  {https://doi.org/10.1080/00107514.2016.1151199} \BibitemShut {NoStop}%
\bibitem [{\citenamefont {Zohar}\ \emph {et~al.}(2015)\citenamefont {Zohar},
  \citenamefont {Cirac},\ and\ \citenamefont {Reznik}}]{Zohar_review}%
  \BibitemOpen
  \bibfield  {author} {\bibinfo {author} {\bibfnamefont {Erez}\ \bibnamefont
  {Zohar}}, \bibinfo {author} {\bibfnamefont {J~Ignacio}\ \bibnamefont
  {Cirac}}, \ and\ \bibinfo {author} {\bibfnamefont {Benni}\ \bibnamefont
  {Reznik}},\ }\bibfield  {title} {\enquote {\bibinfo {title} {Quantum
  simulations of lattice gauge theories using ultracold atoms in optical
  lattices},}\ }\href {\doibase 10.1088/0034-4885/79/1/014401} {\bibfield
  {journal} {\bibinfo  {journal} {Reports on Progress in Physics}\ }\textbf
  {\bibinfo {volume} {79}},\ \bibinfo {pages} {014401} (\bibinfo {year}
  {2015})}\BibitemShut {NoStop}%
\bibitem [{\citenamefont {Aidelsburger}\ \emph {et~al.}(2022)\citenamefont
  {Aidelsburger}, \citenamefont {Barbiero}, \citenamefont {Bermudez},
  \citenamefont {Chanda}, \citenamefont {Dauphin}, \citenamefont
  {González-Cuadra}, \citenamefont {Grzybowski}, \citenamefont {Hands},
  \citenamefont {Jendrzejewski}, \citenamefont {Jünemann}, \citenamefont
  {Juzeliūnas}, \citenamefont {Kasper}, \citenamefont {Piga}, \citenamefont
  {Ran}, \citenamefont {Rizzi}, \citenamefont {Sierra}, \citenamefont
  {Tagliacozzo}, \citenamefont {Tirrito}, \citenamefont {Zache}, \citenamefont
  {Zakrzewski}, \citenamefont {Zohar},\ and\ \citenamefont
  {Lewenstein}}]{aidelsburger2021cold}%
  \BibitemOpen
  \bibfield  {author} {\bibinfo {author} {\bibfnamefont {Monika}\ \bibnamefont
  {Aidelsburger}}, \bibinfo {author} {\bibfnamefont {Luca}\ \bibnamefont
  {Barbiero}}, \bibinfo {author} {\bibfnamefont {Alejandro}\ \bibnamefont
  {Bermudez}}, \bibinfo {author} {\bibfnamefont {Titas}\ \bibnamefont
  {Chanda}}, \bibinfo {author} {\bibfnamefont {Alexandre}\ \bibnamefont
  {Dauphin}}, \bibinfo {author} {\bibfnamefont {Daniel}\ \bibnamefont
  {González-Cuadra}}, \bibinfo {author} {\bibfnamefont {Przemysław~R.}\
  \bibnamefont {Grzybowski}}, \bibinfo {author} {\bibfnamefont {Simon}\
  \bibnamefont {Hands}}, \bibinfo {author} {\bibfnamefont {Fred}\ \bibnamefont
  {Jendrzejewski}}, \bibinfo {author} {\bibfnamefont {Johannes}\ \bibnamefont
  {Jünemann}}, \bibinfo {author} {\bibfnamefont {Gediminas}\ \bibnamefont
  {Juzeliūnas}}, \bibinfo {author} {\bibfnamefont {Valentin}\ \bibnamefont
  {Kasper}}, \bibinfo {author} {\bibfnamefont {Angelo}\ \bibnamefont {Piga}},
  \bibinfo {author} {\bibfnamefont {Shi-Ju}\ \bibnamefont {Ran}}, \bibinfo
  {author} {\bibfnamefont {Matteo}\ \bibnamefont {Rizzi}}, \bibinfo {author}
  {\bibfnamefont {Germán}\ \bibnamefont {Sierra}}, \bibinfo {author}
  {\bibfnamefont {Luca}\ \bibnamefont {Tagliacozzo}}, \bibinfo {author}
  {\bibfnamefont {Emanuele}\ \bibnamefont {Tirrito}}, \bibinfo {author}
  {\bibfnamefont {Torsten~V.}\ \bibnamefont {Zache}}, \bibinfo {author}
  {\bibfnamefont {Jakub}\ \bibnamefont {Zakrzewski}}, \bibinfo {author}
  {\bibfnamefont {Erez}\ \bibnamefont {Zohar}}, \ and\ \bibinfo {author}
  {\bibfnamefont {Maciej}\ \bibnamefont {Lewenstein}},\ }\bibfield  {title}
  {\enquote {\bibinfo {title} {Cold atoms meet lattice gauge theory},}\ }\href
  {\doibase 10.1098/rsta.2021.0064} {\bibfield  {journal} {\bibinfo  {journal}
  {Philosophical Transactions of the Royal Society A: Mathematical, Physical
  and Engineering Sciences}\ }\textbf {\bibinfo {volume} {380}},\ \bibinfo
  {pages} {20210064} (\bibinfo {year} {2022})}\BibitemShut {NoStop}%
\bibitem [{\citenamefont {{Zohar}}(2022)}]{Zohar_NewReview}%
  \BibitemOpen
  \bibfield  {author} {\bibinfo {author} {\bibfnamefont {Erez}\ \bibnamefont
  {{Zohar}}},\ }\bibfield  {title} {\enquote {\bibinfo {title} {{Quantum
  simulation of lattice gauge theories in more than one space
  dimension{\textemdash}requirements, challenges and methods}},}\ }\href
  {\doibase 10.1098/rsta.2021.0069} {\bibfield  {journal} {\bibinfo  {journal}
  {Philosophical Transactions of the Royal Society of London Series A}\
  }\textbf {\bibinfo {volume} {380}},\ \bibinfo {eid} {20210069} (\bibinfo
  {year} {2022})},\ \Eprint {http://arxiv.org/abs/2106.04609} {arXiv:2106.04609
  [quant-ph]} \BibitemShut {NoStop}%
\bibitem [{\citenamefont {Davoudi}\ \emph {et~al.}(2022)\citenamefont
  {Davoudi}, \citenamefont {Balantekin}, \citenamefont {Bhattacharya},
  \citenamefont {Carena}, \citenamefont {de~Jong}, \citenamefont {Draper},
  \citenamefont {El-Khadra}, \citenamefont {Gemelke}, \citenamefont {Hanada},
  \citenamefont {Kharzeev}, \citenamefont {Lamm}, \citenamefont {Li},
  \citenamefont {Liu}, \citenamefont {Lukin}, \citenamefont {Meurice},
  \citenamefont {Monroe}, \citenamefont {Nachman}, \citenamefont {Pagano},
  \citenamefont {Preskill}, \citenamefont {Rinaldi}, \citenamefont {Roggero},
  \citenamefont {Santiago}, \citenamefont {Savage}, \citenamefont {Siddiqi},
  \citenamefont {Siopsis}, \citenamefont {Van~Zanten}, \citenamefont {Wiebe},
  \citenamefont {Yamauchi}, \citenamefont {Yeter-Aydeniz},\ and\ \citenamefont
  {Zorzetti}}]{Bauer_review}%
  \BibitemOpen
  \bibfield  {author} {\bibinfo {author} {\bibfnamefont {Christian W.
  Bauer.~Zohreh}\ \bibnamefont {Davoudi}}, \bibinfo {author} {\bibfnamefont
  {A.~Baha}\ \bibnamefont {Balantekin}}, \bibinfo {author} {\bibfnamefont
  {Tanmoy}\ \bibnamefont {Bhattacharya}}, \bibinfo {author} {\bibfnamefont
  {Marcela}\ \bibnamefont {Carena}}, \bibinfo {author} {\bibfnamefont
  {Wibe~A.}\ \bibnamefont {de~Jong}}, \bibinfo {author} {\bibfnamefont
  {Patrick}\ \bibnamefont {Draper}}, \bibinfo {author} {\bibfnamefont {Aida}\
  \bibnamefont {El-Khadra}}, \bibinfo {author} {\bibfnamefont {Nate}\
  \bibnamefont {Gemelke}}, \bibinfo {author} {\bibfnamefont {Masanori}\
  \bibnamefont {Hanada}}, \bibinfo {author} {\bibfnamefont {Dmitri}\
  \bibnamefont {Kharzeev}}, \bibinfo {author} {\bibfnamefont {Henry}\
  \bibnamefont {Lamm}}, \bibinfo {author} {\bibfnamefont {Ying-Ying}\
  \bibnamefont {Li}}, \bibinfo {author} {\bibfnamefont {Junyu}\ \bibnamefont
  {Liu}}, \bibinfo {author} {\bibfnamefont {Mikhail}\ \bibnamefont {Lukin}},
  \bibinfo {author} {\bibfnamefont {Yannick}\ \bibnamefont {Meurice}}, \bibinfo
  {author} {\bibfnamefont {Christopher}\ \bibnamefont {Monroe}}, \bibinfo
  {author} {\bibfnamefont {Benjamin}\ \bibnamefont {Nachman}}, \bibinfo
  {author} {\bibfnamefont {Guido}\ \bibnamefont {Pagano}}, \bibinfo {author}
  {\bibfnamefont {John}\ \bibnamefont {Preskill}}, \bibinfo {author}
  {\bibfnamefont {Enrico}\ \bibnamefont {Rinaldi}}, \bibinfo {author}
  {\bibfnamefont {Alessandro}\ \bibnamefont {Roggero}}, \bibinfo {author}
  {\bibfnamefont {David~I.}\ \bibnamefont {Santiago}}, \bibinfo {author}
  {\bibfnamefont {Martin~J.}\ \bibnamefont {Savage}}, \bibinfo {author}
  {\bibfnamefont {Irfan}\ \bibnamefont {Siddiqi}}, \bibinfo {author}
  {\bibfnamefont {George}\ \bibnamefont {Siopsis}}, \bibinfo {author}
  {\bibfnamefont {David}\ \bibnamefont {Van~Zanten}}, \bibinfo {author}
  {\bibfnamefont {Nathan}\ \bibnamefont {Wiebe}}, \bibinfo {author}
  {\bibfnamefont {Yukari}\ \bibnamefont {Yamauchi}}, \bibinfo {author}
  {\bibfnamefont {Kübra}\ \bibnamefont {Yeter-Aydeniz}}, \ and\ \bibinfo
  {author} {\bibfnamefont {Silvia}\ \bibnamefont {Zorzetti}},\ }\bibfield
  {title} {\enquote {\bibinfo {title} {Quantum simulation for high energy
  physics},}\ }\href {\doibase 10.48550/ARXIV.2204.03381} {\  (\bibinfo {year}
  {2022}),\ 10.48550/ARXIV.2204.03381}\BibitemShut {NoStop}%
\bibitem [{\citenamefont {'t~Hooft}(1976)}]{tHooft1976}%
  \BibitemOpen
  \bibfield  {author} {\bibinfo {author} {\bibfnamefont {G.}~\bibnamefont
  {'t~Hooft}},\ }\bibfield  {title} {\enquote {\bibinfo {title} {Computation of
  the quantum effects due to a four-dimensional pseudoparticle},}\ }\href
  {\doibase 10.1103/PhysRevD.14.3432} {\bibfield  {journal} {\bibinfo
  {journal} {Phys. Rev. D}\ }\textbf {\bibinfo {volume} {14}},\ \bibinfo
  {pages} {3432--3450} (\bibinfo {year} {1976})}\BibitemShut {NoStop}%
\bibitem [{\citenamefont {Jackiw}\ and\ \citenamefont
  {Rebbi}(1976)}]{Jackiw1976}%
  \BibitemOpen
  \bibfield  {author} {\bibinfo {author} {\bibfnamefont {R.}~\bibnamefont
  {Jackiw}}\ and\ \bibinfo {author} {\bibfnamefont {C.}~\bibnamefont {Rebbi}},\
  }\bibfield  {title} {\enquote {\bibinfo {title} {Vacuum periodicity in a
  yang-mills quantum theory},}\ }\href {\doibase 10.1103/PhysRevLett.37.172}
  {\bibfield  {journal} {\bibinfo  {journal} {Phys. Rev. Lett.}\ }\textbf
  {\bibinfo {volume} {37}},\ \bibinfo {pages} {172--175} (\bibinfo {year}
  {1976})}\BibitemShut {NoStop}%
\bibitem [{\citenamefont {Callan}\ \emph {et~al.}(1979)\citenamefont {Callan},
  \citenamefont {Dashen},\ and\ \citenamefont {Gross}}]{Callan1979}%
  \BibitemOpen
  \bibfield  {author} {\bibinfo {author} {\bibfnamefont {Curtis~G.}\
  \bibnamefont {Callan}}, \bibinfo {author} {\bibfnamefont {Roger~F.}\
  \bibnamefont {Dashen}}, \ and\ \bibinfo {author} {\bibfnamefont {David~J.}\
  \bibnamefont {Gross}},\ }\bibfield  {title} {\enquote {\bibinfo {title}
  {Instantons as a bridge between weak and strong coupling in quantum
  chromodynamics},}\ }\href {\doibase 10.1103/PhysRevD.20.3279} {\bibfield
  {journal} {\bibinfo  {journal} {Phys. Rev. D}\ }\textbf {\bibinfo {volume}
  {20}},\ \bibinfo {pages} {3279--3291} (\bibinfo {year} {1979})}\BibitemShut
  {NoStop}%
\bibitem [{\citenamefont {Chupp}\ \emph {et~al.}(2019)\citenamefont {Chupp},
  \citenamefont {Fierlinger}, \citenamefont {Ramsey-Musolf},\ and\
  \citenamefont {Singh}}]{Chupp_review}%
  \BibitemOpen
  \bibfield  {author} {\bibinfo {author} {\bibfnamefont {T.~E.}\ \bibnamefont
  {Chupp}}, \bibinfo {author} {\bibfnamefont {P.}~\bibnamefont {Fierlinger}},
  \bibinfo {author} {\bibfnamefont {M.~J.}\ \bibnamefont {Ramsey-Musolf}}, \
  and\ \bibinfo {author} {\bibfnamefont {J.~T.}\ \bibnamefont {Singh}},\
  }\bibfield  {title} {\enquote {\bibinfo {title} {Electric dipole moments of
  atoms, molecules, nuclei, and particles},}\ }\href {\doibase
  10.1103/RevModPhys.91.015001} {\bibfield  {journal} {\bibinfo  {journal}
  {Rev. Mod. Phys.}\ }\textbf {\bibinfo {volume} {91}},\ \bibinfo {pages}
  {015001} (\bibinfo {year} {2019})}\BibitemShut {NoStop}%
\bibitem [{\citenamefont {Peccei}\ and\ \citenamefont
  {Quinn}(1977)}]{Peccei1977}%
  \BibitemOpen
  \bibfield  {author} {\bibinfo {author} {\bibfnamefont {R.~D.}\ \bibnamefont
  {Peccei}}\ and\ \bibinfo {author} {\bibfnamefont {Helen~R.}\ \bibnamefont
  {Quinn}},\ }\bibfield  {title} {\enquote {\bibinfo {title} {$\mathrm{CP}$
  conservation in the presence of pseudoparticles},}\ }\href {\doibase
  10.1103/PhysRevLett.38.1440} {\bibfield  {journal} {\bibinfo  {journal}
  {Phys. Rev. Lett.}\ }\textbf {\bibinfo {volume} {38}},\ \bibinfo {pages}
  {1440--1443} (\bibinfo {year} {1977})}\BibitemShut {NoStop}%
\bibitem [{\citenamefont {Weinberg}(1978)}]{Weinberg1978}%
  \BibitemOpen
  \bibfield  {author} {\bibinfo {author} {\bibfnamefont {Steven}\ \bibnamefont
  {Weinberg}},\ }\bibfield  {title} {\enquote {\bibinfo {title} {A new light
  boson?}}\ }\href {\doibase 10.1103/PhysRevLett.40.223} {\bibfield  {journal}
  {\bibinfo  {journal} {Phys. Rev. Lett.}\ }\textbf {\bibinfo {volume} {40}},\
  \bibinfo {pages} {223--226} (\bibinfo {year} {1978})}\BibitemShut {NoStop}%
\bibitem [{\citenamefont {Wilczek}(1978)}]{Wilczek1978}%
  \BibitemOpen
  \bibfield  {author} {\bibinfo {author} {\bibfnamefont {F.}~\bibnamefont
  {Wilczek}},\ }\bibfield  {title} {\enquote {\bibinfo {title} {Problem of
  strong $p$ and $t$ invariance in the presence of instantons},}\ }\href
  {\doibase 10.1103/PhysRevLett.40.279} {\bibfield  {journal} {\bibinfo
  {journal} {Phys. Rev. Lett.}\ }\textbf {\bibinfo {volume} {40}},\ \bibinfo
  {pages} {279--282} (\bibinfo {year} {1978})}\BibitemShut {NoStop}%
\bibitem [{\citenamefont {Graham}\ \emph {et~al.}(2015)\citenamefont {Graham},
  \citenamefont {Irastorza}, \citenamefont {Lamoreaux}, \citenamefont
  {Lindner},\ and\ \citenamefont {van Bibber}}]{Graham2015}%
  \BibitemOpen
  \bibfield  {author} {\bibinfo {author} {\bibfnamefont {Peter~W.}\
  \bibnamefont {Graham}}, \bibinfo {author} {\bibfnamefont {Igor~G.}\
  \bibnamefont {Irastorza}}, \bibinfo {author} {\bibfnamefont {Steven~K.}\
  \bibnamefont {Lamoreaux}}, \bibinfo {author} {\bibfnamefont {Axel}\
  \bibnamefont {Lindner}}, \ and\ \bibinfo {author} {\bibfnamefont {Karl~A.}\
  \bibnamefont {van Bibber}},\ }\bibfield  {title} {\enquote {\bibinfo {title}
  {Experimental searches for the axion and axion-like particles},}\ }\href
  {\doibase 10.1146/annurev-nucl-102014-022120} {\bibfield  {journal} {\bibinfo
   {journal} {Annual Review of Nuclear and Particle Science}\ }\textbf
  {\bibinfo {volume} {65}},\ \bibinfo {pages} {485--514} (\bibinfo {year}
  {2015})},\ \Eprint
  {http://arxiv.org/abs/https://doi.org/10.1146/annurev-nucl-102014-022120}
  {https://doi.org/10.1146/annurev-nucl-102014-022120} \BibitemShut {NoStop}%
\bibitem [{\citenamefont {\"Unsal}(2012)}]{Unsal2012}%
  \BibitemOpen
  \bibfield  {author} {\bibinfo {author} {\bibfnamefont {Mithat}\ \bibnamefont
  {\"Unsal}},\ }\bibfield  {title} {\enquote {\bibinfo {title} {Theta
  dependence, sign problems, and topological interference},}\ }\href {\doibase
  10.1103/PhysRevD.86.105012} {\bibfield  {journal} {\bibinfo  {journal} {Phys.
  Rev. D}\ }\textbf {\bibinfo {volume} {86}},\ \bibinfo {pages} {105012}
  (\bibinfo {year} {2012})}\BibitemShut {NoStop}%
\bibitem [{\citenamefont {Byrnes}\ \emph {et~al.}(2002)\citenamefont {Byrnes},
  \citenamefont {Sriganesh}, \citenamefont {Bursill},\ and\ \citenamefont
  {Hamer}}]{Byrnes2002}%
  \BibitemOpen
  \bibfield  {author} {\bibinfo {author} {\bibfnamefont {T.~M.~R.}\
  \bibnamefont {Byrnes}}, \bibinfo {author} {\bibfnamefont {P.}~\bibnamefont
  {Sriganesh}}, \bibinfo {author} {\bibfnamefont {R.~J.}\ \bibnamefont
  {Bursill}}, \ and\ \bibinfo {author} {\bibfnamefont {C.~J.}\ \bibnamefont
  {Hamer}},\ }\bibfield  {title} {\enquote {\bibinfo {title} {Density matrix
  renormalization group approach to the massive schwinger model},}\ }\href
  {\doibase 10.1103/PhysRevD.66.013002} {\bibfield  {journal} {\bibinfo
  {journal} {Phys. Rev. D}\ }\textbf {\bibinfo {volume} {66}},\ \bibinfo
  {pages} {013002} (\bibinfo {year} {2002})}\BibitemShut {NoStop}%
\bibitem [{\citenamefont {Buyens}\ \emph {et~al.}(2014)\citenamefont {Buyens},
  \citenamefont {Haegeman}, \citenamefont {Van~Acoleyen}, \citenamefont
  {Verschelde},\ and\ \citenamefont {Verstraete}}]{Buyens2014}%
  \BibitemOpen
  \bibfield  {author} {\bibinfo {author} {\bibfnamefont {Boye}\ \bibnamefont
  {Buyens}}, \bibinfo {author} {\bibfnamefont {Jutho}\ \bibnamefont
  {Haegeman}}, \bibinfo {author} {\bibfnamefont {Karel}\ \bibnamefont
  {Van~Acoleyen}}, \bibinfo {author} {\bibfnamefont {Henri}\ \bibnamefont
  {Verschelde}}, \ and\ \bibinfo {author} {\bibfnamefont {Frank}\ \bibnamefont
  {Verstraete}},\ }\bibfield  {title} {\enquote {\bibinfo {title} {Matrix
  product states for gauge field theories},}\ }\href {\doibase
  10.1103/PhysRevLett.113.091601} {\bibfield  {journal} {\bibinfo  {journal}
  {Phys. Rev. Lett.}\ }\textbf {\bibinfo {volume} {113}},\ \bibinfo {pages}
  {091601} (\bibinfo {year} {2014})}\BibitemShut {NoStop}%
\bibitem [{\citenamefont {Shimizu}\ and\ \citenamefont
  {Kuramashi}(2014)}]{Shimizu2014}%
  \BibitemOpen
  \bibfield  {author} {\bibinfo {author} {\bibfnamefont {Yuya}\ \bibnamefont
  {Shimizu}}\ and\ \bibinfo {author} {\bibfnamefont {Yoshinobu}\ \bibnamefont
  {Kuramashi}},\ }\bibfield  {title} {\enquote {\bibinfo {title} {Critical
  behavior of the lattice schwinger model with a topological term at
  $\ensuremath{\theta}=\ensuremath{\pi}$ using the grassmann tensor
  renormalization group},}\ }\href {\doibase 10.1103/PhysRevD.90.074503}
  {\bibfield  {journal} {\bibinfo  {journal} {Phys. Rev. D}\ }\textbf {\bibinfo
  {volume} {90}},\ \bibinfo {pages} {074503} (\bibinfo {year}
  {2014})}\BibitemShut {NoStop}%
\bibitem [{\citenamefont {Buyens}\ \emph {et~al.}(2016)\citenamefont {Buyens},
  \citenamefont {Haegeman}, \citenamefont {Verschelde}, \citenamefont
  {Verstraete},\ and\ \citenamefont {Van~Acoleyen}}]{Buyens2016}%
  \BibitemOpen
  \bibfield  {author} {\bibinfo {author} {\bibfnamefont {Boye}\ \bibnamefont
  {Buyens}}, \bibinfo {author} {\bibfnamefont {Jutho}\ \bibnamefont
  {Haegeman}}, \bibinfo {author} {\bibfnamefont {Henri}\ \bibnamefont
  {Verschelde}}, \bibinfo {author} {\bibfnamefont {Frank}\ \bibnamefont
  {Verstraete}}, \ and\ \bibinfo {author} {\bibfnamefont {Karel}\ \bibnamefont
  {Van~Acoleyen}},\ }\bibfield  {title} {\enquote {\bibinfo {title}
  {Confinement and string breaking for ${\mathrm{qed}}_{2}$ in the hamiltonian
  picture},}\ }\href {\doibase 10.1103/PhysRevX.6.041040} {\bibfield  {journal}
  {\bibinfo  {journal} {Phys. Rev. X}\ }\textbf {\bibinfo {volume} {6}},\
  \bibinfo {pages} {041040} (\bibinfo {year} {2016})}\BibitemShut {NoStop}%
\bibitem [{\citenamefont {Surace}\ \emph {et~al.}(2020)\citenamefont {Surace},
  \citenamefont {Mazza}, \citenamefont {Giudici}, \citenamefont {Lerose},
  \citenamefont {Gambassi},\ and\ \citenamefont {Dalmonte}}]{Surace2020}%
  \BibitemOpen
  \bibfield  {author} {\bibinfo {author} {\bibfnamefont {Federica~M.}\
  \bibnamefont {Surace}}, \bibinfo {author} {\bibfnamefont {Paolo~P.}\
  \bibnamefont {Mazza}}, \bibinfo {author} {\bibfnamefont {Giuliano}\
  \bibnamefont {Giudici}}, \bibinfo {author} {\bibfnamefont {Alessio}\
  \bibnamefont {Lerose}}, \bibinfo {author} {\bibfnamefont {Andrea}\
  \bibnamefont {Gambassi}}, \ and\ \bibinfo {author} {\bibfnamefont {Marcello}\
  \bibnamefont {Dalmonte}},\ }\bibfield  {title} {\enquote {\bibinfo {title}
  {Lattice gauge theories and string dynamics in {Rydberg} atom quantum
  simulators},}\ }\href {\doibase 10.1103/PhysRevX.10.021041} {\bibfield
  {journal} {\bibinfo  {journal} {Phys. Rev. X}\ }\textbf {\bibinfo {volume}
  {10}},\ \bibinfo {pages} {021041} (\bibinfo {year} {2020})}\BibitemShut
  {NoStop}%
\bibitem [{\citenamefont {Coleman}(1976)}]{Coleman1976}%
  \BibitemOpen
  \bibfield  {author} {\bibinfo {author} {\bibfnamefont {Sidney}\ \bibnamefont
  {Coleman}},\ }\bibfield  {title} {\enquote {\bibinfo {title} {More about the
  massive schwinger model},}\ }\href {\doibase
  https://doi.org/10.1016/0003-4916(76)90280-3} {\bibfield  {journal} {\bibinfo
   {journal} {Annals of Physics}\ }\textbf {\bibinfo {volume} {101}},\ \bibinfo
  {pages} {239 -- 267} (\bibinfo {year} {1976})}\BibitemShut {NoStop}%
\bibitem [{\citenamefont {Zache}\ \emph {et~al.}(2019)\citenamefont {Zache},
  \citenamefont {Mueller}, \citenamefont {Schneider}, \citenamefont
  {Jendrzejewski}, \citenamefont {Berges},\ and\ \citenamefont
  {Hauke}}]{Zache2019}%
  \BibitemOpen
  \bibfield  {author} {\bibinfo {author} {\bibfnamefont {T.~V.}\ \bibnamefont
  {Zache}}, \bibinfo {author} {\bibfnamefont {N.}~\bibnamefont {Mueller}},
  \bibinfo {author} {\bibfnamefont {J.~T.}\ \bibnamefont {Schneider}}, \bibinfo
  {author} {\bibfnamefont {F.}~\bibnamefont {Jendrzejewski}}, \bibinfo {author}
  {\bibfnamefont {J.}~\bibnamefont {Berges}}, \ and\ \bibinfo {author}
  {\bibfnamefont {P.}~\bibnamefont {Hauke}},\ }\bibfield  {title} {\enquote
  {\bibinfo {title} {Dynamical topological transitions in the massive schwinger
  model with a $\ensuremath{\theta}$ term},}\ }\href {\doibase
  10.1103/PhysRevLett.122.050403} {\bibfield  {journal} {\bibinfo  {journal}
  {Phys. Rev. Lett.}\ }\textbf {\bibinfo {volume} {122}},\ \bibinfo {pages}
  {050403} (\bibinfo {year} {2019})}\BibitemShut {NoStop}%
\bibitem [{\citenamefont {Kharzeev}\ \emph {et~al.}(2008)\citenamefont
  {Kharzeev}, \citenamefont {McLerran},\ and\ \citenamefont
  {Warringa}}]{Kharzeev2008}%
  \BibitemOpen
  \bibfield  {author} {\bibinfo {author} {\bibfnamefont {Dmitri~E.}\
  \bibnamefont {Kharzeev}}, \bibinfo {author} {\bibfnamefont {Larry~D.}\
  \bibnamefont {McLerran}}, \ and\ \bibinfo {author} {\bibfnamefont
  {Harmen~J.}\ \bibnamefont {Warringa}},\ }\bibfield  {title} {\enquote
  {\bibinfo {title} {The effects of topological charge change in heavy ion
  collisions: “event by event p and cp violation”},}\ }\href {\doibase
  https://doi.org/10.1016/j.nuclphysa.2008.02.298} {\bibfield  {journal}
  {\bibinfo  {journal} {Nuclear Physics A}\ }\textbf {\bibinfo {volume}
  {803}},\ \bibinfo {pages} {227--253} (\bibinfo {year} {2008})}\BibitemShut
  {NoStop}%
\bibitem [{\citenamefont {Fukushima}\ \emph {et~al.}(2008)\citenamefont
  {Fukushima}, \citenamefont {Kharzeev},\ and\ \citenamefont
  {Warringa}}]{Fukushima2008}%
  \BibitemOpen
  \bibfield  {author} {\bibinfo {author} {\bibfnamefont {Kenji}\ \bibnamefont
  {Fukushima}}, \bibinfo {author} {\bibfnamefont {Dmitri~E.}\ \bibnamefont
  {Kharzeev}}, \ and\ \bibinfo {author} {\bibfnamefont {Harmen~J.}\
  \bibnamefont {Warringa}},\ }\bibfield  {title} {\enquote {\bibinfo {title}
  {Chiral magnetic effect},}\ }\href {\doibase 10.1103/PhysRevD.78.074033}
  {\bibfield  {journal} {\bibinfo  {journal} {Phys. Rev. D}\ }\textbf {\bibinfo
  {volume} {78}},\ \bibinfo {pages} {074033} (\bibinfo {year}
  {2008})}\BibitemShut {NoStop}%
\bibitem [{\citenamefont {Kharzeev}\ \emph {et~al.}(2016)\citenamefont
  {Kharzeev}, \citenamefont {Liao}, \citenamefont {Voloshin},\ and\
  \citenamefont {Wang}}]{Kharzeev2016}%
  \BibitemOpen
  \bibfield  {author} {\bibinfo {author} {\bibfnamefont {D.E.}\ \bibnamefont
  {Kharzeev}}, \bibinfo {author} {\bibfnamefont {J.}~\bibnamefont {Liao}},
  \bibinfo {author} {\bibfnamefont {S.A.}\ \bibnamefont {Voloshin}}, \ and\
  \bibinfo {author} {\bibfnamefont {G.}~\bibnamefont {Wang}},\ }\bibfield
  {title} {\enquote {\bibinfo {title} {Chiral magnetic and vortical effects in
  high-energy nuclear collisions—a status report},}\ }\href {\doibase
  https://doi.org/10.1016/j.ppnp.2016.01.001} {\bibfield  {journal} {\bibinfo
  {journal} {Progress in Particle and Nuclear Physics}\ }\textbf {\bibinfo
  {volume} {88}},\ \bibinfo {pages} {1--28} (\bibinfo {year}
  {2016})}\BibitemShut {NoStop}%
\bibitem [{\citenamefont {Koch}\ \emph {et~al.}(2017)\citenamefont {Koch},
  \citenamefont {Schlichting}, \citenamefont {Skokov}, \citenamefont
  {Sorensen}, \citenamefont {Thomas}, \citenamefont {Voloshin}, \citenamefont
  {Wang},\ and\ \citenamefont {Yee}}]{Koch2017}%
  \BibitemOpen
  \bibfield  {author} {\bibinfo {author} {\bibfnamefont {Volker}\ \bibnamefont
  {Koch}}, \bibinfo {author} {\bibfnamefont {Soeren}\ \bibnamefont
  {Schlichting}}, \bibinfo {author} {\bibfnamefont {Vladimir}\ \bibnamefont
  {Skokov}}, \bibinfo {author} {\bibfnamefont {Paul}\ \bibnamefont {Sorensen}},
  \bibinfo {author} {\bibfnamefont {Jim}\ \bibnamefont {Thomas}}, \bibinfo
  {author} {\bibfnamefont {Sergei}\ \bibnamefont {Voloshin}}, \bibinfo {author}
  {\bibfnamefont {Gang}\ \bibnamefont {Wang}}, \ and\ \bibinfo {author}
  {\bibfnamefont {Ho-Ung}\ \bibnamefont {Yee}},\ }\bibfield  {title} {\enquote
  {\bibinfo {title} {Status of the chiral magnetic effect and collisions of
  isobars},}\ }\href {\doibase 10.1088/1674-1137/41/7/072001} {\bibfield
  {journal} {\bibinfo  {journal} {Chinese Physics C}\ }\textbf {\bibinfo
  {volume} {41}},\ \bibinfo {pages} {072001} (\bibinfo {year}
  {2017})}\BibitemShut {NoStop}%
\bibitem [{\citenamefont {Kharzeev}\ and\ \citenamefont
  {Kikuchi}(2020)}]{Kharzeev2020}%
  \BibitemOpen
  \bibfield  {author} {\bibinfo {author} {\bibfnamefont {Dmitri~E.}\
  \bibnamefont {Kharzeev}}\ and\ \bibinfo {author} {\bibfnamefont {Yuta}\
  \bibnamefont {Kikuchi}},\ }\bibfield  {title} {\enquote {\bibinfo {title}
  {Real-time chiral dynamics from a digital quantum simulation},}\ }\href
  {\doibase 10.1103/PhysRevResearch.2.023342} {\bibfield  {journal} {\bibinfo
  {journal} {Phys. Rev. Research}\ }\textbf {\bibinfo {volume} {2}},\ \bibinfo
  {pages} {023342} (\bibinfo {year} {2020})}\BibitemShut {NoStop}%
\bibitem [{\citenamefont {ADAM}(1999)}]{Adam1999}%
  \BibitemOpen
  \bibfield  {author} {\bibinfo {author} {\bibfnamefont {C.}~\bibnamefont
  {ADAM}},\ }\bibfield  {title} {\enquote {\bibinfo {title} {Theta vacuum in
  different gauges},}\ }\href {\doibase 10.1142/S0217732399000225} {\bibfield
  {journal} {\bibinfo  {journal} {Modern Physics Letters A}\ }\textbf {\bibinfo
  {volume} {14}},\ \bibinfo {pages} {185--197} (\bibinfo {year} {1999})},\
  \Eprint {http://arxiv.org/abs/https://doi.org/10.1142/S0217732399000225}
  {https://doi.org/10.1142/S0217732399000225} \BibitemShut {NoStop}%
\bibitem [{\citenamefont {Tong}(2018)}]{Tong_LectureNotes}%
  \BibitemOpen
  \bibfield  {author} {\bibinfo {author} {\bibfnamefont {David}\ \bibnamefont
  {Tong}},\ }\href@noop {} {\enquote {\bibinfo {title} {{Gauge Theory}},}\
  }\bibinfo {howpublished}
  {\url{https://www.damtp.cam.ac.uk/user/tong/gaugetheory.html}} (\bibinfo
  {year} {2018})\BibitemShut {NoStop}%
\bibitem [{\citenamefont {Yang}\ \emph
  {et~al.}(2020{\natexlab{a}})\citenamefont {Yang}, \citenamefont {Sun},
  \citenamefont {Ott}, \citenamefont {Wang}, \citenamefont {Zache},
  \citenamefont {Halimeh}, \citenamefont {Yuan}, \citenamefont {Hauke},\ and\
  \citenamefont {Pan}}]{Yang2020}%
  \BibitemOpen
  \bibfield  {author} {\bibinfo {author} {\bibfnamefont {Bing}\ \bibnamefont
  {Yang}}, \bibinfo {author} {\bibfnamefont {Hui}\ \bibnamefont {Sun}},
  \bibinfo {author} {\bibfnamefont {Robert}\ \bibnamefont {Ott}}, \bibinfo
  {author} {\bibfnamefont {Han-Yi}\ \bibnamefont {Wang}}, \bibinfo {author}
  {\bibfnamefont {Torsten~V.}\ \bibnamefont {Zache}}, \bibinfo {author}
  {\bibfnamefont {Jad~C.}\ \bibnamefont {Halimeh}}, \bibinfo {author}
  {\bibfnamefont {Zhen-Sheng}\ \bibnamefont {Yuan}}, \bibinfo {author}
  {\bibfnamefont {Philipp}\ \bibnamefont {Hauke}}, \ and\ \bibinfo {author}
  {\bibfnamefont {Jian-Wei}\ \bibnamefont {Pan}},\ }\bibfield  {title}
  {\enquote {\bibinfo {title} {Observation of gauge invariance in a 71-site
  {Bose--Hubbard} quantum simulator},}\ }\href {\doibase
  10.1038/s41586-020-2910-8} {\bibfield  {journal} {\bibinfo  {journal}
  {Nature}\ }\textbf {\bibinfo {volume} {587}},\ \bibinfo {pages} {392--396}
  (\bibinfo {year} {2020}{\natexlab{a}})}\BibitemShut {NoStop}%
\bibitem [{\citenamefont {Halimeh}\ \emph
  {et~al.}(2021{\natexlab{a}})\citenamefont {Halimeh}, \citenamefont {Lang},
  \citenamefont {Mildenberger}, \citenamefont {Jiang},\ and\ \citenamefont
  {Hauke}}]{Halimeh2020e}%
  \BibitemOpen
  \bibfield  {author} {\bibinfo {author} {\bibfnamefont {Jad~C.}\ \bibnamefont
  {Halimeh}}, \bibinfo {author} {\bibfnamefont {Haifeng}\ \bibnamefont {Lang}},
  \bibinfo {author} {\bibfnamefont {Julius}\ \bibnamefont {Mildenberger}},
  \bibinfo {author} {\bibfnamefont {Zhang}\ \bibnamefont {Jiang}}, \ and\
  \bibinfo {author} {\bibfnamefont {Philipp}\ \bibnamefont {Hauke}},\
  }\bibfield  {title} {\enquote {\bibinfo {title} {Gauge-symmetry protection
  using single-body terms},}\ }\href {\doibase 10.1103/PRXQuantum.2.040311}
  {\bibfield  {journal} {\bibinfo  {journal} {PRX Quantum}\ }\textbf {\bibinfo
  {volume} {2}},\ \bibinfo {pages} {040311} (\bibinfo {year}
  {2021}{\natexlab{a}})}\BibitemShut {NoStop}%
\bibitem [{\citenamefont {Zhou}\ \emph {et~al.}(2021)\citenamefont {Zhou},
  \citenamefont {Su}, \citenamefont {Halimeh}, \citenamefont {Ott},
  \citenamefont {Sun}, \citenamefont {Hauke}, \citenamefont {Yang},
  \citenamefont {Yuan}, \citenamefont {Berges},\ and\ \citenamefont
  {Pan}}]{Zhou2021}%
  \BibitemOpen
  \bibfield  {author} {\bibinfo {author} {\bibfnamefont {Zhao-Yu}\ \bibnamefont
  {Zhou}}, \bibinfo {author} {\bibfnamefont {Guo-Xian}\ \bibnamefont {Su}},
  \bibinfo {author} {\bibfnamefont {Jad~C.}\ \bibnamefont {Halimeh}}, \bibinfo
  {author} {\bibfnamefont {Robert}\ \bibnamefont {Ott}}, \bibinfo {author}
  {\bibfnamefont {Hui}\ \bibnamefont {Sun}}, \bibinfo {author} {\bibfnamefont
  {Philipp}\ \bibnamefont {Hauke}}, \bibinfo {author} {\bibfnamefont {Bing}\
  \bibnamefont {Yang}}, \bibinfo {author} {\bibfnamefont {Zhen-Sheng}\
  \bibnamefont {Yuan}}, \bibinfo {author} {\bibfnamefont {Jürgen}\
  \bibnamefont {Berges}}, \ and\ \bibinfo {author} {\bibfnamefont {Jian-Wei}\
  \bibnamefont {Pan}},\ }\bibfield  {title} {\enquote {\bibinfo {title}
  {Thermalization dynamics of a gauge theory on a quantum simulator},}\
  }\href@noop {} {\bibfield  {journal} {\bibinfo  {journal} {arXiv preprint}\ }
  (\bibinfo {year} {2021})},\ \Eprint {http://arxiv.org/abs/2107.13563}
  {arXiv:2107.13563 [cond-mat.quant-gas]} \BibitemShut {NoStop}%
\bibitem [{\citenamefont {Vidal}(2004)}]{Vidal2004}%
  \BibitemOpen
  \bibfield  {author} {\bibinfo {author} {\bibfnamefont {Guifr\'e}\
  \bibnamefont {Vidal}},\ }\bibfield  {title} {\enquote {\bibinfo {title}
  {Efficient simulation of one-dimensional quantum many-body systems},}\ }\href
  {\doibase 10.1103/PhysRevLett.93.040502} {\bibfield  {journal} {\bibinfo
  {journal} {Phys. Rev. Lett.}\ }\textbf {\bibinfo {volume} {93}},\ \bibinfo
  {pages} {040502} (\bibinfo {year} {2004})}\BibitemShut {NoStop}%
\bibitem [{\citenamefont {White}\ and\ \citenamefont
  {Feiguin}(2004)}]{White2004}%
  \BibitemOpen
  \bibfield  {author} {\bibinfo {author} {\bibfnamefont {Steven~R.}\
  \bibnamefont {White}}\ and\ \bibinfo {author} {\bibfnamefont {Adrian~E.}\
  \bibnamefont {Feiguin}},\ }\bibfield  {title} {\enquote {\bibinfo {title}
  {Real-time evolution using the density matrix renormalization group},}\
  }\href {\doibase 10.1103/PhysRevLett.93.076401} {\bibfield  {journal}
  {\bibinfo  {journal} {Phys. Rev. Lett.}\ }\textbf {\bibinfo {volume} {93}},\
  \bibinfo {pages} {076401} (\bibinfo {year} {2004})}\BibitemShut {NoStop}%
\bibitem [{\citenamefont {Daley}\ \emph {et~al.}(2004)\citenamefont {Daley},
  \citenamefont {Kollath}, \citenamefont {Schollwöck},\ and\ \citenamefont
  {Vidal}}]{Daley2004}%
  \BibitemOpen
  \bibfield  {author} {\bibinfo {author} {\bibfnamefont {A~J}\ \bibnamefont
  {Daley}}, \bibinfo {author} {\bibfnamefont {C}~\bibnamefont {Kollath}},
  \bibinfo {author} {\bibfnamefont {U}~\bibnamefont {Schollwöck}}, \ and\
  \bibinfo {author} {\bibfnamefont {G}~\bibnamefont {Vidal}},\ }\bibfield
  {title} {\enquote {\bibinfo {title} {Time-dependent density-matrix
  renormalization-group using adaptive effective hilbert spaces},}\ }\href
  {\doibase 10.1088/1742-5468/2004/04/p04005} {\bibfield  {journal} {\bibinfo
  {journal} {Journal of Statistical Mechanics: Theory and Experiment}\ }\textbf
  {\bibinfo {volume} {2004}},\ \bibinfo {pages} {P04005} (\bibinfo {year}
  {2004})}\BibitemShut {NoStop}%
\bibitem [{\citenamefont {Bernien}\ \emph {et~al.}(2017)\citenamefont
  {Bernien}, \citenamefont {Schwartz}, \citenamefont {Keesling}, \citenamefont
  {Levine}, \citenamefont {Omran}, \citenamefont {Pichler}, \citenamefont
  {Choi}, \citenamefont {Zibrov}, \citenamefont {Endres}, \citenamefont
  {Greiner}, \citenamefont {Vuleti{\'c}},\ and\ \citenamefont
  {Lukin}}]{Bernien2017}%
  \BibitemOpen
  \bibfield  {author} {\bibinfo {author} {\bibfnamefont {Hannes}\ \bibnamefont
  {Bernien}}, \bibinfo {author} {\bibfnamefont {Sylvain}\ \bibnamefont
  {Schwartz}}, \bibinfo {author} {\bibfnamefont {Alexander}\ \bibnamefont
  {Keesling}}, \bibinfo {author} {\bibfnamefont {Harry}\ \bibnamefont
  {Levine}}, \bibinfo {author} {\bibfnamefont {Ahmed}\ \bibnamefont {Omran}},
  \bibinfo {author} {\bibfnamefont {Hannes}\ \bibnamefont {Pichler}}, \bibinfo
  {author} {\bibfnamefont {Soonwon}\ \bibnamefont {Choi}}, \bibinfo {author}
  {\bibfnamefont {Alexander~S.}\ \bibnamefont {Zibrov}}, \bibinfo {author}
  {\bibfnamefont {Manuel}\ \bibnamefont {Endres}}, \bibinfo {author}
  {\bibfnamefont {Markus}\ \bibnamefont {Greiner}}, \bibinfo {author}
  {\bibfnamefont {Vladan}\ \bibnamefont {Vuleti{\'c}}}, \ and\ \bibinfo
  {author} {\bibfnamefont {Mikhail~D.}\ \bibnamefont {Lukin}},\ }\bibfield
  {title} {\enquote {\bibinfo {title} {Probing many-body dynamics on a 51-atom
  quantum simulator},}\ }\href {\doibase 10.1038/nature24622} {\bibfield
  {journal} {\bibinfo  {journal} {Nature}\ }\textbf {\bibinfo {volume} {551}},\
  \bibinfo {pages} {579--584} (\bibinfo {year} {2017})}\BibitemShut {NoStop}%
\bibitem [{\citenamefont {Kokail}\ \emph {et~al.}(2019)\citenamefont {Kokail},
  \citenamefont {Maier}, \citenamefont {van Bijnen}, \citenamefont {Brydges},
  \citenamefont {Joshi}, \citenamefont {Jurcevic}, \citenamefont {Muschik},
  \citenamefont {Silvi}, \citenamefont {Blatt}, \citenamefont {Roos},\ and\
  \citenamefont {Zoller}}]{Kokail2019}%
  \BibitemOpen
  \bibfield  {author} {\bibinfo {author} {\bibfnamefont {C.}~\bibnamefont
  {Kokail}}, \bibinfo {author} {\bibfnamefont {C.}~\bibnamefont {Maier}},
  \bibinfo {author} {\bibfnamefont {R.}~\bibnamefont {van Bijnen}}, \bibinfo
  {author} {\bibfnamefont {T.}~\bibnamefont {Brydges}}, \bibinfo {author}
  {\bibfnamefont {M.~K.}\ \bibnamefont {Joshi}}, \bibinfo {author}
  {\bibfnamefont {P.}~\bibnamefont {Jurcevic}}, \bibinfo {author}
  {\bibfnamefont {C.~A.}\ \bibnamefont {Muschik}}, \bibinfo {author}
  {\bibfnamefont {P.}~\bibnamefont {Silvi}}, \bibinfo {author} {\bibfnamefont
  {R.}~\bibnamefont {Blatt}}, \bibinfo {author} {\bibfnamefont {C.~F.}\
  \bibnamefont {Roos}}, \ and\ \bibinfo {author} {\bibfnamefont
  {P.}~\bibnamefont {Zoller}},\ }\bibfield  {title} {\enquote {\bibinfo {title}
  {Self-verifying variational quantum simulation of lattice models},}\ }\href
  {\doibase 10.1038/s41586-019-1177-4} {\bibfield  {journal} {\bibinfo
  {journal} {Nature}\ }\textbf {\bibinfo {volume} {569}},\ \bibinfo {pages}
  {355--360} (\bibinfo {year} {2019})}\BibitemShut {NoStop}%
\bibitem [{\citenamefont {Martinez}\ \emph {et~al.}(2016)\citenamefont
  {Martinez}, \citenamefont {Muschik}, \citenamefont {Schindler}, \citenamefont
  {Nigg}, \citenamefont {Erhard}, \citenamefont {Heyl}, \citenamefont {Hauke},
  \citenamefont {Dalmonte}, \citenamefont {Monz}, \citenamefont {Zoller},\ and\
  \citenamefont {Blatt}}]{Martinez2016}%
  \BibitemOpen
  \bibfield  {author} {\bibinfo {author} {\bibfnamefont {Esteban~A.}\
  \bibnamefont {Martinez}}, \bibinfo {author} {\bibfnamefont {Christine~A.}\
  \bibnamefont {Muschik}}, \bibinfo {author} {\bibfnamefont {Philipp}\
  \bibnamefont {Schindler}}, \bibinfo {author} {\bibfnamefont {Daniel}\
  \bibnamefont {Nigg}}, \bibinfo {author} {\bibfnamefont {Alexander}\
  \bibnamefont {Erhard}}, \bibinfo {author} {\bibfnamefont {Markus}\
  \bibnamefont {Heyl}}, \bibinfo {author} {\bibfnamefont {Philipp}\
  \bibnamefont {Hauke}}, \bibinfo {author} {\bibfnamefont {Marcello}\
  \bibnamefont {Dalmonte}}, \bibinfo {author} {\bibfnamefont {Thomas}\
  \bibnamefont {Monz}}, \bibinfo {author} {\bibfnamefont {Peter}\ \bibnamefont
  {Zoller}}, \ and\ \bibinfo {author} {\bibfnamefont {Rainer}\ \bibnamefont
  {Blatt}},\ }\bibfield  {title} {\enquote {\bibinfo {title} {Real-time
  dynamics of lattice gauge theories with a few-qubit quantum computer},}\
  }\href {\doibase 10.1038/nature18318} {\bibfield  {journal} {\bibinfo
  {journal} {Nature}\ }\textbf {\bibinfo {volume} {534}},\ \bibinfo {pages}
  {516--519} (\bibinfo {year} {2016})}\BibitemShut {NoStop}%
\bibitem [{\citenamefont {Muschik}\ \emph {et~al.}(2017)\citenamefont
  {Muschik}, \citenamefont {Heyl}, \citenamefont {Martinez}, \citenamefont
  {Monz}, \citenamefont {Schindler}, \citenamefont {Vogell}, \citenamefont
  {Dalmonte}, \citenamefont {Hauke}, \citenamefont {Blatt},\ and\ \citenamefont
  {Zoller}}]{Muschik2017}%
  \BibitemOpen
  \bibfield  {author} {\bibinfo {author} {\bibfnamefont {Christine}\
  \bibnamefont {Muschik}}, \bibinfo {author} {\bibfnamefont {Markus}\
  \bibnamefont {Heyl}}, \bibinfo {author} {\bibfnamefont {Esteban}\
  \bibnamefont {Martinez}}, \bibinfo {author} {\bibfnamefont {Thomas}\
  \bibnamefont {Monz}}, \bibinfo {author} {\bibfnamefont {Philipp}\
  \bibnamefont {Schindler}}, \bibinfo {author} {\bibfnamefont {Berit}\
  \bibnamefont {Vogell}}, \bibinfo {author} {\bibfnamefont {Marcello}\
  \bibnamefont {Dalmonte}}, \bibinfo {author} {\bibfnamefont {Philipp}\
  \bibnamefont {Hauke}}, \bibinfo {author} {\bibfnamefont {Rainer}\
  \bibnamefont {Blatt}}, \ and\ \bibinfo {author} {\bibfnamefont {Peter}\
  \bibnamefont {Zoller}},\ }\bibfield  {title} {\enquote {\bibinfo {title}
  {U(1) {Wilson} lattice gauge theories in digital quantum simulators},}\
  }\href {\doibase 10.1088/1367-2630/aa89ab} {\bibfield  {journal} {\bibinfo
  {journal} {New Journal of Physics}\ }\textbf {\bibinfo {volume} {19}},\
  \bibinfo {pages} {103020} (\bibinfo {year} {2017})}\BibitemShut {NoStop}%
\bibitem [{\citenamefont {Klco}\ \emph {et~al.}(2018)\citenamefont {Klco},
  \citenamefont {Dumitrescu}, \citenamefont {McCaskey}, \citenamefont {Morris},
  \citenamefont {Pooser}, \citenamefont {Sanz}, \citenamefont {Solano},
  \citenamefont {Lougovski},\ and\ \citenamefont {Savage}}]{Klco2018}%
  \BibitemOpen
  \bibfield  {author} {\bibinfo {author} {\bibfnamefont {N.}~\bibnamefont
  {Klco}}, \bibinfo {author} {\bibfnamefont {E.~F.}\ \bibnamefont
  {Dumitrescu}}, \bibinfo {author} {\bibfnamefont {A.~J.}\ \bibnamefont
  {McCaskey}}, \bibinfo {author} {\bibfnamefont {T.~D.}\ \bibnamefont
  {Morris}}, \bibinfo {author} {\bibfnamefont {R.~C.}\ \bibnamefont {Pooser}},
  \bibinfo {author} {\bibfnamefont {M.}~\bibnamefont {Sanz}}, \bibinfo {author}
  {\bibfnamefont {E.}~\bibnamefont {Solano}}, \bibinfo {author} {\bibfnamefont
  {P.}~\bibnamefont {Lougovski}}, \ and\ \bibinfo {author} {\bibfnamefont
  {M.~J.}\ \bibnamefont {Savage}},\ }\bibfield  {title} {\enquote {\bibinfo
  {title} {Quantum-classical computation of {Schwinger} model dynamics using
  quantum computers},}\ }\href {\doibase 10.1103/PhysRevA.98.032331} {\bibfield
   {journal} {\bibinfo  {journal} {Phys. Rev. A}\ }\textbf {\bibinfo {volume}
  {98}},\ \bibinfo {pages} {032331} (\bibinfo {year} {2018})}\BibitemShut
  {NoStop}%
\bibitem [{\citenamefont {Schweizer}\ \emph {et~al.}(2019)\citenamefont
  {Schweizer}, \citenamefont {Grusdt}, \citenamefont {Berngruber},
  \citenamefont {Barbiero}, \citenamefont {Demler}, \citenamefont {Goldman},
  \citenamefont {Bloch},\ and\ \citenamefont {Aidelsburger}}]{Schweizer2019}%
  \BibitemOpen
  \bibfield  {author} {\bibinfo {author} {\bibfnamefont {Christian}\
  \bibnamefont {Schweizer}}, \bibinfo {author} {\bibfnamefont {Fabian}\
  \bibnamefont {Grusdt}}, \bibinfo {author} {\bibfnamefont {Moritz}\
  \bibnamefont {Berngruber}}, \bibinfo {author} {\bibfnamefont {Luca}\
  \bibnamefont {Barbiero}}, \bibinfo {author} {\bibfnamefont {Eugene}\
  \bibnamefont {Demler}}, \bibinfo {author} {\bibfnamefont {Nathan}\
  \bibnamefont {Goldman}}, \bibinfo {author} {\bibfnamefont {Immanuel}\
  \bibnamefont {Bloch}}, \ and\ \bibinfo {author} {\bibfnamefont {Monika}\
  \bibnamefont {Aidelsburger}},\ }\bibfield  {title} {\enquote {\bibinfo
  {title} {Floquet approach to $\mathbb{Z}$2 lattice gauge theories with
  ultracold atoms in optical lattices},}\ }\href {\doibase
  10.1038/s41567-019-0649-7} {\bibfield  {journal} {\bibinfo  {journal} {Nature
  Physics}\ }\textbf {\bibinfo {volume} {15}},\ \bibinfo {pages} {1168--1173}
  (\bibinfo {year} {2019})}\BibitemShut {NoStop}%
\bibitem [{\citenamefont {G{\"o}rg}\ \emph {et~al.}(2019)\citenamefont
  {G{\"o}rg}, \citenamefont {Sandholzer}, \citenamefont {Minguzzi},
  \citenamefont {Desbuquois}, \citenamefont {Messer},\ and\ \citenamefont
  {Esslinger}}]{Goerg2019}%
  \BibitemOpen
  \bibfield  {author} {\bibinfo {author} {\bibfnamefont {Frederik}\
  \bibnamefont {G{\"o}rg}}, \bibinfo {author} {\bibfnamefont {Kilian}\
  \bibnamefont {Sandholzer}}, \bibinfo {author} {\bibfnamefont {Joaqu{\'\i}n}\
  \bibnamefont {Minguzzi}}, \bibinfo {author} {\bibfnamefont {R{\'e}mi}\
  \bibnamefont {Desbuquois}}, \bibinfo {author} {\bibfnamefont {Michael}\
  \bibnamefont {Messer}}, \ and\ \bibinfo {author} {\bibfnamefont {Tilman}\
  \bibnamefont {Esslinger}},\ }\bibfield  {title} {\enquote {\bibinfo {title}
  {Realization of density-dependent {Peierls} phases to engineer quantized
  gauge fields coupled to ultracold matter},}\ }\href {\doibase
  10.1038/s41567-019-0615-4} {\bibfield  {journal} {\bibinfo  {journal} {Nature
  Physics}\ }\textbf {\bibinfo {volume} {15}},\ \bibinfo {pages} {1161--1167}
  (\bibinfo {year} {2019})}\BibitemShut {NoStop}%
\bibitem [{\citenamefont {Mil}\ \emph {et~al.}(2020)\citenamefont {Mil},
  \citenamefont {Zache}, \citenamefont {Hegde}, \citenamefont {Xia},
  \citenamefont {Bhatt}, \citenamefont {Oberthaler}, \citenamefont {Hauke},
  \citenamefont {Berges},\ and\ \citenamefont {Jendrzejewski}}]{Mil2020}%
  \BibitemOpen
  \bibfield  {author} {\bibinfo {author} {\bibfnamefont {Alexander}\
  \bibnamefont {Mil}}, \bibinfo {author} {\bibfnamefont {Torsten~V.}\
  \bibnamefont {Zache}}, \bibinfo {author} {\bibfnamefont {Apoorva}\
  \bibnamefont {Hegde}}, \bibinfo {author} {\bibfnamefont {Andy}\ \bibnamefont
  {Xia}}, \bibinfo {author} {\bibfnamefont {Rohit~P.}\ \bibnamefont {Bhatt}},
  \bibinfo {author} {\bibfnamefont {Markus~K.}\ \bibnamefont {Oberthaler}},
  \bibinfo {author} {\bibfnamefont {Philipp}\ \bibnamefont {Hauke}}, \bibinfo
  {author} {\bibfnamefont {J{\"u}rgen}\ \bibnamefont {Berges}}, \ and\ \bibinfo
  {author} {\bibfnamefont {Fred}\ \bibnamefont {Jendrzejewski}},\ }\bibfield
  {title} {\enquote {\bibinfo {title} {A scalable realization of local {U(1)}
  gauge invariance in cold atomic mixtures},}\ }\href {\doibase
  10.1126/science.aaz5312} {\bibfield  {journal} {\bibinfo  {journal}
  {Science}\ }\textbf {\bibinfo {volume} {367}},\ \bibinfo {pages} {1128--1130}
  (\bibinfo {year} {2020})}\BibitemShut {NoStop}%
\bibitem [{\citenamefont {Klco}\ \emph {et~al.}(2020)\citenamefont {Klco},
  \citenamefont {Savage},\ and\ \citenamefont {Stryker}}]{Klco2020}%
  \BibitemOpen
  \bibfield  {author} {\bibinfo {author} {\bibfnamefont {Natalie}\ \bibnamefont
  {Klco}}, \bibinfo {author} {\bibfnamefont {Martin~J.}\ \bibnamefont
  {Savage}}, \ and\ \bibinfo {author} {\bibfnamefont {Jesse~R.}\ \bibnamefont
  {Stryker}},\ }\bibfield  {title} {\enquote {\bibinfo {title} {{SU(2)}
  {non-Abelian} gauge field theory in one dimension on digital quantum
  computers},}\ }\href {\doibase 10.1103/PhysRevD.101.074512} {\bibfield
  {journal} {\bibinfo  {journal} {Phys. Rev. D}\ }\textbf {\bibinfo {volume}
  {101}},\ \bibinfo {pages} {074512} (\bibinfo {year} {2020})}\BibitemShut
  {NoStop}%
\bibitem [{\citenamefont {Nguyen}\ \emph {et~al.}(2021)\citenamefont {Nguyen},
  \citenamefont {Tran}, \citenamefont {Zhu}, \citenamefont {Green},
  \citenamefont {Alderete}, \citenamefont {Davoudi},\ and\ \citenamefont
  {Linke}}]{Nguyen2021}%
  \BibitemOpen
  \bibfield  {author} {\bibinfo {author} {\bibfnamefont {Nhung~H.}\
  \bibnamefont {Nguyen}}, \bibinfo {author} {\bibfnamefont {Minh~C.}\
  \bibnamefont {Tran}}, \bibinfo {author} {\bibfnamefont {Yingyue}\
  \bibnamefont {Zhu}}, \bibinfo {author} {\bibfnamefont {Alaina~M.}\
  \bibnamefont {Green}}, \bibinfo {author} {\bibfnamefont {C.~Huerta}\
  \bibnamefont {Alderete}}, \bibinfo {author} {\bibfnamefont {Zohreh}\
  \bibnamefont {Davoudi}}, \ and\ \bibinfo {author} {\bibfnamefont
  {Norbert~M.}\ \bibnamefont {Linke}},\ }\bibfield  {title} {\enquote {\bibinfo
  {title} {Digital quantum simulation of the schwinger model and symmetry
  protection with trapped ions},}\ }\href {\doibase 10.48550/ARXIV.2112.14262}
  {\  (\bibinfo {year} {2021}),\ 10.48550/ARXIV.2112.14262}\BibitemShut
  {NoStop}%
\bibitem [{\citenamefont {Wang}\ \emph {et~al.}(2021)\citenamefont {Wang},
  \citenamefont {Ge}, \citenamefont {Xiang}, \citenamefont {Song},
  \citenamefont {Huang}, \citenamefont {Song}, \citenamefont {Guo},
  \citenamefont {Su}, \citenamefont {Xu}, \citenamefont {Zheng},\ and\
  \citenamefont {Fan}}]{Wang2021}%
  \BibitemOpen
  \bibfield  {author} {\bibinfo {author} {\bibfnamefont {Zhan}\ \bibnamefont
  {Wang}}, \bibinfo {author} {\bibfnamefont {Zi-Yong}\ \bibnamefont {Ge}},
  \bibinfo {author} {\bibfnamefont {Zhongcheng}\ \bibnamefont {Xiang}},
  \bibinfo {author} {\bibfnamefont {Xiaohui}\ \bibnamefont {Song}}, \bibinfo
  {author} {\bibfnamefont {Rui-Zhen}\ \bibnamefont {Huang}}, \bibinfo {author}
  {\bibfnamefont {Pengtao}\ \bibnamefont {Song}}, \bibinfo {author}
  {\bibfnamefont {Xue-Yi}\ \bibnamefont {Guo}}, \bibinfo {author}
  {\bibfnamefont {Luhong}\ \bibnamefont {Su}}, \bibinfo {author} {\bibfnamefont
  {Kai}\ \bibnamefont {Xu}}, \bibinfo {author} {\bibfnamefont {Dongning}\
  \bibnamefont {Zheng}}, \ and\ \bibinfo {author} {\bibfnamefont {Heng}\
  \bibnamefont {Fan}},\ }\bibfield  {title} {\enquote {\bibinfo {title}
  {Observation of emergent $\mathbb{Z}_2$ gauge invariance in a superconducting
  circuit},}\ }\href {\doibase 10.48550/ARXIV.2111.05048} {\  (\bibinfo {year}
  {2021}),\ 10.48550/ARXIV.2111.05048}\BibitemShut {NoStop}%
\bibitem [{\citenamefont {Mildenberger}\ \emph {et~al.}(2022)\citenamefont
  {Mildenberger}, \citenamefont {Mruczkiewicz}, \citenamefont {Halimeh},
  \citenamefont {Jiang},\ and\ \citenamefont {Hauke}}]{Mildenberger2022}%
  \BibitemOpen
  \bibfield  {author} {\bibinfo {author} {\bibfnamefont {Julius}\ \bibnamefont
  {Mildenberger}}, \bibinfo {author} {\bibfnamefont {Wojciech}\ \bibnamefont
  {Mruczkiewicz}}, \bibinfo {author} {\bibfnamefont {Jad~C.}\ \bibnamefont
  {Halimeh}}, \bibinfo {author} {\bibfnamefont {Zhang}\ \bibnamefont {Jiang}},
  \ and\ \bibinfo {author} {\bibfnamefont {Philipp}\ \bibnamefont {Hauke}},\
  }\bibfield  {title} {\enquote {\bibinfo {title} {Probing confinement in a
  $\mathbb{Z}_2$ lattice gauge theory on a quantum computer},}\ }\href
  {\doibase 10.48550/ARXIV.2203.08905} {\  (\bibinfo {year} {2022}),\
  10.48550/ARXIV.2203.08905}\BibitemShut {NoStop}%
\bibitem [{\citenamefont {Hamer}\ \emph {et~al.}(1982)\citenamefont {Hamer},
  \citenamefont {Kogut}, \citenamefont {Crewther},\ and\ \citenamefont
  {Mazzolini}}]{Hamer1982}%
  \BibitemOpen
  \bibfield  {author} {\bibinfo {author} {\bibfnamefont {C.J.}\ \bibnamefont
  {Hamer}}, \bibinfo {author} {\bibfnamefont {J.}~\bibnamefont {Kogut}},
  \bibinfo {author} {\bibfnamefont {D.P.}\ \bibnamefont {Crewther}}, \ and\
  \bibinfo {author} {\bibfnamefont {M.M.}\ \bibnamefont {Mazzolini}},\
  }\bibfield  {title} {\enquote {\bibinfo {title} {The massive schwinger model
  on a lattice: Background field, chiral symmetry and the string tension},}\
  }\href {\doibase https://doi.org/10.1016/0550-3213(82)90229-2} {\bibfield
  {journal} {\bibinfo  {journal} {Nuclear Physics B}\ }\textbf {\bibinfo
  {volume} {208}},\ \bibinfo {pages} {413--438} (\bibinfo {year}
  {1982})}\BibitemShut {NoStop}%
\bibitem [{\citenamefont {Chandrasekharan}\ and\ \citenamefont
  {Wiese}(1997)}]{Chandrasekharan1997}%
  \BibitemOpen
  \bibfield  {author} {\bibinfo {author} {\bibfnamefont {S}~\bibnamefont
  {Chandrasekharan}}\ and\ \bibinfo {author} {\bibfnamefont {U.-J}\
  \bibnamefont {Wiese}},\ }\bibfield  {title} {\enquote {\bibinfo {title}
  {Quantum link models: A discrete approach to gauge theories},}\ }\href
  {\doibase https://doi.org/10.1016/S0550-3213(97)80041-7} {\bibfield
  {journal} {\bibinfo  {journal} {Nuclear Physics B}\ }\textbf {\bibinfo
  {volume} {492}},\ \bibinfo {pages} {455 -- 471} (\bibinfo {year}
  {1997})}\BibitemShut {NoStop}%
\bibitem [{\citenamefont {Wiese}(2013)}]{Wiese_review}%
  \BibitemOpen
  \bibfield  {author} {\bibinfo {author} {\bibfnamefont {U.-J.}\ \bibnamefont
  {Wiese}},\ }\bibfield  {title} {\enquote {\bibinfo {title} {Ultracold quantum
  gases and lattice systems: quantum simulation of lattice gauge theories},}\
  }\href {\doibase 10.1002/andp.201300104} {\bibfield  {journal} {\bibinfo
  {journal} {Annalen der Physik}\ }\textbf {\bibinfo {volume} {525}},\ \bibinfo
  {pages} {777--796} (\bibinfo {year} {2013})}\BibitemShut {NoStop}%
\bibitem [{\citenamefont {Yang}\ \emph {et~al.}(2016)\citenamefont {Yang},
  \citenamefont {Giri}, \citenamefont {Johanning}, \citenamefont {Wunderlich},
  \citenamefont {Zoller},\ and\ \citenamefont {Hauke}}]{Yang2016}%
  \BibitemOpen
  \bibfield  {author} {\bibinfo {author} {\bibfnamefont {Dayou}\ \bibnamefont
  {Yang}}, \bibinfo {author} {\bibfnamefont {Gouri~Shankar}\ \bibnamefont
  {Giri}}, \bibinfo {author} {\bibfnamefont {Michael}\ \bibnamefont
  {Johanning}}, \bibinfo {author} {\bibfnamefont {Christof}\ \bibnamefont
  {Wunderlich}}, \bibinfo {author} {\bibfnamefont {Peter}\ \bibnamefont
  {Zoller}}, \ and\ \bibinfo {author} {\bibfnamefont {Philipp}\ \bibnamefont
  {Hauke}},\ }\bibfield  {title} {\enquote {\bibinfo {title} {Analog quantum
  simulation of $(1+1)$-dimensional lattice qed with trapped ions},}\ }\href
  {\doibase 10.1103/PhysRevA.94.052321} {\bibfield  {journal} {\bibinfo
  {journal} {Phys. Rev. A}\ }\textbf {\bibinfo {volume} {94}},\ \bibinfo
  {pages} {052321} (\bibinfo {year} {2016})}\BibitemShut {NoStop}%
\bibitem [{\citenamefont {Kogut}\ and\ \citenamefont
  {Susskind}(1975)}]{Kogut1975}%
  \BibitemOpen
  \bibfield  {author} {\bibinfo {author} {\bibfnamefont {John}\ \bibnamefont
  {Kogut}}\ and\ \bibinfo {author} {\bibfnamefont {Leonard}\ \bibnamefont
  {Susskind}},\ }\bibfield  {title} {\enquote {\bibinfo {title} {Hamiltonian
  formulation of wilson's lattice gauge theories},}\ }\href {\doibase
  10.1103/PhysRevD.11.395} {\bibfield  {journal} {\bibinfo  {journal} {Phys.
  Rev. D}\ }\textbf {\bibinfo {volume} {11}},\ \bibinfo {pages} {395--408}
  (\bibinfo {year} {1975})}\BibitemShut {NoStop}%
\bibitem [{\citenamefont {Buyens}\ \emph {et~al.}(2017)\citenamefont {Buyens},
  \citenamefont {Montangero}, \citenamefont {Haegeman}, \citenamefont
  {Verstraete},\ and\ \citenamefont {Van~Acoleyen}}]{Buyens2017}%
  \BibitemOpen
  \bibfield  {author} {\bibinfo {author} {\bibfnamefont {Boye}\ \bibnamefont
  {Buyens}}, \bibinfo {author} {\bibfnamefont {Simone}\ \bibnamefont
  {Montangero}}, \bibinfo {author} {\bibfnamefont {Jutho}\ \bibnamefont
  {Haegeman}}, \bibinfo {author} {\bibfnamefont {Frank}\ \bibnamefont
  {Verstraete}}, \ and\ \bibinfo {author} {\bibfnamefont {Karel}\ \bibnamefont
  {Van~Acoleyen}},\ }\bibfield  {title} {\enquote {\bibinfo {title}
  {Finite-representation approximation of lattice gauge theories at the
  continuum limit with tensor networks},}\ }\href {\doibase
  10.1103/PhysRevD.95.094509} {\bibfield  {journal} {\bibinfo  {journal} {Phys.
  Rev. D}\ }\textbf {\bibinfo {volume} {95}},\ \bibinfo {pages} {094509}
  (\bibinfo {year} {2017})}\BibitemShut {NoStop}%
\bibitem [{\citenamefont {Banuls}\ \emph {et~al.}(2019)\citenamefont {Banuls},
  \citenamefont {Cichy}, \citenamefont {Cirac}, \citenamefont {Jansen},\ and\
  \citenamefont {Kühn}}]{Banuls2018}%
  \BibitemOpen
  \bibfield  {author} {\bibinfo {author} {\bibfnamefont {Mari~Carmen}\
  \bibnamefont {Banuls}}, \bibinfo {author} {\bibfnamefont {Krzysztof}\
  \bibnamefont {Cichy}}, \bibinfo {author} {\bibfnamefont {J.~Ignacio}\
  \bibnamefont {Cirac}}, \bibinfo {author} {\bibfnamefont {Karl}\ \bibnamefont
  {Jansen}}, \ and\ \bibinfo {author} {\bibfnamefont {Stefan}\ \bibnamefont
  {Kühn}},\ }\bibfield  {title} {\enquote {\bibinfo {title} {{Tensor Networks
  and their use for Lattice Gauge Theories}},}\ }\href {\doibase
  10.22323/1.334.0022} {\bibfield  {journal} {\bibinfo  {journal} {PoS}\
  }\textbf {\bibinfo {volume} {LATTICE2018}},\ \bibinfo {pages} {022} (\bibinfo
  {year} {2019})}\BibitemShut {NoStop}%
\bibitem [{\citenamefont {Ba{\~{n}}uls}\ and\ \citenamefont
  {Cichy}(2020)}]{Banuls2020}%
  \BibitemOpen
  \bibfield  {author} {\bibinfo {author} {\bibfnamefont {Mari~Carmen}\
  \bibnamefont {Ba{\~{n}}uls}}\ and\ \bibinfo {author} {\bibfnamefont
  {Krzysztof}\ \bibnamefont {Cichy}},\ }\bibfield  {title} {\enquote {\bibinfo
  {title} {Review on novel methods for lattice gauge theories},}\ }\href
  {\doibase 10.1088/1361-6633/ab6311} {\bibfield  {journal} {\bibinfo
  {journal} {Reports on Progress in Physics}\ }\textbf {\bibinfo {volume}
  {83}},\ \bibinfo {pages} {024401} (\bibinfo {year} {2020})}\BibitemShut
  {NoStop}%
\bibitem [{\citenamefont {Zache}\ \emph {et~al.}(2021)\citenamefont {Zache},
  \citenamefont {Damme}, \citenamefont {Halimeh}, \citenamefont {Hauke},\ and\
  \citenamefont {Banerjee}}]{Zache2021achieving}%
  \BibitemOpen
  \bibfield  {author} {\bibinfo {author} {\bibfnamefont {Torsten~V.}\
  \bibnamefont {Zache}}, \bibinfo {author} {\bibfnamefont {Maarten~Van}\
  \bibnamefont {Damme}}, \bibinfo {author} {\bibfnamefont {Jad~C.}\
  \bibnamefont {Halimeh}}, \bibinfo {author} {\bibfnamefont {Philipp}\
  \bibnamefont {Hauke}}, \ and\ \bibinfo {author} {\bibfnamefont {Debasish}\
  \bibnamefont {Banerjee}},\ }\bibfield  {title} {\enquote {\bibinfo {title}
  {Achieving the continuum limit of quantum link lattice gauge theories on
  quantum devices},}\ }\href@noop {} {\bibfield  {journal} {\bibinfo  {journal}
  {arXiv preprint}\ } (\bibinfo {year} {2021})},\ \Eprint
  {http://arxiv.org/abs/2104.00025} {arXiv:2104.00025 [hep-lat]} \BibitemShut
  {NoStop}%
\bibitem [{\citenamefont {Halimeh}\ \emph
  {et~al.}(2021{\natexlab{b}})\citenamefont {Halimeh}, \citenamefont {Damme},
  \citenamefont {Zache}, \citenamefont {Banerjee},\ and\ \citenamefont
  {Hauke}}]{Halimeh2021achieving}%
  \BibitemOpen
  \bibfield  {author} {\bibinfo {author} {\bibfnamefont {Jad~C.}\ \bibnamefont
  {Halimeh}}, \bibinfo {author} {\bibfnamefont {Maarten~Van}\ \bibnamefont
  {Damme}}, \bibinfo {author} {\bibfnamefont {Torsten~V.}\ \bibnamefont
  {Zache}}, \bibinfo {author} {\bibfnamefont {Debasish}\ \bibnamefont
  {Banerjee}}, \ and\ \bibinfo {author} {\bibfnamefont {Philipp}\ \bibnamefont
  {Hauke}},\ }\bibfield  {title} {\enquote {\bibinfo {title} {Achieving the
  quantum field theory limit in far-from-equilibrium quantum link models},}\
  }\href@noop {} {\bibfield  {journal} {\bibinfo  {journal} {arXiv preprint}\ }
  (\bibinfo {year} {2021}{\natexlab{b}})},\ \Eprint
  {http://arxiv.org/abs/2112.04501} {arXiv:2112.04501 [cond-mat.quant-gas]}
  \BibitemShut {NoStop}%
\bibitem [{\citenamefont {Banerjee}\ \emph {et~al.}(2012)\citenamefont
  {Banerjee}, \citenamefont {Dalmonte}, \citenamefont {Müller}, \citenamefont
  {Rico}, \citenamefont {Stebler}, \citenamefont {Wiese},\ and\ \citenamefont
  {Zoller}}]{Banerjee2012}%
  \BibitemOpen
  \bibfield  {author} {\bibinfo {author} {\bibfnamefont {D.}~\bibnamefont
  {Banerjee}}, \bibinfo {author} {\bibfnamefont {M.}~\bibnamefont {Dalmonte}},
  \bibinfo {author} {\bibfnamefont {M.}~\bibnamefont {Müller}}, \bibinfo
  {author} {\bibfnamefont {E.}~\bibnamefont {Rico}}, \bibinfo {author}
  {\bibfnamefont {P.}~\bibnamefont {Stebler}}, \bibinfo {author} {\bibfnamefont
  {U.-J.}\ \bibnamefont {Wiese}}, \ and\ \bibinfo {author} {\bibfnamefont
  {P.}~\bibnamefont {Zoller}},\ }\bibfield  {title} {\enquote {\bibinfo {title}
  {Atomic quantum simulation of dynamical gauge fields coupled to fermionic
  matter: From string breaking to evolution after a quench},}\ }\href {\doibase
  10.1103/physrevlett.109.175302} {\bibfield  {journal} {\bibinfo  {journal}
  {Physical Review Letters}\ }\textbf {\bibinfo {volume} {109}} (\bibinfo
  {year} {2012}),\ 10.1103/physrevlett.109.175302}\BibitemShut {NoStop}%
\bibitem [{\citenamefont {Su}\ \emph {et~al.}(2022)\citenamefont {Su},
  \citenamefont {Sun}, \citenamefont {Hudomal}, \citenamefont {Desaules},
  \citenamefont {Zhou}, \citenamefont {Yang}, \citenamefont {Halimeh},
  \citenamefont {Yuan}, \citenamefont {Papi{\'{c}}},\ and\ \citenamefont
  {Pan}}]{Su2022}%
  \BibitemOpen
  \bibfield  {author} {\bibinfo {author} {\bibfnamefont {Guo-Xian}\
  \bibnamefont {Su}}, \bibinfo {author} {\bibfnamefont {Hui}\ \bibnamefont
  {Sun}}, \bibinfo {author} {\bibfnamefont {Ana}\ \bibnamefont {Hudomal}},
  \bibinfo {author} {\bibfnamefont {Jean-Yves}\ \bibnamefont {Desaules}},
  \bibinfo {author} {\bibfnamefont {Zhao-Yu}\ \bibnamefont {Zhou}}, \bibinfo
  {author} {\bibfnamefont {Bing}\ \bibnamefont {Yang}}, \bibinfo {author}
  {\bibfnamefont {Jad~C.}\ \bibnamefont {Halimeh}}, \bibinfo {author}
  {\bibfnamefont {Zhen-Sheng}\ \bibnamefont {Yuan}}, \bibinfo {author}
  {\bibfnamefont {Zlatko}\ \bibnamefont {Papi{\'{c}}}}, \ and\ \bibinfo
  {author} {\bibfnamefont {Jian-Wei}\ \bibnamefont {Pan}},\ }\bibfield  {title}
  {\enquote {\bibinfo {title} {{Observation of unconventional many-body
  scarring in a quantum simulator}},}\ }\href@noop {} {\bibfield  {journal}
  {\bibinfo  {journal} {arXiv preprint}\ } (\bibinfo {year} {2022})},\ \Eprint
  {http://arxiv.org/abs/2201.00821} {arXiv:2201.00821} \BibitemShut {NoStop}%
\bibitem [{\citenamefont {Turner}\ \emph {et~al.}(2018)\citenamefont {Turner},
  \citenamefont {Michailidis}, \citenamefont {Abanin}, \citenamefont {Serbyn},\
  and\ \citenamefont {Papi{\'c}}}]{Turner2018}%
  \BibitemOpen
  \bibfield  {author} {\bibinfo {author} {\bibfnamefont {C.~J.}\ \bibnamefont
  {Turner}}, \bibinfo {author} {\bibfnamefont {A.~A.}\ \bibnamefont
  {Michailidis}}, \bibinfo {author} {\bibfnamefont {D.~A.}\ \bibnamefont
  {Abanin}}, \bibinfo {author} {\bibfnamefont {M.}~\bibnamefont {Serbyn}}, \
  and\ \bibinfo {author} {\bibfnamefont {Z.}~\bibnamefont {Papi{\'c}}},\
  }\bibfield  {title} {\enquote {\bibinfo {title} {Weak ergodicity breaking
  from quantum many-body scars},}\ }\href {\doibase 10.1038/s41567-018-0137-5}
  {\bibfield  {journal} {\bibinfo  {journal} {Nature Physics}\ }\textbf
  {\bibinfo {volume} {14}},\ \bibinfo {pages} {745--749} (\bibinfo {year}
  {2018})}\BibitemShut {NoStop}%
\bibitem [{\citenamefont {{Desaules}}\ \emph {et~al.}(2022)\citenamefont
  {{Desaules}}, \citenamefont {{Banerjee}}, \citenamefont {{Hudomal}},
  \citenamefont {{Papi{\'c}}}, \citenamefont {{Sen}},\ and\ \citenamefont
  {{Halimeh}}}]{Desaules2022weak}%
  \BibitemOpen
  \bibfield  {author} {\bibinfo {author} {\bibfnamefont {Jean-Yves}\
  \bibnamefont {{Desaules}}}, \bibinfo {author} {\bibfnamefont {Debasish}\
  \bibnamefont {{Banerjee}}}, \bibinfo {author} {\bibfnamefont {Ana}\
  \bibnamefont {{Hudomal}}}, \bibinfo {author} {\bibfnamefont {Zlatko}\
  \bibnamefont {{Papi{\'c}}}}, \bibinfo {author} {\bibfnamefont {Arnab}\
  \bibnamefont {{Sen}}}, \ and\ \bibinfo {author} {\bibfnamefont {Jad~C.}\
  \bibnamefont {{Halimeh}}},\ }\bibfield  {title} {\enquote {\bibinfo {title}
  {{Weak Ergodicity Breaking in the Schwinger Model}},}\ }\href@noop {}
  {\bibfield  {journal} {\bibinfo  {journal} {arXiv preprint}\ } (\bibinfo
  {year} {2022})},\ \Eprint {http://arxiv.org/abs/2203.08830} {arXiv:2203.08830
  [cond-mat.str-el]} \BibitemShut {NoStop}%
\bibitem [{\citenamefont {Desaules}\ \emph {et~al.}(2022)\citenamefont
  {Desaules}, \citenamefont {Hudomal}, \citenamefont {Banerjee}, \citenamefont
  {Sen}, \citenamefont {Papić},\ and\ \citenamefont
  {Halimeh}}]{Desaules2022prominent}%
  \BibitemOpen
  \bibfield  {author} {\bibinfo {author} {\bibfnamefont {Jean-Yves}\
  \bibnamefont {Desaules}}, \bibinfo {author} {\bibfnamefont {Ana}\
  \bibnamefont {Hudomal}}, \bibinfo {author} {\bibfnamefont {Debasish}\
  \bibnamefont {Banerjee}}, \bibinfo {author} {\bibfnamefont {Arnab}\
  \bibnamefont {Sen}}, \bibinfo {author} {\bibfnamefont {Zlatko}\ \bibnamefont
  {Papić}}, \ and\ \bibinfo {author} {\bibfnamefont {Jad~C.}\ \bibnamefont
  {Halimeh}},\ }\bibfield  {title} {\enquote {\bibinfo {title} {Prominent
  quantum many-body scars in a truncated schwinger model},}\ }\href {\doibase
  10.48550/ARXIV.2204.01745} {\  (\bibinfo {year} {2022}),\
  10.48550/ARXIV.2204.01745}\BibitemShut {NoStop}%
\bibitem [{\citenamefont {Lang}\ \emph {et~al.}(2022)\citenamefont {Lang},
  \citenamefont {Hauke}, \citenamefont {Knolle}, \citenamefont {Grusdt},\ and\
  \citenamefont {Halimeh}}]{Lang2022SGP}%
  \BibitemOpen
  \bibfield  {author} {\bibinfo {author} {\bibfnamefont {Haifeng}\ \bibnamefont
  {Lang}}, \bibinfo {author} {\bibfnamefont {Philipp}\ \bibnamefont {Hauke}},
  \bibinfo {author} {\bibfnamefont {Johannes}\ \bibnamefont {Knolle}}, \bibinfo
  {author} {\bibfnamefont {Fabian}\ \bibnamefont {Grusdt}}, \ and\ \bibinfo
  {author} {\bibfnamefont {Jad~C.}\ \bibnamefont {Halimeh}},\ }\bibfield
  {title} {\enquote {\bibinfo {title} {Disorder-free localization with stark
  gauge protection},}\ }\href {\doibase 10.48550/ARXIV.2203.01338} {\
  (\bibinfo {year} {2022}),\ 10.48550/ARXIV.2203.01338}\BibitemShut {NoStop}%
\bibitem [{\citenamefont {Yang}\ \emph
  {et~al.}(2020{\natexlab{b}})\citenamefont {Yang}, \citenamefont {Sun},
  \citenamefont {Huang}, \citenamefont {Wang}, \citenamefont {Deng},
  \citenamefont {Dai}, \citenamefont {Yuan},\ and\ \citenamefont
  {Pan}}]{Yang:2020science}%
  \BibitemOpen
  \bibfield  {author} {\bibinfo {author} {\bibfnamefont {Bing}\ \bibnamefont
  {Yang}}, \bibinfo {author} {\bibfnamefont {Hui}\ \bibnamefont {Sun}},
  \bibinfo {author} {\bibfnamefont {Chun-Jiong}\ \bibnamefont {Huang}},
  \bibinfo {author} {\bibfnamefont {Han-Yi}\ \bibnamefont {Wang}}, \bibinfo
  {author} {\bibfnamefont {Youjin}\ \bibnamefont {Deng}}, \bibinfo {author}
  {\bibfnamefont {Han-Ning}\ \bibnamefont {Dai}}, \bibinfo {author}
  {\bibfnamefont {Zhen-Sheng}\ \bibnamefont {Yuan}}, \ and\ \bibinfo {author}
  {\bibfnamefont {Jian-Wei}\ \bibnamefont {Pan}},\ }\bibfield  {title}
  {\enquote {\bibinfo {title} {Cooling and entangling ultracold atoms in
  optical lattices},}\ }\href {\doibase 10.1126/science.aaz6801} {\bibfield
  {journal} {\bibinfo  {journal} {Science}\ }\textbf {\bibinfo {volume}
  {369}},\ \bibinfo {pages} {550--553} (\bibinfo {year}
  {2020}{\natexlab{b}})}\BibitemShut {NoStop}%
\bibitem [{\citenamefont {Yang}\ \emph {et~al.}(2017)\citenamefont {Yang},
  \citenamefont {Dai}, \citenamefont {Sun}, \citenamefont {Reingruber},
  \citenamefont {Yuan},\ and\ \citenamefont {Pan}}]{Yang:2017pra}%
  \BibitemOpen
  \bibfield  {author} {\bibinfo {author} {\bibfnamefont {Bing}\ \bibnamefont
  {Yang}}, \bibinfo {author} {\bibfnamefont {Han-Ning}\ \bibnamefont {Dai}},
  \bibinfo {author} {\bibfnamefont {Hui}\ \bibnamefont {Sun}}, \bibinfo
  {author} {\bibfnamefont {Andreas}\ \bibnamefont {Reingruber}}, \bibinfo
  {author} {\bibfnamefont {Zhen-Sheng}\ \bibnamefont {Yuan}}, \ and\ \bibinfo
  {author} {\bibfnamefont {Jian-Wei}\ \bibnamefont {Pan}},\ }\bibfield  {title}
  {\enquote {\bibinfo {title} {Spin-dependent optical superlattice},}\ }\href
  {\doibase 10.1103/PhysRevA.96.011602} {\bibfield  {journal} {\bibinfo
  {journal} {Phys. Rev. A}\ }\textbf {\bibinfo {volume} {96}},\ \bibinfo
  {pages} {011602} (\bibinfo {year} {2017})}\BibitemShut {NoStop}%
\bibitem [{\citenamefont {Schmitteckert}(2004)}]{Schmitteckert2004}%
  \BibitemOpen
  \bibfield  {author} {\bibinfo {author} {\bibfnamefont {Peter}\ \bibnamefont
  {Schmitteckert}},\ }\bibfield  {title} {\enquote {\bibinfo {title}
  {Nonequilibrium electron transport using the density matrix renormalization
  group method},}\ }\href {\doibase 10.1103/PhysRevB.70.121302} {\bibfield
  {journal} {\bibinfo  {journal} {Phys. Rev. B}\ }\textbf {\bibinfo {volume}
  {70}},\ \bibinfo {pages} {121302} (\bibinfo {year} {2004})}\BibitemShut
  {NoStop}%
\bibitem [{\citenamefont {Feiguin}\ and\ \citenamefont
  {White}(2005)}]{Feiguin2005}%
  \BibitemOpen
  \bibfield  {author} {\bibinfo {author} {\bibfnamefont {Adrian~E.}\
  \bibnamefont {Feiguin}}\ and\ \bibinfo {author} {\bibfnamefont {Steven~R.}\
  \bibnamefont {White}},\ }\bibfield  {title} {\enquote {\bibinfo {title}
  {Time-step targeting methods for real-time dynamics using the density matrix
  renormalization group},}\ }\href {\doibase 10.1103/PhysRevB.72.020404}
  {\bibfield  {journal} {\bibinfo  {journal} {Phys. Rev. B}\ }\textbf {\bibinfo
  {volume} {72}},\ \bibinfo {pages} {020404} (\bibinfo {year}
  {2005})}\BibitemShut {NoStop}%
\bibitem [{\citenamefont {Garc{\'{\i}}a-Ripoll}(2006)}]{Ripoll2006}%
  \BibitemOpen
  \bibfield  {author} {\bibinfo {author} {\bibfnamefont {Juan~Jos{\'{e}}}\
  \bibnamefont {Garc{\'{\i}}a-Ripoll}},\ }\bibfield  {title} {\enquote
  {\bibinfo {title} {Time evolution of matrix product states},}\ }\href
  {\doibase 10.1088/1367-2630/8/12/305} {\bibfield  {journal} {\bibinfo
  {journal} {New Journal of Physics}\ }\textbf {\bibinfo {volume} {8}},\
  \bibinfo {pages} {305--305} (\bibinfo {year} {2006})}\BibitemShut {NoStop}%
\bibitem [{\citenamefont {McCulloch}(2007)}]{McCulloch2007}%
  \BibitemOpen
  \bibfield  {author} {\bibinfo {author} {\bibfnamefont {Ian~P}\ \bibnamefont
  {McCulloch}},\ }\bibfield  {title} {\enquote {\bibinfo {title} {From
  density-matrix renormalization group to matrix product states},}\ }\href
  {\doibase 10.1088/1742-5468/2007/10/p10014} {\bibfield  {journal} {\bibinfo
  {journal} {Journal of Statistical Mechanics: Theory and Experiment}\ }\textbf
  {\bibinfo {volume} {2007}},\ \bibinfo {pages} {P10014--P10014} (\bibinfo
  {year} {2007})}\BibitemShut {NoStop}%
\bibitem [{\citenamefont {Schollwöck}(2011)}]{Uli_review}%
  \BibitemOpen
  \bibfield  {author} {\bibinfo {author} {\bibfnamefont {Ulrich}\ \bibnamefont
  {Schollwöck}},\ }\bibfield  {title} {\enquote {\bibinfo {title} {The
  density-matrix renormalization group in the age of matrix product states},}\
  }\href {\doibase https://doi.org/10.1016/j.aop.2010.09.012} {\bibfield
  {journal} {\bibinfo  {journal} {Annals of Physics}\ }\textbf {\bibinfo
  {volume} {326}},\ \bibinfo {pages} {96--192} (\bibinfo {year} {2011})},\
  \bibinfo {note} {january 2011 Special Issue}\BibitemShut {NoStop}%
\bibitem [{\citenamefont {Paeckel}\ \emph {et~al.}(2019)\citenamefont
  {Paeckel}, \citenamefont {Köhler}, \citenamefont {Swoboda}, \citenamefont
  {Manmana}, \citenamefont {Schollwöck},\ and\ \citenamefont
  {Hubig}}]{Paeckel_review}%
  \BibitemOpen
  \bibfield  {author} {\bibinfo {author} {\bibfnamefont {Sebastian}\
  \bibnamefont {Paeckel}}, \bibinfo {author} {\bibfnamefont {Thomas}\
  \bibnamefont {Köhler}}, \bibinfo {author} {\bibfnamefont {Andreas}\
  \bibnamefont {Swoboda}}, \bibinfo {author} {\bibfnamefont {Salvatore~R.}\
  \bibnamefont {Manmana}}, \bibinfo {author} {\bibfnamefont {Ulrich}\
  \bibnamefont {Schollwöck}}, \ and\ \bibinfo {author} {\bibfnamefont
  {Claudius}\ \bibnamefont {Hubig}},\ }\bibfield  {title} {\enquote {\bibinfo
  {title} {Time-evolution methods for matrix-product states},}\ }\href
  {\doibase https://doi.org/10.1016/j.aop.2019.167998} {\bibfield  {journal}
  {\bibinfo  {journal} {Annals of Physics}\ }\textbf {\bibinfo {volume}
  {411}},\ \bibinfo {pages} {167998} (\bibinfo {year} {2019})}\BibitemShut
  {NoStop}%
\bibitem [{\citenamefont {Halimeh}\ \emph
  {et~al.}(2020{\natexlab{a}})\citenamefont {Halimeh}, \citenamefont {Ott},
  \citenamefont {McCulloch}, \citenamefont {Yang},\ and\ \citenamefont
  {Hauke}}]{Halimeh2020d}%
  \BibitemOpen
  \bibfield  {author} {\bibinfo {author} {\bibfnamefont {Jad~C.}\ \bibnamefont
  {Halimeh}}, \bibinfo {author} {\bibfnamefont {Robert}\ \bibnamefont {Ott}},
  \bibinfo {author} {\bibfnamefont {Ian~P.}\ \bibnamefont {McCulloch}},
  \bibinfo {author} {\bibfnamefont {Bing}\ \bibnamefont {Yang}}, \ and\
  \bibinfo {author} {\bibfnamefont {Philipp}\ \bibnamefont {Hauke}},\
  }\bibfield  {title} {\enquote {\bibinfo {title} {Robustness of
  gauge-invariant dynamics against defects in ultracold-atom gauge theories},}\
  }\href {\doibase 10.1103/PhysRevResearch.2.033361} {\bibfield  {journal}
  {\bibinfo  {journal} {Phys. Rev. Research}\ }\textbf {\bibinfo {volume}
  {2}},\ \bibinfo {pages} {033361} (\bibinfo {year}
  {2020}{\natexlab{a}})}\BibitemShut {NoStop}%
\bibitem [{\citenamefont {D'Emidio}\ \emph {et~al.}(2021)\citenamefont
  {D'Emidio}, \citenamefont {Eberharter},\ and\ \citenamefont
  {Läuchli}}]{demidio2021}%
  \BibitemOpen
  \bibfield  {author} {\bibinfo {author} {\bibfnamefont {Jonathan}\
  \bibnamefont {D'Emidio}}, \bibinfo {author} {\bibfnamefont {Alexander~A.}\
  \bibnamefont {Eberharter}}, \ and\ \bibinfo {author} {\bibfnamefont
  {Andreas~M.}\ \bibnamefont {Läuchli}},\ }\bibfield  {title} {\enquote
  {\bibinfo {title} {Diagnosing weakly first-order phase transitions by
  coupling to order parameters},}\ }\href {\doibase 10.48550/ARXIV.2106.15462}
  {\  (\bibinfo {year} {2021}),\ 10.48550/ARXIV.2106.15462}\BibitemShut
  {NoStop}%
\bibitem [{\citenamefont {Moudgalya}\ \emph {et~al.}(2018)\citenamefont
  {Moudgalya}, \citenamefont {Rachel}, \citenamefont {Bernevig},\ and\
  \citenamefont {Regnault}}]{Moudgalya2018}%
  \BibitemOpen
  \bibfield  {author} {\bibinfo {author} {\bibfnamefont {Sanjay}\ \bibnamefont
  {Moudgalya}}, \bibinfo {author} {\bibfnamefont {Stephan}\ \bibnamefont
  {Rachel}}, \bibinfo {author} {\bibfnamefont {B.~Andrei}\ \bibnamefont
  {Bernevig}}, \ and\ \bibinfo {author} {\bibfnamefont {Nicolas}\ \bibnamefont
  {Regnault}},\ }\bibfield  {title} {\enquote {\bibinfo {title} {Exact excited
  states of nonintegrable models},}\ }\href {\doibase
  10.1103/PhysRevB.98.235155} {\bibfield  {journal} {\bibinfo  {journal} {Phys.
  Rev. B}\ }\textbf {\bibinfo {volume} {98}},\ \bibinfo {pages} {235155}
  (\bibinfo {year} {2018})}\BibitemShut {NoStop}%
\bibitem [{\citenamefont {Smith}\ \emph {et~al.}(2017)\citenamefont {Smith},
  \citenamefont {Knolle}, \citenamefont {Kovrizhin},\ and\ \citenamefont
  {Moessner}}]{Smith2017}%
  \BibitemOpen
  \bibfield  {author} {\bibinfo {author} {\bibfnamefont {A.}~\bibnamefont
  {Smith}}, \bibinfo {author} {\bibfnamefont {J.}~\bibnamefont {Knolle}},
  \bibinfo {author} {\bibfnamefont {D.~L.}\ \bibnamefont {Kovrizhin}}, \ and\
  \bibinfo {author} {\bibfnamefont {R.}~\bibnamefont {Moessner}},\ }\bibfield
  {title} {\enquote {\bibinfo {title} {Disorder-free localization},}\ }\href
  {\doibase 10.1103/PhysRevLett.118.266601} {\bibfield  {journal} {\bibinfo
  {journal} {Phys. Rev. Lett.}\ }\textbf {\bibinfo {volume} {118}},\ \bibinfo
  {pages} {266601} (\bibinfo {year} {2017})}\BibitemShut {NoStop}%
\bibitem [{\citenamefont {Brenes}\ \emph {et~al.}(2018)\citenamefont {Brenes},
  \citenamefont {Dalmonte}, \citenamefont {Heyl},\ and\ \citenamefont
  {Scardicchio}}]{Brenes2018}%
  \BibitemOpen
  \bibfield  {author} {\bibinfo {author} {\bibfnamefont {Marlon}\ \bibnamefont
  {Brenes}}, \bibinfo {author} {\bibfnamefont {Marcello}\ \bibnamefont
  {Dalmonte}}, \bibinfo {author} {\bibfnamefont {Markus}\ \bibnamefont {Heyl}},
  \ and\ \bibinfo {author} {\bibfnamefont {Antonello}\ \bibnamefont
  {Scardicchio}},\ }\bibfield  {title} {\enquote {\bibinfo {title} {Many-body
  localization dynamics from gauge invariance},}\ }\href {\doibase
  10.1103/PhysRevLett.120.030601} {\bibfield  {journal} {\bibinfo  {journal}
  {Phys. Rev. Lett.}\ }\textbf {\bibinfo {volume} {120}},\ \bibinfo {pages}
  {030601} (\bibinfo {year} {2018})}\BibitemShut {NoStop}%
\bibitem [{\citenamefont {Berges}\ \emph {et~al.}(2021)\citenamefont {Berges},
  \citenamefont {Heller}, \citenamefont {Mazeliauskas},\ and\ \citenamefont
  {Venugopalan}}]{Berges_review}%
  \BibitemOpen
  \bibfield  {author} {\bibinfo {author} {\bibfnamefont {J\"urgen}\
  \bibnamefont {Berges}}, \bibinfo {author} {\bibfnamefont {Michal~P.}\
  \bibnamefont {Heller}}, \bibinfo {author} {\bibfnamefont {Aleksas}\
  \bibnamefont {Mazeliauskas}}, \ and\ \bibinfo {author} {\bibfnamefont {Raju}\
  \bibnamefont {Venugopalan}},\ }\bibfield  {title} {\enquote {\bibinfo {title}
  {Qcd thermalization: Ab initio approaches and interdisciplinary
  connections},}\ }\href {\doibase 10.1103/RevModPhys.93.035003} {\bibfield
  {journal} {\bibinfo  {journal} {Rev. Mod. Phys.}\ }\textbf {\bibinfo {volume}
  {93}},\ \bibinfo {pages} {035003} (\bibinfo {year} {2021})}\BibitemShut
  {NoStop}%
\bibitem [{\citenamefont {Hebenstreit}\ \emph
  {et~al.}(2013{\natexlab{a}})\citenamefont {Hebenstreit}, \citenamefont
  {Berges},\ and\ \citenamefont {Gelfand}}]{Hebenstreit2013}%
  \BibitemOpen
  \bibfield  {author} {\bibinfo {author} {\bibfnamefont {F.}~\bibnamefont
  {Hebenstreit}}, \bibinfo {author} {\bibfnamefont {J.}~\bibnamefont {Berges}},
  \ and\ \bibinfo {author} {\bibfnamefont {D.}~\bibnamefont {Gelfand}},\
  }\bibfield  {title} {\enquote {\bibinfo {title} {Real-time dynamics of string
  breaking},}\ }\href {\doibase 10.1103/PhysRevLett.111.201601} {\bibfield
  {journal} {\bibinfo  {journal} {Phys. Rev. Lett.}\ }\textbf {\bibinfo
  {volume} {111}},\ \bibinfo {pages} {201601} (\bibinfo {year}
  {2013}{\natexlab{a}})}\BibitemShut {NoStop}%
\bibitem [{\citenamefont {Serbyn}\ \emph {et~al.}(2021)\citenamefont {Serbyn},
  \citenamefont {Abanin},\ and\ \citenamefont {Papi{\'c}}}]{Serbyn2020}%
  \BibitemOpen
  \bibfield  {author} {\bibinfo {author} {\bibfnamefont {Maksym}\ \bibnamefont
  {Serbyn}}, \bibinfo {author} {\bibfnamefont {Dmitry~A.}\ \bibnamefont
  {Abanin}}, \ and\ \bibinfo {author} {\bibfnamefont {Zlatko}\ \bibnamefont
  {Papi{\'c}}},\ }\bibfield  {title} {\enquote {\bibinfo {title} {Quantum
  many-body scars and weak breaking of ergodicity},}\ }\href {\doibase
  10.1038/s41567-021-01230-2} {\bibfield  {journal} {\bibinfo  {journal}
  {Nature Physics}\ }\textbf {\bibinfo {volume} {17}},\ \bibinfo {pages}
  {675--685} (\bibinfo {year} {2021})}\BibitemShut {NoStop}%
\bibitem [{\citenamefont {Moudgalya}\ \emph {et~al.}(2021)\citenamefont
  {Moudgalya}, \citenamefont {Bernevig},\ and\ \citenamefont
  {Regnault}}]{MoudgalyaReview}%
  \BibitemOpen
  \bibfield  {author} {\bibinfo {author} {\bibfnamefont {Sanjay}\ \bibnamefont
  {Moudgalya}}, \bibinfo {author} {\bibfnamefont {B.~Andrei}\ \bibnamefont
  {Bernevig}}, \ and\ \bibinfo {author} {\bibfnamefont {Nicolas}\ \bibnamefont
  {Regnault}},\ }\bibfield  {title} {\enquote {\bibinfo {title} {Quantum
  many-body scars and {Hilbert} space fragmentation: A review of exact
  results},}\ }\href@noop {} {\bibfield  {journal} {\bibinfo  {journal} {arXiv
  preprint}\ } (\bibinfo {year} {2021})},\ \Eprint
  {http://arxiv.org/abs/2109.00548} {arXiv:2109.00548 [cond-mat.str-el]}
  \BibitemShut {NoStop}%
\bibitem [{\citenamefont {Shiraishi}\ and\ \citenamefont
  {Mori}(2017)}]{ShiraishiMori}%
  \BibitemOpen
  \bibfield  {author} {\bibinfo {author} {\bibfnamefont {Naoto}\ \bibnamefont
  {Shiraishi}}\ and\ \bibinfo {author} {\bibfnamefont {Takashi}\ \bibnamefont
  {Mori}},\ }\bibfield  {title} {\enquote {\bibinfo {title} {Systematic
  construction of counterexamples to the eigenstate thermalization
  hypothesis},}\ }\href {\doibase 10.1103/PhysRevLett.119.030601} {\bibfield
  {journal} {\bibinfo  {journal} {Phys. Rev. Lett.}\ }\textbf {\bibinfo
  {volume} {119}},\ \bibinfo {pages} {030601} (\bibinfo {year}
  {2017})}\BibitemShut {NoStop}%
\bibitem [{\citenamefont {Lin}\ and\ \citenamefont
  {Motrunich}(2019)}]{lin2018exact}%
  \BibitemOpen
  \bibfield  {author} {\bibinfo {author} {\bibfnamefont {Cheng-Ju}\
  \bibnamefont {Lin}}\ and\ \bibinfo {author} {\bibfnamefont {Olexei~I.}\
  \bibnamefont {Motrunich}},\ }\bibfield  {title} {\enquote {\bibinfo {title}
  {Exact quantum many-body scar states in the {Rydberg}-blockaded atom
  chain},}\ }\href {\doibase 10.1103/PhysRevLett.122.173401} {\bibfield
  {journal} {\bibinfo  {journal} {Phys. Rev. Lett.}\ }\textbf {\bibinfo
  {volume} {122}},\ \bibinfo {pages} {173401} (\bibinfo {year}
  {2019})}\BibitemShut {NoStop}%
\bibitem [{\citenamefont {Christandl}\ \emph {et~al.}(2004)\citenamefont
  {Christandl}, \citenamefont {Datta}, \citenamefont {Ekert},\ and\
  \citenamefont {Landahl}}]{Christandl2004}%
  \BibitemOpen
  \bibfield  {author} {\bibinfo {author} {\bibfnamefont {Matthias}\
  \bibnamefont {Christandl}}, \bibinfo {author} {\bibfnamefont {Nilanjana}\
  \bibnamefont {Datta}}, \bibinfo {author} {\bibfnamefont {Artur}\ \bibnamefont
  {Ekert}}, \ and\ \bibinfo {author} {\bibfnamefont {Andrew~J.}\ \bibnamefont
  {Landahl}},\ }\bibfield  {title} {\enquote {\bibinfo {title} {Perfect state
  transfer in quantum spin networks},}\ }\href {\doibase
  10.1103/PhysRevLett.92.187902} {\bibfield  {journal} {\bibinfo  {journal}
  {Phys. Rev. Lett.}\ }\textbf {\bibinfo {volume} {92}},\ \bibinfo {pages}
  {187902} (\bibinfo {year} {2004})}\BibitemShut {NoStop}%
\bibitem [{\citenamefont {Halimeh}\ \emph {et~al.}(2017)\citenamefont
  {Halimeh}, \citenamefont {Zauner-Stauber}, \citenamefont {McCulloch},
  \citenamefont {de~Vega}, \citenamefont {Schollw\"ock},\ and\ \citenamefont
  {Kastner}}]{Halimeh2017}%
  \BibitemOpen
  \bibfield  {author} {\bibinfo {author} {\bibfnamefont {Jad~C.}\ \bibnamefont
  {Halimeh}}, \bibinfo {author} {\bibfnamefont {Valentin}\ \bibnamefont
  {Zauner-Stauber}}, \bibinfo {author} {\bibfnamefont {Ian~P.}\ \bibnamefont
  {McCulloch}}, \bibinfo {author} {\bibfnamefont {In\'es}\ \bibnamefont
  {de~Vega}}, \bibinfo {author} {\bibfnamefont {Ulrich}\ \bibnamefont
  {Schollw\"ock}}, \ and\ \bibinfo {author} {\bibfnamefont {Michael}\
  \bibnamefont {Kastner}},\ }\bibfield  {title} {\enquote {\bibinfo {title}
  {Prethermalization and persistent order in the absence of a thermal phase
  transition},}\ }\href {\doibase 10.1103/PhysRevB.95.024302} {\bibfield
  {journal} {\bibinfo  {journal} {Phys. Rev. B}\ }\textbf {\bibinfo {volume}
  {95}},\ \bibinfo {pages} {024302} (\bibinfo {year} {2017})}\BibitemShut
  {NoStop}%
\bibitem [{\citenamefont {Liu}\ \emph {et~al.}(2019)\citenamefont {Liu},
  \citenamefont {Lundgren}, \citenamefont {Titum}, \citenamefont {Pagano},
  \citenamefont {Zhang}, \citenamefont {Monroe},\ and\ \citenamefont
  {Gorshkov}}]{Liu2019}%
  \BibitemOpen
  \bibfield  {author} {\bibinfo {author} {\bibfnamefont {Fangli}\ \bibnamefont
  {Liu}}, \bibinfo {author} {\bibfnamefont {Rex}\ \bibnamefont {Lundgren}},
  \bibinfo {author} {\bibfnamefont {Paraj}\ \bibnamefont {Titum}}, \bibinfo
  {author} {\bibfnamefont {Guido}\ \bibnamefont {Pagano}}, \bibinfo {author}
  {\bibfnamefont {Jiehang}\ \bibnamefont {Zhang}}, \bibinfo {author}
  {\bibfnamefont {Christopher}\ \bibnamefont {Monroe}}, \ and\ \bibinfo
  {author} {\bibfnamefont {Alexey~V.}\ \bibnamefont {Gorshkov}},\ }\bibfield
  {title} {\enquote {\bibinfo {title} {Confined quasiparticle dynamics in
  long-range interacting quantum spin chains},}\ }\href {\doibase
  10.1103/PhysRevLett.122.150601} {\bibfield  {journal} {\bibinfo  {journal}
  {Phys. Rev. Lett.}\ }\textbf {\bibinfo {volume} {122}},\ \bibinfo {pages}
  {150601} (\bibinfo {year} {2019})}\BibitemShut {NoStop}%
\bibitem [{\citenamefont {Halimeh}\ \emph
  {et~al.}(2020{\natexlab{b}})\citenamefont {Halimeh}, \citenamefont
  {Van~Damme}, \citenamefont {Zauner-Stauber},\ and\ \citenamefont
  {Vanderstraeten}}]{Halimeh2020quasiparticle}%
  \BibitemOpen
  \bibfield  {author} {\bibinfo {author} {\bibfnamefont {Jad~C.}\ \bibnamefont
  {Halimeh}}, \bibinfo {author} {\bibfnamefont {Maarten}\ \bibnamefont
  {Van~Damme}}, \bibinfo {author} {\bibfnamefont {Valentin}\ \bibnamefont
  {Zauner-Stauber}}, \ and\ \bibinfo {author} {\bibfnamefont {Laurens}\
  \bibnamefont {Vanderstraeten}},\ }\bibfield  {title} {\enquote {\bibinfo
  {title} {Quasiparticle origin of dynamical quantum phase transitions},}\
  }\href {\doibase 10.1103/PhysRevResearch.2.033111} {\bibfield  {journal}
  {\bibinfo  {journal} {Phys. Rev. Research}\ }\textbf {\bibinfo {volume}
  {2}},\ \bibinfo {pages} {033111} (\bibinfo {year}
  {2020}{\natexlab{b}})}\BibitemShut {NoStop}%
\bibitem [{\citenamefont {Kormos}\ \emph {et~al.}(2017)\citenamefont {Kormos},
  \citenamefont {Collura}, \citenamefont {Tak{\'a}cs},\ and\ \citenamefont
  {Calabrese}}]{Kormos2017}%
  \BibitemOpen
  \bibfield  {author} {\bibinfo {author} {\bibfnamefont {Marton}\ \bibnamefont
  {Kormos}}, \bibinfo {author} {\bibfnamefont {Mario}\ \bibnamefont {Collura}},
  \bibinfo {author} {\bibfnamefont {Gabor}\ \bibnamefont {Tak{\'a}cs}}, \ and\
  \bibinfo {author} {\bibfnamefont {Pasquale}\ \bibnamefont {Calabrese}},\
  }\bibfield  {title} {\enquote {\bibinfo {title} {Real-time confinement
  following a quantum quench to a non-integrable model},}\ }\href {\doibase
  10.1038/nphys3934} {\bibfield  {journal} {\bibinfo  {journal} {Nature
  Physics}\ }\textbf {\bibinfo {volume} {13}},\ \bibinfo {pages} {246--249}
  (\bibinfo {year} {2017})}\BibitemShut {NoStop}%
\bibitem [{\citenamefont {Halimeh}\ and\ \citenamefont
  {Hauke}(2020)}]{Halimeh2020a}%
  \BibitemOpen
  \bibfield  {author} {\bibinfo {author} {\bibfnamefont {Jad~C.}\ \bibnamefont
  {Halimeh}}\ and\ \bibinfo {author} {\bibfnamefont {Philipp}\ \bibnamefont
  {Hauke}},\ }\bibfield  {title} {\enquote {\bibinfo {title} {Reliability of
  lattice gauge theories},}\ }\href {\doibase 10.1103/PhysRevLett.125.030503}
  {\bibfield  {journal} {\bibinfo  {journal} {Phys. Rev. Lett.}\ }\textbf
  {\bibinfo {volume} {125}},\ \bibinfo {pages} {030503} (\bibinfo {year}
  {2020})}\BibitemShut {NoStop}%
\bibitem [{\citenamefont {Hebenstreit}\ \emph
  {et~al.}(2013{\natexlab{b}})\citenamefont {Hebenstreit}, \citenamefont
  {Berges},\ and\ \citenamefont {Gelfand}}]{Hebenstreit2013simulating}%
  \BibitemOpen
  \bibfield  {author} {\bibinfo {author} {\bibfnamefont {F.}~\bibnamefont
  {Hebenstreit}}, \bibinfo {author} {\bibfnamefont {J.}~\bibnamefont {Berges}},
  \ and\ \bibinfo {author} {\bibfnamefont {D.}~\bibnamefont {Gelfand}},\
  }\bibfield  {title} {\enquote {\bibinfo {title} {Simulating fermion
  production in $1\mathbf{+}1$ dimensional qed},}\ }\href {\doibase
  10.1103/PhysRevD.87.105006} {\bibfield  {journal} {\bibinfo  {journal} {Phys.
  Rev. D}\ }\textbf {\bibinfo {volume} {87}},\ \bibinfo {pages} {105006}
  (\bibinfo {year} {2013}{\natexlab{b}})}\BibitemShut {NoStop}%
\bibitem [{\citenamefont {Bakr}\ \emph {et~al.}(2009)\citenamefont {Bakr},
  \citenamefont {Gillen}, \citenamefont {Peng}, \citenamefont {F{\"o}lling},\
  and\ \citenamefont {Greiner}}]{Bakr2009}%
  \BibitemOpen
  \bibfield  {author} {\bibinfo {author} {\bibfnamefont {Waseem~S.}\
  \bibnamefont {Bakr}}, \bibinfo {author} {\bibfnamefont {Jonathon~I.}\
  \bibnamefont {Gillen}}, \bibinfo {author} {\bibfnamefont {Amy}\ \bibnamefont
  {Peng}}, \bibinfo {author} {\bibfnamefont {Simon}\ \bibnamefont
  {F{\"o}lling}}, \ and\ \bibinfo {author} {\bibfnamefont {Markus}\
  \bibnamefont {Greiner}},\ }\bibfield  {title} {\enquote {\bibinfo {title} {A
  quantum gas microscope for detecting single atoms in a {Hubbard}-regime
  optical lattice},}\ }\href {\doibase 10.1038/nature08482} {\bibfield
  {journal} {\bibinfo  {journal} {Nature}\ }\textbf {\bibinfo {volume} {462}},\
  \bibinfo {pages} {74--77} (\bibinfo {year} {2009})}\BibitemShut {NoStop}%
\bibitem [{\citenamefont {Cheng}\ \emph {et~al.}(2022)\citenamefont {Cheng},
  \citenamefont {Liu}, \citenamefont {Zheng}, \citenamefont {Zhang},\ and\
  \citenamefont {Zhai}}]{Cheng2022tunable}%
  \BibitemOpen
  \bibfield  {author} {\bibinfo {author} {\bibfnamefont {Yanting}\ \bibnamefont
  {Cheng}}, \bibinfo {author} {\bibfnamefont {Shang}\ \bibnamefont {Liu}},
  \bibinfo {author} {\bibfnamefont {Wei}\ \bibnamefont {Zheng}}, \bibinfo
  {author} {\bibfnamefont {Pengfei}\ \bibnamefont {Zhang}}, \ and\ \bibinfo
  {author} {\bibfnamefont {Hui}\ \bibnamefont {Zhai}},\ }\bibfield  {title}
  {\enquote {\bibinfo {title} {Tunable confinement-deconfinement transition in
  an ultracold atom quantum simulator},}\ }\href {\doibase
  10.48550/ARXIV.2204.06586} {\  (\bibinfo {year} {2022}),\
  10.48550/ARXIV.2204.06586}\BibitemShut {NoStop}%
\bibitem [{\citenamefont {Ba\~nuls}\ \emph {et~al.}(2017)\citenamefont
  {Ba\~nuls}, \citenamefont {Cichy}, \citenamefont {Cirac}, \citenamefont
  {Jansen},\ and\ \citenamefont {K\"uhn}}]{Banuls2017}%
  \BibitemOpen
  \bibfield  {author} {\bibinfo {author} {\bibfnamefont {Mari~Carmen}\
  \bibnamefont {Ba\~nuls}}, \bibinfo {author} {\bibfnamefont {Krzysztof}\
  \bibnamefont {Cichy}}, \bibinfo {author} {\bibfnamefont {J.~Ignacio}\
  \bibnamefont {Cirac}}, \bibinfo {author} {\bibfnamefont {Karl}\ \bibnamefont
  {Jansen}}, \ and\ \bibinfo {author} {\bibfnamefont {Stefan}\ \bibnamefont
  {K\"uhn}},\ }\bibfield  {title} {\enquote {\bibinfo {title} {Density induced
  phase transitions in the schwinger model: A study with matrix product
  states},}\ }\href {\doibase 10.1103/PhysRevLett.118.071601} {\bibfield
  {journal} {\bibinfo  {journal} {Phys. Rev. Lett.}\ }\textbf {\bibinfo
  {volume} {118}},\ \bibinfo {pages} {071601} (\bibinfo {year}
  {2017})}\BibitemShut {NoStop}%
\bibitem [{\citenamefont {Hauke}\ \emph {et~al.}(2013)\citenamefont {Hauke},
  \citenamefont {Marcos}, \citenamefont {Dalmonte},\ and\ \citenamefont
  {Zoller}}]{Hauke2013}%
  \BibitemOpen
  \bibfield  {author} {\bibinfo {author} {\bibfnamefont {P.}~\bibnamefont
  {Hauke}}, \bibinfo {author} {\bibfnamefont {D.}~\bibnamefont {Marcos}},
  \bibinfo {author} {\bibfnamefont {M.}~\bibnamefont {Dalmonte}}, \ and\
  \bibinfo {author} {\bibfnamefont {P.}~\bibnamefont {Zoller}},\ }\bibfield
  {title} {\enquote {\bibinfo {title} {Quantum simulation of a lattice
  schwinger model in a chain of trapped ions},}\ }\href {\doibase
  10.1103/PhysRevX.3.041018} {\bibfield  {journal} {\bibinfo  {journal} {Phys.
  Rev. X}\ }\textbf {\bibinfo {volume} {3}},\ \bibinfo {pages} {041018}
  (\bibinfo {year} {2013})}\BibitemShut {NoStop}%
\end{thebibliography}%
\end{document}